\documentclass[12pt,english,american]{article}
\usepackage[T1]{fontenc}
\usepackage[latin9]{inputenc}
\usepackage{geometry}
\usepackage{color}
\usepackage{babel}
\usepackage{float}
\usepackage{amsmath}
\usepackage{amssymb}
\usepackage{graphicx}
\usepackage{setspace}
\usepackage{esint}
\usepackage[authoryear]{natbib}
\usepackage[unicode=true,pdfusetitle,
 bookmarks=true,bookmarksnumbered=false,bookmarksopen=false,
 breaklinks=false,pdfborder={0 0 0},pdfborderstyle={},backref=false,colorlinks=false]
 {hyperref}

\makeatletter

\providecommand{\tabularnewline}{\\}

\usepackage{multirow}%\usepackage{babel}
\usepackage{lscape}\usepackage{rotating}\usepackage{rotating}

\usepackage{babel}
\usepackage{ragged2e}

\usepackage{chapterbib}

\makeatother

\begin{document}
\title{Sectoral Labor Mobility and Optimal Monetary Policy\thanks{{\footnotesize{}The views expressed in this paper are those of the
authors and do not necessarily represent those of the International
Monetary Fund or IMF policy or Banca d'Italia. We are grateful to
Michael Ben-Gad, Cristiano Cantore, Romain Duval, Davide Furceri,
Paul Levine, Prakash Loungani, Chris Papageorgiou, Joseph Pearlman,
Ivan Petrella, Carlo Pizzinelli, Francesco Zanetti, two anonymous
referees and participants at the 2016 IMF-OCP Policy Center-Brunel
University Workshop on Global Labor Markets, the 48$^{\text{th}}$
Money Macro and Finance Annual Conference, the ASSET 2016 Annual Meeting
and seminars at Banca d'Italia, Banco de Espa$\tilde{\text{n}}$a,
City, University of London and the IMF for useful comments.}}}
\author{Alessandro Cantelmo\thanks{{\footnotesize{}Corresponding author. Banca d'Italia. Via Nazionale
91, Rome 00184, Italy. E-mail: alessandro.cantelmo@esterni.bancaditalia.it.}}\\
Fellow at\\
Banca d'Italia\and Giovanni Melina\thanks{{\footnotesize{}International Monetary Fund, 700 19th Street N.W.,
Washington, D.C. 20431, United States; and CESifo Group, Munich, Germany.
E-mail: gmelina@imf.org.}}\\
International Monetary Fund\\
CESifo}
\date{\today{\footnotesize{}\vspace{-0.5cm}}}
\maketitle
\begin{abstract}
\noindent How should central banks optimally aggregate sectoral inflation
rates in the presence of imperfect labor mobility across sectors?
We study this issue in a two-sector New-Keynesian model and show that
a lower degree of sectoral labor mobility,\textit{ ceteris paribus},
increases the optimal weight on inflation in a sector that would otherwise
receive a lower weight. We analytically and numerically find that,
with limited labor mobility, adjustment to asymmetric shocks cannot
fully occur through the reallocation of labor, thus putting more pressure
on wages, causing inefficient movements in relative prices, and creating
scope for central bank\textquoteright s intervention. These findings
challenge standard central banks\textquoteright{} practice of computing
sectoral inflation weights based solely on sector size, and unveil
a significant role for the degree of sectoral labor mobility to play
in the optimal computation. In an extended estimated model of the
U.S. economy, featuring customary frictions and shocks, the estimated
inflation weights imply a decrease in welfare up to 10 percent relative
to the case of optimal weights.

{\footnotesize{}\vspace{1cm}
}{\footnotesize\par}

{\footnotesize{}\noindent }JEL classification: E52, E58.

{\footnotesize{}\noindent }Keywords: Optimal monetary policy, durable
goods, labor mobility.\pagebreak{}
\end{abstract}

\section{Introduction \label{sec:Introduction}}

What inflation measure should central banks target? This question
arises when a New-Keynesian model is extended to include more than
one sector. In fact, with only one instrument available, the central
bank has to choose how to weight sectoral inflation rates. The literature
has studied this important issue from many angles, but has so far
overlooked the role of the degree of sectoral labor mobility for optimal
monetary policy. Many constraints creates barriers to perfect labor
mobility, including the regulation of labor markets, namely hiring
and firing laws and unemployment benefits (as shown by Botero et al.,
2004), specific human capital skills (Ashournia, 2018), psychological
costs and preference for the status quo (Dix-Carneiro, 2014), the
capital and energy intensity in production and durability of final
goods (Davis and Haltiwanger, 2001), among others. In this paper we
show that the extent to which labor can move across sectors is crucial
in the determination of the optimal inflation composite, especially
in the presence of durable goods.

Central banks generally target a measure of inflation constructed
by weighting sectors according to their economic size. In particular,
the U.S. Federal Reserve targets the Price Index for Personal Consumption
Expenditure (PCE), in which sectors are weighted by their consumption
expenditure shares.\footnote{See the ``FOMC statement of longer-run goals and policy strategy''
released on January 25, 2012 (\href{https://www.federalreserve.gov/newsevents/pressreleases/monetary20120125c.htm}{link here}).
The PCE price index is constructed by the Bureau of Economic Analysis
(see the NIPA Handbook, 2017) and differs from another popular measure
of inflation, the Consumer Price Index (CPI) prepared by the Bureau
of Labor Statistics, as regards the data sources and the way the indices
are calculated. Nevertheless, in both cases sectoral weights correspond
to the economic size of each sector, see McCully et al. (2007) for
more details. Similarly, the European Central Bank stabilizes the
Euro Area Harmonized Index of Consumer Prices (HICP) in which sectoral
and country weights reflect their share in total expenditure, see
Bragoli et al. (2016) for a more detailed discussion.} This practice, however, stands in contrast to the prescription of
optimal monetary policy, which suggests that sectoral weights should
reflect the relative degree of price stickiness and goods durability,
besides economic size (see, e.g. Aoki, 2001; Benigno, 2004; Erceg
and Levin, 2006; Bragoli et al., 2016; Petrella et al., 2019; and
the literature we discuss below).

This paper shows that the degree of sectoral labor mobility should
also be included because, first, micro- and macro-econometric evidence
suggests that sectoral labor mobility is limited (see Horvath, 2000;
Davis and Haltiwanger, 2001; Lee and Wolpin, 2006; Iacoviello and
Neri, 2010; Caliendo et al., 2019; Cardi and Restout, 2015; Cantelmo
and Melina, 2018; Katayama and Kim, 2018, among others)\footnote{See Gallipoli and Pelloni (2013) for a more extensive review on the
micro-macro evidence of limited sectoral labor mobility.} and, second, because sectoral shocks have become relatively more
important than aggregate shocks since the Great Moderation (see Foerster
et al., 2011 for evidence on the U.S.). If labor were perfectly mobile,
it could immediately switch sectors to allow the economy to absorb
asymmetric shocks. Conversely, with limited labor mobility, the adjustment
to these shocks cannot fully occur through the reallocation of labor.
This puts more pressure on wages, which in turn triggers inefficient
movements in relative prices, generating scope for central bank's
intervention.

We illustrate this point by computing optimized simple rules in a
two-sector New-Keynesian model. We start from a small perfectly symmetric
model, in which the two sectors are subject to the same shocks, share
the same price stickiness, the same economic size, and both produce
nondurable goods. Then, we introduce sectoral heterogeneity along
these three dimensions, one at a time. Although other forms of heterogeneity
are possible, we consider the most common in the literature because
of their empirical and theoretical relevance.

The importance of asymmetric price stickiness is well established
both in normative and positive analyses of multi-sector models (see
e.g. Bils and Klenow, 2004; Nakamura and Steinsson, 2008; and Bouakez
et al., 2014 for positive analyses, and Aoki, 2001; Benigno, 2004;
and Bragoli et al., 2016 for normative prescriptions). Emphasis on
heterogeneity in sectoral size has highlighted a contrast between
the standard practice of central banks to weigh sectors by their expenditure
share and the theoretical prescriptions suggesting that price stickiness
is the most important feature to consider (see e.g. Aoki, 2001; and
Benigno, 2004). Finally, the importance of durable goods also deserves
some discussion. Empirical evidence reported by Bernanke and Gertler
(1995), Erceg and Levin (2006), Monacelli (2009) and Sterk and Tenreyro
(2018) suggests that durables goods are important for the transmission
of monetary policy. Moreover, given their inherent characteristics
as investment goods, their relevance arises both in positive (see
e.g. Barsky et al., 2007; Monacelli, 2009; Iacoviello and Neri, 2010;
Bouakez et al., 2011; among many others) and normative (see e.g. Erceg
and Levin, 2006; Barsky et al., 2016; Petrella et al., 2019) analyses
of monetary policy. In particular, the fact that durable goods are
more sensitive to interest rate movements than nondurables (see Erceg
and Levin, 2006), makes them react much more to macroeconomic shocks.

As expected, in the benchmark hypothetical symmetric economy, the
degree of labor mobility does not play any role and the central bank
optimally places an equal weight to inflation in each sector. When
we allow for sectoral heterogeneity, in accordance with previous studies
(discussed below), the central bank optimally assigns less weight
to inflation in the sector with (i) lower degree of price stickiness;
or (ii) smaller economic size; or (iii) producing durable goods. Our
contribution shows that, in each of the three cases, a lower degree
of sectoral labor mobility, \emph{ceteris paribus}, increases the
optimal weight of the sector that would otherwise receive less weight.
This property relies on the fact that lower degrees of labor mobility
amplify the volatility of wage differentials, which translates on
the volatility of the relative price, a result for which we provide
a simple analytical intuition. We furthermore note that the effect
of the degree of sectoral labor mobility is stronger when the two
sectors differ in their goods' durability, given that durability amplifies
relative price fluctuations, and would call for a larger reallocation
of labor across sectors in response to asymmetric shocks.

An important conclusion that can be drawn from these results is that,
in general, the \emph{optimal} weights assigned to sectoral inflation
rates differ from the shares of the sectors in total expenditures,
the usual \emph{suboptimal} criterion adopted by central banks. A
natural question is then: What is the welfare loss suffered by the
economy because of the adoption of suboptimal weights? To answer this
question, we employ an estimated fully-fledged two-sector New-Keynesian
model of the U.S. economy with durable and nondurable goods, imperfect
sectoral labor mobility, and the customary real and nominal frictions.
Estimating the model prior to designing optimized monetary policy
rules is important because such rules heavily depend on the persistence
and the variance of shocks (see, e.g., Cantore et al., 2012 and Melina
and Villa, 2018, among others). The two sectors differ in size, goods'
durability and in the degree of wage and price stickiness, although
the latter turns out not to be significantly different across the
two sectors, in line with our prior macroeconometric estimates (Cantelmo
and Melina, 2018) and other microeconometric studies (see Bils and
Klenow, 2004 and Nakamura and Steinsson, 2008, among others). The
estimation, \emph{inter alia}, confirms a limited degree of labor
mobility across sectors. Consistent with the findings obtained in
the smaller calibrated model, the central bank optimally assigns a
lower weight to inflation in the durables sector. Importantly, we
also confirm that in the fully-fledged model this weight increases
the more limited labor mobility is across sectors. The analysis unveils
that the observed inflation weights imply a decrease in welfare of
up to 10 percent relative to the case of optimal weights. The results
survive a number of robustness checks involving alternative calibrations
and monetary policy rules, including those that entail a feedback
on wages.

Our study is related to a number of key contributions in literature.
In a seminal paper, Aoki (2001) studies a two-sector economy with
sticky- and flexible-price sectors, but no wage stickiness or limited
labor mobility, and finds that the central bank should assign zero
weight to the flexible-price sector. A similar result is attained
by Benigno (2004) in a two-country New-Keynesian model of a currency
union. Here, more weight is attached to inflation in the region displaying
a higher degree of price stickiness. Mankiw and Reis (2003) enrich
these results by showing that, in order to construct a price index
that\textendash if kept on target\textendash stabilizes economic activity,
the sectoral weights should depend on the degree of price stickiness,
the responsiveness to business cycles and the tendency to experience
idiosyncratic shocks. Furthermore, Bragoli et al. (2016) study a multi-country
and a multi-sector model of the Euro Area with price stickiness heterogeneity
across regions (or sectors). They conclude that the optimal weight
to assign to inflation in each country (sector) depends on the interaction
of country's (sector's) price stickiness, economic size and distribution
of the relative price shock. Erceg and Levin (2006), Barsky et al.
(2016) and Petrella et al. (2019) characterize the sectors by their
durability. Abstracting from heterogeneity in price and wage stickiness,
Erceg and Levin (2006) show that the degree of durability of goods
plays an important role for the conduct of monetary policy. Indeed,
as durable goods are more sensitive to the interest rate than nondurables,
the central bank faces a severe trade-off in stabilizing output and
prices across the two sectors. Finally, Huang and Liu (2005), Petrella
and Santoro (2011) and Petrella et al. (2019) explore the role of
input-output interactions (I-O) between intermediate and final goods
firms, a feature we leave aside in the interest of parsimony.\footnote{I-O interactions imply that the two sectoral inflations reflect the
difference between a consumer price index (CPI) and a producer price
index (PPI). In such context, Huang and Liu (2005), Gerberding et
al. (2012) and Strum (2009) conclude that targeting hybrid measures
of inflation delivers desirable welfare results but the weight assigned
to each sectoral inflation reflects their size. Within production
networks, La\textquoteright O and Tahbaz-Salehi (2020) and Rubbo (2020)
show the importance of accounting for heterogenous price stickiness,
while Pasten et al. (2020) show how the I-O structure and sectoral
prices stickiness interact with heterogenous size. Similar conclusions
are drawn when, neglecting I-O interactions, durable goods are used
as collateral by households to borrow (Monacelli, 2008); sectors differ
by factor intensities (Jeske and Liu, 2013); or the length of wage
contracts differs across sectors (Kara, 2010). However, Kara (2010)
assumes prices to be flexible and the only source of nominal rigidities
to be wage stickiness.} Petrella et al. (2019) also assume limited sectoral labor mobility
and compute the optimal weight attached to durables inflation, but
do not isolate the impact that the degree of labor mobility has on
the weight itself.

The bottom line of our analysis is similar in spirit to that of Bragoli
et al. (2016) because it highlights that it is a combination of elements
that the central bank has to take into account to determine the optimal
inflation weights. The novel perspective we add to the debate is that
the degree of sectoral labor mobility should also be part of the central
bank's decision factors, given the observed sector heterogeneity and
the increased importance of sector-specific shocks.

The remainder of the paper is organized as follows. Section \ref{sec:Model_Ch2}
presents the two-sector New-Keynesian model. Section \ref{sec:Optimal-monetary-policy}
shows the results of the optimal monetary policy analysis. Section
\ref{sec:The-fully-fledged-two-sectors} describes the extensions
needed to obtain the fully-fledged two-sector model and discusses
the results of the Bayesian estimation and optimal monetary policy.
Finally, Section \ref{sec:Conclusion} concludes. More details about
the model's equilibrium conditions, the data, the Bayesian estimation,
the Ramsey problem, the role of durable goods and robustness checks
are provided in the Appendix.

\section{The two-sectors model\label{sec:Model_Ch2}}

We start our analysis by constructing a simple two-sector New-Keynesian
model in the spirit of Aoki (2001), with the addition of imperfect
labor mobility across sectors. First, we lay out a framework in which
both sectors produce nondurables goods and asymmetries in price stickiness
and size of each sector are achieved by an appropriate calibration.
Then Section \ref{subsec:durables} describes what modifications are
needed to allow for heterogeneity in goods' durability.

\subsection{Households}

There is a continuum $i\in\left[0,1\right]$ of identical and infinitely-lived
households consuming goods produced in the two sectors $j=\left\{ C,D\right\} $
and supplying labor, whose lifetime utility is
\begin{equation}
E_{0}\sum_{t=0}^{\infty}e_{t}^{B}\beta^{t}U\left(X_{i,t},N{}_{i,t}\right),\label{Utility}
\end{equation}
where $\beta\in\left[0,1\right]$ is the subjective discount factor,
$e_{t}^{B}$ is a preference shock, $X_{i,t}=C_{i,t}^{1-\alpha}D_{i,t}^{\alpha}$
is a Cobb-Douglas consumption aggregator of the goods produced in
sectors $C$ and $D$, respectively, with $\alpha\in\left[0,1\right]$
representing the share of good $D$ consumption on total expenditure
(as in Aoki, 2001; Benigno, 2004; and Bragoli et al., 2016; among
others), and $N_{i,t}$ being the household's labor supply. We assume
that the utility function is additively separable and logarithmic
in consumption: $U\left(X_{t},N{}_{t}\right)=\log\left(X_{t}\right)-\nu\frac{N_{t}^{1+\varphi}}{1+\varphi}$,
where $\nu$ is a scaling parameter and $\varphi$ is the inverse
of the Frisch elasticity of labor supply.

Members of each household supply labor to firms in both sectors according
to:
\begin{equation}
N_{i,t}=\left[\left(\chi^{C}\right)^{-\frac{1}{\lambda}}\left(N_{i,t}^{C}\right)^{\frac{1+\lambda}{\lambda}}+\left(1-\chi^{C}\right)^{-\frac{1}{\lambda}}\left(N_{i,t}^{D}\right)^{\frac{1+\lambda}{\lambda}}\right]^{\frac{\lambda}{1+\lambda}},\label{eq:agg}
\end{equation}
where $\chi^{C}\equiv N^{C}/N$ represents the steady-state share
of labor supply in sector $C$. Following Horvath (2000) and a growing
literature,\footnote{Bouakez et al. (2009), Iacoviello and Neri (2010), Petrella and Santoro
(2011), Bouakez et al. (2011), Cardi and Restout (2015), Petrella
et al. (2019), Cantelmo and Melina (2018) and Katayama and Kim (2018)
likewise employ the CES labor aggregator to model imperfect sectoral
labor mobility.} this constant-elasticity-of-substitution (CES) specification of aggregate
labor allows us to capture the degree of labor market mobility without
deviating from the representative agent assumption; it is a reduced-form
way to model imperfect labor mobility regardless of its root causes;
it is useful to derive analytical results; and allows for different
degrees of sectoral labor mobility by means of just one parameter:
$\lambda>0$, i.e. the intra-temporal elasticity of substitution of
labor across sectors (on this see also Cardi and Restout, 2015). Moreover,
as noted by Petrella and Santoro (2011), equation \eqref{eq:agg}
implies that the labor market friction is neutralized at the steady
state. Perfect labor mobility is achieved for $\lambda\rightarrow\infty$.
In this case sectoral labor services are perfect substitutes. If $\lambda<\infty$,
the economy displays a limited degree of labor mobility. Finally,
as $\lambda\rightarrow0$, labor becomes virtually not substitutable
across sectors.\footnote{In macroeconomic models, CES aggregators are widely employed, e.g.,
to aggregate capital and labor in the production function (see, e.g.,
Cantore and Levine, 2012; Cantore et al., 2014, 2015; Di Pace and
Villa, 2016 and Cantore et al., 2017, among others).}

Each household consumes $C_{i,t}$, purchases nominal bonds $B_{i,t}$,
earn nominal wages $W_{i,t}^{j}$ from working in each sector, receives
profits $\Omega_{t}$ from firms and pays a lump-sum tax $T_{t}$.
We assume sector $C$ to be the numeraire of the economy, hence $Q_{t}\equiv\frac{P_{D,t}}{P_{C,t}}$
denotes the relative price of sector $D$, while $w_{i,t}^{j}=\frac{W_{i,t}^{j}}{P_{t}^{C}}$
is the real wage in sector $j$. The period-by-period real budget
constraint reads as follows:
\begin{align}
C_{i,t}+Q_{t}D_{i,t}+\frac{B_{i,t}}{P_{t}^{C}}=\sum_{j=\left\{ C,D\right\} }\frac{W_{i,t}^{j}}{P_{t}^{C}}N_{i,t}^{j}+R_{t-1}\frac{B_{i,t-1}}{P_{t}^{C}}+\Omega_{t}-T_{t}.\label{BC}
\end{align}
Households choose $C_{i,t}$, $B_{i,t}$, $D_{i,t}$, $N_{i,t}^{C}$,
$N_{i,t}^{D}$ to maximize (\ref{Utility}) subject to (\ref{eq:agg})
and (\ref{BC}). At the symmetric equilibrium, the household's optimality
conditions are:

\begin{eqnarray}
1 & = & E_{t}\left[\Lambda_{t,t+1}\frac{R_{t}}{\Pi_{t+1}^{C}}\right],\label{EulEq1}\\
Q_{t} & = & U_{D,t}\setminus U_{C,t},\label{RelPrice}\\
w_{t}^{C} & = & \nu\left(\chi^{C}\right)^{-\dfrac{1}{\lambda}}\left(N_{t}^{C}\right)^{\dfrac{1}{\lambda}}N_{t}^{\varphi-\frac{1}{\lambda}}\setminus U_{C,t},\label{wage}\\
w_{t}^{D} & = & \nu\left(1-\chi^{C}\right)^{-\dfrac{1}{\lambda}}\left(N_{t}^{D}\right)^{\dfrac{1}{\lambda}}N_{t}^{\varphi-\frac{1}{\lambda}}\setminus U_{C,t}.\label{WageC}
\end{eqnarray}
Equation (\ref{EulEq1}) is a standard Euler equation with $\Lambda_{t,t+1}\equiv\beta\frac{e_{t+1}^{B}U_{C,t+1}}{e_{t}^{B}U_{C,t}}$
representing the stochastic discount factor. $U_{C,t}=\frac{1-\alpha}{C_{i,t}}$
and $U_{D,t}=\frac{\alpha}{D_{i,t}}$ denote the marginal utilities
of consumption of goods produced in each sector. Equation (\ref{RelPrice})
indicates that the relative demand of the two goods depends on the
relative price $Q_{t}$. Finally, equations \eqref{wage} and \eqref{WageC}
define the sectoral labor supply schedules that, combined, yield an
intuitive relationship between sectoral labor supplies and relative
wages:
\begin{equation}
\frac{w_{t}^{C}}{w_{t}^{D}}=\left(\frac{\chi^{C}}{1-\chi^{C}}\right)^{-\frac{1}{\lambda}}\left(\frac{N_{t}^{C}}{N_{t}^{D}}\right)^{\frac{1}{\lambda}}.\label{rel wages}
\end{equation}
According to \eqref{rel wages}, higher substitutability of sectoral
hours (larger $\lambda)$ reduces sectoral wage differentials. Conversely,
lower substitutability (smaller $\lambda)$ implies larger wage differentials.

\subsection{Firms}

A continuum $\omega\in\left[0,1\right]$ of firms in each sector $j=\left\{ C,D\right\} $
operates in monopolistic competition and face quadratic costs of changing
prices $\frac{\vartheta_{j}}{2}\left(\frac{P_{\omega,t}^{j}}{P_{\omega,t-1}^{j}}-1\right)^{2}Y_{t}^{j}$,
where $\vartheta_{j}$ is the parameter of sectoral price stickiness.
Each firm produces differentiated goods according to a linear production
function,
\begin{equation}
Y_{\omega,t}^{j}=e_{t}^{A}e_{t}^{A,j}N_{\omega,t}^{j},\label{Prod func}
\end{equation}
where $e_{t}^{A}$ and $e_{t}^{A,j}$ are aggregate and sector-specific
labor-augmenting productivity shocks, respectively. Firms maximize
the present discounted value of profits,
\begin{equation}
E_{t}\left\{ \sum_{t=0}^{\infty}\Lambda_{t,t+1}\left[\frac{P_{\omega,t}^{j}}{P_{t}^{j}}Y_{\omega,t}^{j}-\frac{W_{\omega t}^{j}}{P_{t}^{j}}N_{\omega,t}^{j}-\frac{\vartheta_{j}}{2}\left(\frac{P_{\omega,t}^{j}}{P_{\omega,t-1}^{j}}-1\right)^{2}Y_{t}^{j}\right]\right\} ,\label{Firm's Profits}
\end{equation}
subject to production function (\ref{Prod func}) and a standard Dixit-Stiglitz
demand equation $Y_{\omega,t}^{j}=\left(\frac{P_{\omega,t}^{j}}{P_{t}^{j}}\right)^{-\epsilon_{j}}Y_{t}^{j},$
where $\epsilon_{j}$ is the sectoral intratemporal elasticity of
substitution across goods. At the symmetric equilibrium, the price
setting equations for the two sectors read as
\begin{flalign}
\left(1-\epsilon_{c}\right)+\epsilon_{c}MC_{t}^{C} & =\vartheta_{c}\left(\Pi_{t}^{C}-\Pi^{C}\right)\Pi_{t}^{C}-\nonumber \\
 & -\vartheta_{c}E_{t}\left[\Lambda_{t,t+1}\frac{Y_{t+1}^{C}}{Y_{t}^{C}}\left(\Pi_{t+1}^{C}-\Pi^{C}\right)\Pi_{t+1}^{C}\right],\label{pricesettC}\\
\left(1-\epsilon_{d}\right)+\epsilon_{d}MC_{t}^{D} & =\vartheta_{d}\left(\Pi_{t}^{D}-\Pi^{D}\right)\Pi_{t}^{D}-\nonumber \\
 & -\vartheta_{d}E_{t}\left[\Lambda_{t,t+1}\frac{Q_{t+1}Y_{t+1}^{D}}{Q_{t}Y_{t}^{D}}\left(\Pi_{t+1}^{D}-\Pi^{D}\right)\Pi_{t+1}^{D}\right],\label{Price_settD-2}
\end{flalign}
where $MC_{t}^{C}=\frac{w_{t}^{C}}{e_{t}^{A}e_{t}^{A,C}}$ and $MC_{t}^{D}=\frac{w_{t}^{D}}{e_{t}^{A}e_{t}^{A,D}Q_{t}}$
are the sectoral marginal costs. When $\vartheta_{j}=0$ prices are
fully flexible and are set as constant markups over the marginal costs.

\subsection{Monetary policy\label{subsec:Fiscal-and-Monetary}}

Monetary policy is conducted by an independent central bank via the
following interest rate rule:
\begin{eqnarray}
\log\left(\frac{R_{t}}{\bar{R}}\right) & = & \rho_{r}\log\left(\frac{R_{t-1}}{\bar{R}}\right)+\alpha_{\pi}\log\left(\frac{\tilde{\Pi}_{t}}{\tilde{\Pi}}\right)+\nonumber \\
 & + & \alpha_{y}\log\left(\frac{Y_{t}}{Y_{t}^{f}}\right)+\alpha_{\Delta y}\left[\log\left(\frac{Y_{t}}{Y_{t}^{f}}\right)-\log\left(\frac{Y_{t-1}}{Y_{t-1}^{f}}\right)\right],\label{SW Rule-1-1-2}
\end{eqnarray}
which has been popularized by Smets and Wouters (2007) and implies
that the central bank reacts to inflation, the output gap and the
output gap growth to an extent determined by parameters $\alpha_{\pi},\alpha_{y}$
and $\alpha_{\Delta y}$, respectively. The output gap is defined
as the deviation of output from the level that would prevail with
flexible prices, $Y_{t}^{f}$, and $\rho_{r}$ is the degree of interest
rate smoothing. The flexible price equilibrium features the same degree
of sectoral labor mobility as the sticky price equilibrium. This rule
is flexible in that it also includes the case of a price-level rule
when $\rho_{r}=1$ or a superinertial rule when $\rho_{r}>1$ (see,
e.g., Woodford, 2003; Cantore et al., 2012; Giannoni, 2014; Melina
and Villa, 2018; and Cantore et al., 2019, among others).

The aggregator of the gross rates of sectoral inflation is
\begin{equation}
\widetilde{\Pi}_{t}\equiv\left(\Pi_{t}^{C}\right)^{1-\tau}\left(\Pi_{t}^{D}\right)^{\tau},\label{eq:infl_comp}
\end{equation}
where $\tau\in\left[0,1\right]$ represents the weight assigned by
the central bank to sector $D$'s inflation in the composite.

\subsection{Market clearing conditions and exogenous processes}

In equilibrium all markets clear and the model is closed by the following
identities:
\begin{align}
Y_{t}^{C} & =C_{t}+\frac{\vartheta_{c}}{2}\left(\Pi_{t}^{C}-\Pi^{C}\right)^{2}Y_{t}^{C},\label{MC1}\\
Y_{t}^{D} & =D_{t}+\frac{\vartheta_{d}}{2}\left(\Pi_{t}^{D}-\Pi^{D}\right)^{2}Y_{t}^{D},\label{MC2}\\
Y_{t} & =Y_{t}^{C}+Q_{t}Y_{t}^{D}.\label{AggY}
\end{align}
 We let the shocks follow an AR(1) process:
\begin{equation}
\log\left(\frac{\kappa_{t}}{\bar{\kappa}}\right)=\rho_{\kappa}\log\left(\frac{\kappa_{t-1}}{\bar{\kappa}}\right)+\epsilon_{t}^{\kappa},
\end{equation}
where $\kappa=\left[e^{B},e^{A},e^{A,C},e^{A,D}\right]$ is a vector
of exogenous variables, $\rho_{\kappa}$ are the autoregressive parameters,
and $\epsilon_{t}^{\kappa}$ are i.i.d shocks with zero mean and standard
deviations $\sigma_{\kappa}$.

\subsection{Extension of the two-sector model to account for durable goods\label{subsec:durables}}

One of the popular dimensions of heterogeneity in a two-sector model
that turns out to be very relevant for optimal monetary policy, consists
of allowing one sector to produce durable goods (see, e.g., Erceg
and Levin, 2006). In some of our exercises, we therefore extend the
model and define sector $D$ to be the durable goods sector while
sector $C$ continues to produce nondurable goods, approximating the
models of Barsky et al. (2007) and Petrella et al. (2019). While the
supply side remains unchanged, introducing durable goods requires
a slight modification of the demand side of the economy. In particular,
household $i$ demands and consumes the stock of durables, $D_{i,t}$,
which evolves according to the following law of motion: 
\begin{equation}
D_{i,t+1}=(1-\delta)D_{i,t}+I_{i,t}^{D},\label{Durables LOM}
\end{equation}
where $\delta$ is the depreciation rate and $I_{i,t}^{D}$ is investment
in durable goods. Each period the household decides the stock of durables
to hold and therefore determines the required investment. Thus the
budget constraint \eqref{BC} now reads as
\begin{align}
C_{i,t}+Q_{t}I_{i,t}^{D}+\frac{B_{i,t}}{P_{t}^{C}}=\sum_{j=\left\{ C,D\right\} }\frac{W_{i,t}^{j}}{P_{t}^{C}}N_{i,t}^{j}+R_{t-1}\frac{B_{i,t-1}}{P_{t}^{C}}+\Omega_{t}-T_{t}.\label{BC-1}
\end{align}
Maximizing utility (\ref{Utility}) subject to (\ref{eq:agg}), \eqref{Durables LOM}
and (\ref{BC-1}) implies that (at the symmetric equilibrium) the
first-order condition \eqref{RelPrice} becomes
\begin{equation}
Q_{t}=\frac{U_{D,t}}{U_{C,t}}+\left(1-\delta\right)E_{t}\left[\Lambda_{t,t+1}Q_{t+1}\right].\label{eq:RelPriceD}
\end{equation}
Equation \eqref{eq:RelPriceD}, whose right-hand side is usually referred
to as the \textit{shadow value of durable goods}, exhibits an additional
term accounting for the discounted expected value of the undepreciated
stock of durables. In particular, it represents the future utility
stemming from selling the durable good the following period, i.e.,
the capital gain. Note that as the depreciation rate of durables increases
(higher $\delta$), durability of goods produced in sector $D$ decreases.
The model collapses to the model outlined in the previous section
in case of full depreciation ($\delta=1$).

While the equations defining the problem of the firms in sector $D$
are unaffected by the presence of durable goods, the market clearing
condition \eqref{MC2} now implies that the period expenditure in
sector $D$ is determined by the flow of durables $I_{t}^{D}$:
\begin{equation}
Y_{t}^{D}=I_{t}^{D}+\frac{\vartheta_{d}}{2}\left(\Pi_{t}^{D}-\Pi^{D}\right)^{2}Y_{t}^{D}.\label{eq:MCD2}
\end{equation}
All the remaining equations of the model, including the monetary policy
rule and the inflation aggregator remain unaltered.

\subsection{Analytical intuition of the impact of the degree of sectoral labor
mobility on fluctuations of the relative price\label{subsec:The-impact-of-labmob}}

We already anticipated in Section \ref{sec:Introduction} that lower
degrees of sectoral labor mobility amplify the volatility of the relative
price, especially in the presence of durable goods. To provide an
analytical intuition of the mechanism, we log-linearize and combine
the relative labor supply schedule \eqref{rel wages}, the definition
of inflation in sector \textit{D $(\Pi_{t}^{D}=\Pi_{t}^{C}Q_{t}\setminus Q_{t-1})$,}
and the sectoral pricing equations \eqref{pricesettC} and \eqref{Price_settD-2}
around the steady state (variables with $\hat{}$ denote percent deviations
from their respective steady state). This procedure is convenient
as the algebraic expressions will exhibit cyclical fluctuations of
the relevant macroeconomic variables, and larger cyclical fluctuations
imply higher volatilities of the underlying variables at business
cycle frequencies. The log-linearized expressions of the above-mentioned
equations read as follow:
\begin{align}
\hat{w}_{t}^{C}-\hat{w}_{t}^{D} & =\frac{1}{\lambda}\left(\hat{N}_{t}^{C}-\hat{N}_{t}^{D}\right),\label{rel_wage_dev}\\
\hat{Q_{t}} & =\hat{\Pi}_{t}^{D}-\hat{\Pi}_{t}^{C}+\hat{Q}_{t-1},\\
\hat{\Pi}_{t}^{D} & =\frac{1-\epsilon_{d}}{\vartheta_{d}}\left(\hat{w}_{t}^{D}-\hat{Q}_{t}-\hat{e}_{t}^{A,D}\right)+\beta E_{t}\hat{\Pi}_{t+1}^{D},\\
\hat{\Pi}_{t}^{C} & =\frac{1-\epsilon_{c}}{\vartheta_{c}}\left(\hat{w}_{t}^{C}-\hat{e}_{t}^{A,C}\right)+\beta E_{t}\hat{\Pi}_{t+1}^{C}.\label{piec_dev}
\end{align}
For ease of exposition, and without loss of generality, assume that
price stickiness and monopolistic competition are equal across sectors
(i.e. $\vartheta_{c}=\vartheta_{d}=\vartheta$ and $\epsilon_{c}=\epsilon_{d}=\epsilon$),
it is then possible to show that combining equations \eqref{rel_wage_dev}-\eqref{piec_dev}
yields:
\begin{align}
\hat{Q_{t}} & =\varpi_{1}\frac{1}{\lambda}\left(\hat{N}_{t}^{D}-\hat{N}_{t}^{C}\right)-\varpi_{1}\left(\hat{e}_{t}^{A,D}-\hat{e}_{t}^{A,C}\right)+\varpi_{2}\beta E_{t}\left[\hat{\Pi}_{t+1}^{D}-\hat{\Pi}_{t+1}^{C}\right]+\varpi_{2}\hat{Q}_{t-1},\label{qhat}
\end{align}
where $\varpi_{1}=\frac{1-\epsilon}{\vartheta+1-\epsilon},\varpi_{2}=\frac{\vartheta}{\vartheta+1-\epsilon}$.
Equation \eqref{qhat} shows that for $\lambda\rightarrow\infty$,
the first summand on the right-hand side $\left(\varpi_{1}\frac{1}{\lambda}\left[\hat{N}_{t}^{D}-\hat{N}_{t}^{C}\right]\right)$
approaches zero, that is, perfect labor mobility removes a source
of volatility in the cyclical fluctuations of the relative price.
Conversely, lower degrees of labor mobility (lower $\lambda$) imply
a higher impact of fluctuations of sectoral wage differentials on
relative price fluctuations (note that, via equation \eqref{rel_wage_dev},
$\hat{w}_{t}^{C}-\hat{w}_{t}^{D}=\frac{1}{\lambda}\left[\hat{N}_{t}^{C}-\hat{N}_{t}^{D}\right]$).
\foreignlanguage{english}{In the sticky price equilibrium, imperfect
labor mobility thus adds a further source of inefficiency in addition
to price stickiness. Intuitively, when prices cannot immediately adjust,
the limited ability of workers to switch sectors in response to sectoral
shocks exerts pressure on wages and hence on firms' marginal costs.
Since firms cannot fully reset their prices, the response of the relative
price is larger than what would be with perfect labor mobility, inducing
the central bank to put relatively more weight on inflation in the
sector that would otherwise receive a lower weight.}

To see the role of goods' durability, as detailed in Appendix \ref{sec:The-role-of-durables},
we log-linearize equation \eqref{eq:RelPriceD} around the steady
state and obtain:
\begin{equation}
\hat{Q}_{t}=\left(\hat{C}_{t}-\hat{D}_{t}\right)\left[1-\left(1-\delta\right)\beta\right]+\left(1-\delta\right)\beta E_{t}\left[\hat{R}_{r,t}-\hat{Q}_{t+1}\right].\label{User cost-loglin-1-1-2}
\end{equation}
Equation \eqref{User cost-loglin-1-1-2} shows that, for any positive
discount factor ($\beta$), lower values of $\delta$ (i.e. higher
good's durability) decrease the effect of the first summand of the
right-hand-side of the equation $\left(\left[\hat{C}_{t}-\hat{D}_{t}\right]\left[1-\left(1-\delta\right)\beta\right]\right)$,
that is, the marginal rate of substitution between goods produced
in sectors \textit{C} and \textit{D}, and increase the importance
of the second summand $\left(\left[1-\delta\right]\beta E_{t}\left[\hat{R}_{r,t}-\hat{Q}_{t+1}\right]\right)$,
which depends on the fluctuations of next period's relative price
$\hat{Q}_{t+1}$. Iterating \eqref{qhat} one period forward shows
that $\hat{Q}_{t+1}$ is also more volatile when the degree of labor
mobility declines. Putting it differently, the degree of sectoral
labor $\lambda$ affects the period-\textit{t} volatility of the relative
price also through its next period's value, which enters the picture
when $\delta<1$. \foreignlanguage{english}{Durables add further variability
to the relative price because, as explained by Erceg and Levin (2006),
being an investment good, small adjustments in their stock imply large
changes in their flows, making them more responsive to shocks than
nondurables. It follows that, in a model with durables, labor tends
to adjust more in response to shocks than in a model without durables.
Limited labor mobility makes this adjustment harder, generating larger
inefficient fluctuations in the relative price.} In sum, the durability
of final goods produced in sector \textit{D} amplifies the effects
of the degree of sectoral labor mobility on the volatility of the
relative price $Q_{t}$. Clearly, if goods in sector \textit{D} are
nondurables ($\delta=1$) the second summand on the right-hand side
of \eqref{User cost-loglin-1-1-2} disappears and the degree of sectoral
labor mobility affects the volatility of the relative price only via
its current value.

\subsection{Parametrization\label{subsec:Calibration}}

The model is parametrized at a quarterly frequency. The discount factor
$\beta$ is equal to the conventional value of $0.99$, implying an
annual steady-state gross interest rate of $4\%$. The baseline calibration
of the model implies perfect symmetry across sectors and that both
sector $C$ and sector $D$ produce nondurables ($\delta=1$). Therefore,
we set $\alpha=0.50$. The inverse Frisch elasticity of labor supply,
$\varphi$, is set at a standard value of 0.5. The preference parameter
$\nu$ is set to target steady-state labor to a conventional $0.33$.
This assumption is, however, innocuous as results are robust to any
reasonable normalization of steady-state labor. The sectoral elasticities
of substitution across different varieties $\epsilon_{c}$ and $\epsilon_{d}$
equal $6$ in order to target a steady-state gross mark-up of $1.20$
in both sectors, while we assume that prices last four quarters as
in Erceg and Levin (2006), and thus set $\vartheta_{c}=\vartheta_{d}=60$,
following Woodford (2003) and Monacelli (2009) to convert the Rotemberg
parameters to Calvo equivalents. To isolate the role of sectoral labor
mobility, we consider three relevant cases: i) \textit{quasi-immobile
labor} by setting $\lambda=0.10$; ii) \textit{limited labor mobility
}by setting $\lambda=1$; and iii) a case of \textit{perfect mobility}
as $\lambda\rightarrow\infty$.\footnote{In the first case, we approximate the assumption of no mobility made
by Aoki (2001), Benigno (2004), Erceg and Levin (2006) and Bragoli
et al. (2016). Setting $\lambda=1$ is consistent with both the macro-estimates
of Horvath (2000), Iacoviello and Neri (2010), Cantelmo and Melina
(2018), and Katayama and Kim (2018) and with the calibrated models
of Bouakez et al. (2009), Petrella and Santoro (2011) and Petrella
et al. (2019). Finally, $\lambda\rightarrow\infty$ is assumed by
Barsky et al. (2016).} Finally, given that the results for the first part of the paper are
purely illustrative, we set the persistence and standard deviation
of the shocks to $\rho_{\kappa}=0.90$ and $\sigma_{\kappa}=0.01$,
respectively. In the fully-fledged model (Section \ref{sec:The-fully-fledged-two-sectors})
we estimate shock processes together with the rest of structural parameters.

From the baseline parametrization, we achieve sectoral heterogeneity
in three dimensions by means of calibration, one at a time. First,
we allow sectors to differ in the degree of price stickiness. For
sector $D$ we assume flexible prices ($\vartheta_{d}=0$). For sector
$C$, in one exercise, we keep the same price stickiness ($\vartheta_{c}=60$);
in another exercise, we double it ($\vartheta_{c}=120$) \foreignlanguage{english}{to
keep the average price stickiness in the economy constant}, relative
to the symmetric case. Then, we assume that sector $D$ is smaller
than sector $C$ by setting $\alpha=0.30$. Finally, we allow sector
$D$ to produce durable goods, while keeping the same price stickiness
across the two sectors. Following Monacelli (2009), in the this last
case, we calibrate the depreciation rate $\delta$ at $0.010$, amounting
to an annual depreciation of $4$\%.

\subsection{Dynamic impact of sectoral labor mobility\label{subsec:Dynamic-impact-of}}

\begin{figure}[!t]
\centering\includegraphics[width=16cm,height=10cm]{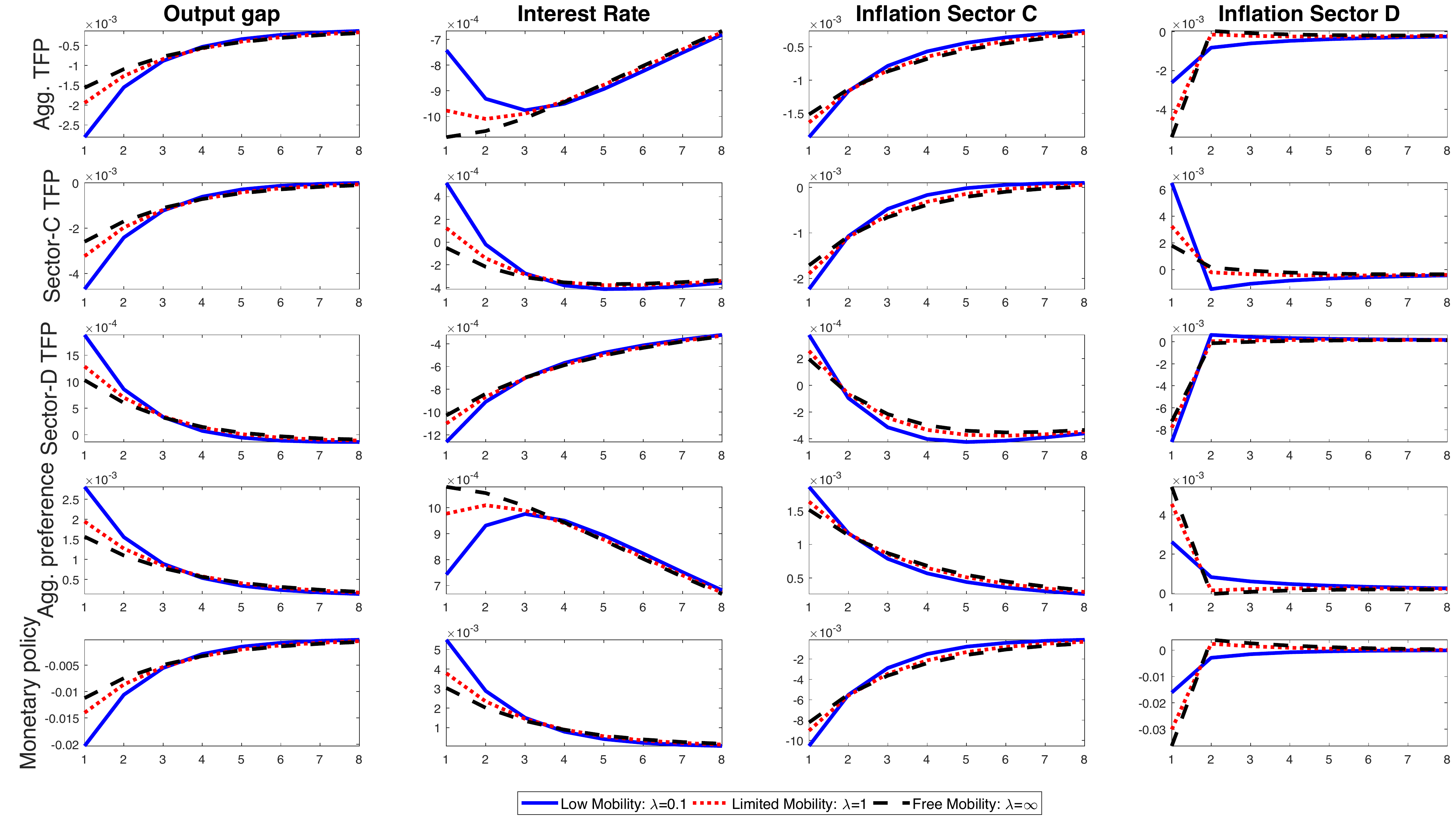}

\caption{\label{IRFs_calibrated}Impulse responses to one standard deviation
shocks in the stylized model with heterogenous price stickiness ($\vartheta_{c}=60,\vartheta_{d}=0$).}
\end{figure}

\selectlanguage{english}%
While various macroeconomic models embed limited labor mobility (see
e.g. Bouakez et al., 2009; Petrella and Santoro, 2011; and Petrella
et al., 2019, among others), there is no systematic investigation
of its role on the dynamic behavior of macroeconomic variables. Therefore,
in this section we show how sectoral labor mobility alters the responses
of the output gap, the interest rate and sectoral inflation rates
to both aggregate and sectoral shocks. We use the model with heterogenous
price stickiness by setting $\vartheta_{d}=0$, to allow both aggregate
and sectoral shocks to generate asymmetric responses. Clearly, in
the symmetric model, sectoral labor mobility plays a role only in
response to sectoral disturbances. In addition to the parametrization
discussed in Section \ref{subsec:Calibration}, we set a simple Taylor
rule with standard values ($\rho_{r}=0.80,\rho_{\pi}=1.50,\rho_{y}=0.125,\rho_{\Delta y}=0$).
Figure \ref{IRFs_calibrated} plots the impulse responses to the shocks
in the model. It is clear that lower degrees of labor mobility entail
larger output gaps and clearly different responses of the interest
rate and sectoral inflation rates. Given the simplicity of the model
and the absence of many frictions that typically make responses more
persistent (e.g. habit in consumption), most of the differences across
the impulse responses are visible on impact. Responses then converge
toward one another after about one year. All in all, larger deviations
of output form the constrained efficient allocation generate scope
for the central bank to take limited labor mobility into account.

\selectlanguage{american}%

\section{Optimal monetary policy\label{sec:Optimal-monetary-policy}}

\subsection{Welfare measure}

The optimal monetary policy analysis serves two purposes: (i) determining
the optimal weights the central bank should assign to sectoral inflations
subject to given degrees of labor mobility, and (ii) seeking parameter
values for interest rate rule \eqref{SW Rule-1-1-2} to minimize the
welfare loss with respect to the Ramsey policy. The flexible price
equilibrium features the same degree of sectoral labor mobility as
the sticky price equilibrium. Monetary policy therefore tries to reach
the constrained efficient equilibrium, i.e. the equilibrium with flexible
prices under the same degree of labor mobility. While in the stylized
model of Section 2 the constrained efficient allocation is characterized
by flexible prices (given that wages are always flexible), in the
fully fledged model of Section \ref{sec:The-fully-fledged-two-sectors},
it requires a flexible-price and flexible-wage equilibrium (given
the presence of sticky wages). The social planner maximizes the present
value of households' utility adjusted for a penalty term to account
for the zero-lower-bound constraint,
\begin{equation}
\Upsilon_{t}=E_{t}\left[\sum_{s=0}^{\infty}e_{s}^{B}\beta^{s}U\left(X_{t+s},N_{t+s}\right)-w_{r}\left(R_{t+s}-R\right)^{2}\right],\label{Welfare-1}
\end{equation}
subject to the equilibrium conditions of the model. This specification,
discussed below, allows avoiding the zero-lower-bound with high probability.
In the analysis, however, welfare losses in consumption-equivalent
terms are calculated excluding the penalty term. Following Schmitt-Grohe
and Uribe (2007), we take a second-order approximation both of the
mean of \emph{$\Upsilon_{t}$} and of the model's equilibrium conditions
around the deterministic steady state. In particular, we take the
approximation around the steady state of the Ramsey equilibrium. Similarly
to many other NK models in the literature (see e.g. Schmitt-Grohe
and Uribe, 2007; Levine et al., 2008; Cantore et al., 2019, among
others), the steady-state value of the gross inflation rate in the
Ramsey equilibrium turns out to be very close to unity, which implies
an almost zero-inflation steady state.\footnote{Nisticò (2007) demonstrates that with zero steady state inflation
and an undistorted steady state, the policy trade-offs the central
bank faces are the same under the Calvo and Rotemberg models. In all
our simulations, steady state inflation is at the most 0.0335\% in
annual terms, that is, very close to zero. Indeed, the impulse responses
of the model solved with a second-order approximation around the fully
optimal steady state and those obtained by solving the model around
a zero-inflation steady state are virtually undistinguishable. This
is in line with Ascari and Ropele (2007), who show that impulse responses
in a model with zero steady-state inflation and those in a model with
a steady-state inflation below 2\% (on an annual basis) are very similar.
Finally, the steady state is undistorted as we employ pruning methods
(see Schmitt-Grohe and Uribe, 2007 and Andreasen et al., 2018). Thus
we expect that assuming Calvo pricing scheme would yield very similar
results.} As anticipated above, since it is not straightforward to account
for the zero-lower-bound (ZLB, henceforth) on the nominal interest
rate when using perturbation methods, we follow Schmitt-Grohe and
Uribe (2007) and Levine et al. (2008) and introduce a term in \eqref{Welfare-1}
that penalizes large deviations of the nominal interest rate from
its steady state. Hence, the imposition of this approximate ZLB constraint
translates into appropriately choosing the weight $w_{r}$ to achieve
an arbitrarily low per-period probability of hitting the ZLB, $Pr\left(ZLB\right)\equiv Pr\left(R_{t}^{n}<1\right)$,
which we set at less than $0.01$ for each calibration.\footnote{Optimal steady state inflation is nearly zero under different parameterizations
of $w_{r}$ and $\lambda$. Using a grid from 0 to 80 for $w_{r}$
and from 0.1 to $\infty$ for $\lambda$, optimal steady state inflation
slightly decreases further (up to 0.02 percentage points, in annual
terms) as $w_{r}$ and/or $\lambda$ increase.} We optimize the interest rate rule \eqref{SW Rule-1-1-2} by numerically
searching for the combination of the policy parameters and the weight
on sector \textit{D}'s inflation $\tau\in\left[0,1\right]$ that maximizes
the present value of households' utility \eqref{Welfare-1}. In doing
so, we follow Schmitt-Grohe and Uribe (2007) and Petrella et al. (2019)
and define the support of $\rho_{r}$ $\left[0,1\right]$ and the
support of $\alpha_{\pi},\alpha_{y}$ and $\alpha_{\Delta y}$ is
$\left[0,5\right]$. Parameter ranges are defined to preserve implementability
of the policy rule. As explained by Schmitt-Grohe and Uribe (2007),
for example, negative or too large positive coefficients would be
difficult to communicate to policymakers or the public. Our ultimate
goal is to unveil how the optimal weight placed on sector \textit{D}'s
inflation ($\tau)$ is affected by the degree of sectoral labor mobility.
We therefore consider three cases of sectoral labor mobility ($\lambda=\left\{ 0.10,1,\infty\right\} )$
and compare the welfare losses in terms of steady-state consumption-equivalent,
$\omega$, with respect to the Ramsey policy, as in Schmitt-Grohe
and Uribe (2007). In particular, for a regime associated to a given
Taylor-type interest rate rule A, the welfare loss is implicitly defined
as

\begin{equation}
E_{0}\left\{ \sum_{t=0}^{\infty}\beta^{t}\left[U\left(\left(1-\omega\right)X_{t}^{R},N_{t}^{R}\right)\right]\right\} =E_{0}\left\{ \sum_{t=0}^{\infty}\beta^{t}\left[U\left(X_{t}^{A},N_{t}^{A}\right)\right]\right\} ,
\end{equation}
where $\omega\times100$ represents the percent permanent loss in
consumption that should occur in the Ramsey regime \textit{(R)} in
order for agents to be as well off in regime \textit{R} as they are
in regime \textit{A}.

\subsection{The impact of the degree of labor mobility}

To discuss the welfare properties of the interest rate rule \eqref{SW Rule-1-1-2},
Table \ref{Optimal_simple_model} reports its optimized parameters
together with the associated welfare costs $\omega$.

\begin{table}[!t]
\setlength\tabcolsep{6 pt}
\renewcommand{\arraystretch}{1.75}

\centering{\small{}}%
\begin{tabular}{ccccccc}
\hline 
{\small{}$\lambda$} & {\small{}$\rho_{r}$} & {\small{}$\alpha_{\pi}$} & {\small{}$\alpha_{y}$} & {\small{}$\alpha_{\Delta y}$} & {\small{}$\tau$} & {\small{}$100\times\omega$}\tabularnewline
\hline 
\multicolumn{7}{c}{\textit{\small{}(i) Symmetric model}}\tabularnewline
{\small{}$\infty$} & {\small{}1.0000} & {\small{}0.0082} & {\small{}0.0217} & {\small{}0.0000} & {\small{}0.5000} & {\small{}0.0002}\tabularnewline
{\small{}1} & {\small{}1.0000} & {\small{}0.0086} & {\small{}0.0214} & {\small{}0.0000} & {\small{}0.5000} & {\small{}0.0004}\tabularnewline
{\small{}0.10} & {\small{}1.0000} & {\small{}0.0101} & {\small{}0.0202} & {\small{}0.0000} & {\small{}0.5000} & {\small{}0.0012}\tabularnewline
\multicolumn{7}{c}{\textit{\small{}(ii) Heterogeneous price stickiness }\foreignlanguage{english}{\textit{\small{}$\vartheta_{c}=60,\vartheta_{d}=0$}}}\tabularnewline
{\small{}$\infty$} & {\small{}1.0000} & {\small{}0.0040} & {\small{}0.0215} & {\small{}0.0000} & {\small{}0.0000} & {\small{}0.0002}\tabularnewline
{\small{}1} & {\small{}1.0000} & {\small{}0.0042} & {\small{}0.0221} & {\small{}0.0000} & {\small{}0.0373} & {\small{}0.0002}\tabularnewline
{\small{}0.10} & {\small{}1.0000} & {\small{}0.0044} & {\small{}0.0231} & {\small{}0.0000} & {\small{}0.0709} & {\small{}0.0003}\tabularnewline
\multicolumn{7}{c}{\textit{\small{}(iii) }\foreignlanguage{english}{\textit{\small{}Heterogeneous
price stickiness $\vartheta_{c}=120,\vartheta_{d}=0$}}}\tabularnewline
\selectlanguage{english}%
{\small{}$\infty$}\selectlanguage{american}%
 & \selectlanguage{english}%
{\small{}1.0000}\selectlanguage{american}%
 & \selectlanguage{english}%
{\small{}0.0076}\selectlanguage{american}%
 & \selectlanguage{english}%
{\small{}0.0197}\selectlanguage{american}%
 & \selectlanguage{english}%
{\small{}0.0000}\selectlanguage{american}%
 & \selectlanguage{english}%
{\small{}0.0000}\selectlanguage{american}%
 & \selectlanguage{english}%
{\small{}0.0002}\selectlanguage{american}%
\tabularnewline
\selectlanguage{english}%
{\small{}1}\selectlanguage{american}%
 & \selectlanguage{english}%
{\small{}1.0000}\selectlanguage{american}%
 & \selectlanguage{english}%
{\small{}0.0079}\selectlanguage{american}%
 & \selectlanguage{english}%
{\small{}0.0202}\selectlanguage{american}%
 & \selectlanguage{english}%
{\small{}0.0000}\selectlanguage{american}%
 & \selectlanguage{english}%
{\small{}0.0184}\selectlanguage{american}%
 & \selectlanguage{english}%
{\small{}0.0003}\selectlanguage{american}%
\tabularnewline
\selectlanguage{english}%
{\small{}0.10}\selectlanguage{american}%
 & \selectlanguage{english}%
{\small{}1.0000}\selectlanguage{american}%
 & \selectlanguage{english}%
{\small{}0.0085}\selectlanguage{american}%
 & \selectlanguage{english}%
{\small{}0.0213}\selectlanguage{american}%
 & \selectlanguage{english}%
{\small{}0.0000}\selectlanguage{american}%
 & \selectlanguage{english}%
{\small{}0.0710}\selectlanguage{american}%
 & \selectlanguage{english}%
{\small{}0.0003}\selectlanguage{american}%
\tabularnewline
\multicolumn{7}{c}{\textit{\small{}(iv) Heterogeneous size}}\tabularnewline
{\small{}$\infty$} & {\small{}1.0000} & {\small{}0.0083} & {\small{}0.0216} & {\small{}0.0000} & {\small{}0.2842} & {\small{}0.0002}\tabularnewline
{\small{}1} & {\small{}1.0000} & {\small{}0.0086} & {\small{}0.0214} & {\small{}0.0000} & {\small{}0.3195} & {\small{}0.0003}\tabularnewline
{\small{}0.10} & {\small{}1.0000} & {\small{}0.0100} & {\small{}0.0203} & {\small{}0.0000} & {\small{}0.3390} & {\small{}0.0008}\tabularnewline
\multicolumn{7}{c}{\textit{\small{}(v) Heterogeneous durability}}\tabularnewline
{\small{}$\infty$} & {\small{}0.8120} & {\small{}0.3847} & {\small{}0.0689} & {\small{}0.0203} & {\small{}0.0538} & {\small{}0.3504}\tabularnewline
{\small{}1} & {\small{}0.8229} & {\small{}0.3802} & {\small{}0.0796} & {\small{}0.0530} & {\small{}0.0638} & {\small{}0.2531}\tabularnewline
{\small{}0.10} & {\small{}0.9521} & {\small{}0.3026} & {\small{}0.1676} & {\small{}0.0496} & {\small{}0.2232} & {\small{}0.2072}\tabularnewline
\hline 
\end{tabular}{\small\par}

\caption{\foreignlanguage{english}{Optimized monetary policy rule in symmetric and asymmetric models}}

\label{Optimal_simple_model}
\end{table}

The primary novel finding of this analysis concerns the inverse relationship
that arises between the optimal weight placed on inflation in sector
$D$, i.e. $\tau$, and the degree of sectoral labor mobility $\lambda$.
The top panel of Table \ref{Optimal_simple_model} shows that obviously
$\lambda$ does not alter $\tau$ in a symmetric model (case \textit{(i)}).
Indeed, the two sectors feature the same price stickiness, size and
durability of the final good produced (both goods are non-durable)
and are subject to symmetric shocks, hence the central bank finds
it optimal to place an equal weight to inflation in each sector ($\tau=0.50$).
To be precise, the two sectors are subject to sectoral disturbances,
but the model's perfect symmetry and the fact that sectoral shocks
are extracted from the same distribution imply that, on average, sector-specific
shocks cancel each other out. In contrast, the remaining panels show
that the degree of sectoral labor mobility affects the optimal weight
placed on inflation in sector \textit{D} whenever the model accounts
for one of the three types of heterogeneity considered. Crucially,
we unveil an inverse relationship between $\lambda$\textendash the
degree of labor mobility\textendash and $\tau$\textendash the optimal
weight on inflation in sector \textit{D}. If sector \textit{D} has
more flexible prices (cases \textit{(ii)} and \textit{(iii)}), or
if it is smaller (case \textit{(iv)}), or if it produces durable goods
(case \textit{(v)}), lower labor mobility implies an increase in the
optimal weight on inflation in sector \textit{D}.

Under perfect labor mobility, when prices in sector \textit{D} are
flexible, the central bank devotes its attention almost entirely to
inflation in (the sticky-price) sector \textit{C}, which is consistent
with previous findings in Aoki (2001) and Benigno (2004). However,
as sectoral labor mobility decreases, the central bank places some
weight on sector \textit{D}'s inflation and $\tau$ rises, regardless
of whether the average price stickiness is halved (by keeping $\vartheta_{c}=60$,
case \textit{(ii)}) or kept constant (by setting $\vartheta_{c}=120$,
case \textit{(iii)}). Interestingly, when the overall price stickiness
is kept constant ($\vartheta_{c}=120,\vartheta_{d}=0$), the optimized
parameters in response to inflation and output gap are very similar
to the symmetric case. In essence, relative to the symmetric model,
the central bank finds it optimal to adjust the measure of inflation
to target (by adjusting $\tau$) while keeping the responses to overall
inflation and output gap virtually unchanged. Table \ref{Table:robust_Stylized}
in Appendix \ref{sec:Robustness-stylized} shows that this holds true
also for an alternative sectoral distribution of the overall price
stickiness ($\vartheta_{c}=90,\vartheta_{d}=30$), and for two additional
(intermediate) degrees of labor mobility (e.g. $\lambda=0.5$ and
$\lambda=3$).

Qualitatively, the same conclusions apply to a model in which sector
\textit{D} holds, as an illustration, a share of 30\% of total consumption
expenditures. In this regard, as labor mobility decreases, the central
bank optimally assigns a higher weight to sector \textit{D}'s inflation,
which exceeds the sector's share in total consumption expenditures.
\foreignlanguage{english}{Intuitively, \textit{ceteris paribus}, the
central bank places a smaller weight on inflation of a smaller sector
(as previously shown by Benigno, 2004). However, as labor becomes
less mobile, the volatilities of sectoral inflation rates and of the
relative price increase (as already explained in Section 2.6). Under
limited labor mobility, optimal monetary policy, by aiming at stabilizing
the relative price, increases the optimal inflation weight associated
to the smaller sector (relative to the weight that would otherwise
be placed under perfect labor mobility).}

Finally, when sector \textit{D} produces durable goods (as outlined
in Section \ref{subsec:durables}), while keeping the same sectoral
price stickiness, we detect the same inverse relationship between
$\lambda$ and $\tau$, and the effects are magnified relative to
the other cases.\footnote{Attaching less weight to the durables sector is in line with Petrella
et al. (2019) and stems from the near constancy of the shadow value
of durable goods, an inherent feature of durables with sufficiently
low depreciation rate, as first noted by Barsky et al. (2007). In
particular, applying repeated forward substitution to \eqref{eq:RelPriceD}
yields $Q_{t}U_{C,t}=\sum_{s=0}^{\infty}\left(1-\delta\right)^{s}\beta^{s}E_{t}\left[U_{D,t+s}\right].$
For a low depreciation rate, the right-hand side (the shadow value
of durables) heavily depends on the marginal utility of durables in
the distant future. Temporary shocks therefore do not influence the
future values of the marginal utility of durables, even if the first
terms of the sum significantly deviate from the steady state. Given
that the shadow value of durables is approximately constant, movements
in the relative price $Q_{t}$ are compensated by movements in the
marginal utility of nondurables. Therefore the central bank achieves
stabilization of nondurables output by stabilizing the relative price,
and viceversa.}

Besides adjusting the optimal weight on inflation in sector $D$,
optimal monetary policy becomes overall more responsive as labor mobility
decreases. In all cases we find that, for lower degrees of labor mobility,
either both the responsiveness to inflation and output gap are larger
(the two cases of heterogenous price stickiness and the model with
durables after accounting for the interest rate inertia), or the increase
in the responsiveness to inflation is larger than the decrease in
the responsiveness to output (symmetric model and case of heterogenous
size). This finding extends also to the case where $\rho_{r}<1$,
once the reparametrization $\gamma_{i}\equiv\frac{\alpha_{i}}{1-\rho_{r}}$,
for $i=\pi,y,\Delta_{y}$ is taken into account. As demonstrated by
the analytical discussion in Section \ref{subsec:The-impact-of-labmob}
and the impulse responses reported in Section \ref{subsec:Dynamic-impact-of},
the increasing severity of limited labor mobility generates larger
fluctuations in relative prices and output gaps, which require stronger
responses of the central bank.

As analytically shown in Section \ref{subsec:The-impact-of-labmob},
the intuition behind our findings is that, with less mobile labor,
adjustments to sectoral shocks cannot easily occur through quantities
(via the reallocation of labor itself) hence wages need to adjust
more. Fluctuations in wage differentials induce higher volatility
of the relative price of goods produced in sector \textit{D,} and
the central bank finds it optimal to place relatively more weight
on sector \textit{D}'s inflation than it would otherwise do. Indeed,
the standard deviation of the relative price $Q_{t}$, under \textit{quasi}
labor immobility, in all cases, is larger than under limited and perfect
labor mobility. As shown analytically in Section \ref{subsec:The-impact-of-labmob},
in the presence of durable goods, the effect of the degree of labor
mobility on the volatility of the relative price is amplified. To
give a quantitative idea, in the illustrative numerical exercises
with durable goods, under \textit{quasi} labor immobility, the relative
price is 1.3 and 1.8 times more volatile than under limited and perfect
mobility, respectively. We also find that in almost all cases (except
when we introduce durable goods), the interest rate smoothing parameter
hits the upper bound of one, thus characterizing equation \eqref{SW Rule-1-1-2}
as a price-level rule.\footnote{As discussed by Giannoni (2014), price-level rules deliver better
welfare results than Taylor-type rules by introducing a sufficient
amount of history dependence in an otherwise entirely forward-looking
behavior of price setters, thus reducing the volatility of inflation.
Similar results hold in other contexts, such as the New-Keynesian
model with financial frictions studied by Melina and Villa (2018),
and in a model with optimal monetary and fiscal policies as in Cantore
et al. (2019). Moreover, McKnight (2018) demonstrates that price-level,
or Wicksellian, rules are desirable even under partially backward-looking
Phillips curves, i.e. due to price indexation.}

All in all, our results reveal that the extent to which labor is able
to reallocate across sectors, by impacting the volatility of the relative
price of goods produced in sector \textit{D}, is important for the
optimal design of monetary policy, whenever sectors display sources
of heterogeneity. In accordance with previous studies, the central
bank optimally assigns less weight to inflation in the sector ($D$)
with lower degree of price stickiness; or smaller economic size; or
producing durable goods. Importantly, our results shows that a lower
degree of sectoral labor mobility, \emph{ceteris paribus}, increases
this optimal weight because it magnifies the volatility of relative
prices, especially in the presence of durable goods. These findings
add another reason to challenge standard central banks\textquoteright{}
practice of computing sectoral inflation weights based solely on economic
size, and unveil a significant role for the degree of sectoral labor
mobility in the optimal computation.

\subsection{Determinacy}

In all cases considered, whether the optimal policy is characterized
by a price-level rule or not, indeterminacy is not an issue. First,
Giannoni (2014) demonstrates that any price level rule with positive
coefficients yields a determinate equilibrium. In addition, Bauducco
and Caputo (2020) show that price-level targeting rules do not require
the Taylor principle to be satisfied for determinacy to hold. Whenever
we find that the optimal monetary policy is characterized by an inertial
rule (with $\rho_{r}<1$), we find that the Taylor principle is satisfied.
Indeed, following Schmitt-Grohe and Uribe (2007), Taylor rule (\ref{SW Rule-1-1-2})
can be reparametrized noting that $\alpha_{i}=\left(1-\rho_{r}\right)\gamma_{i}$,
for $i=\pi,y,\Delta_{y}$. It is therefore possible to recover the
feedback parameters $\gamma_{i}$ given the optimal values of each
$\alpha_{i}$ and $\rho_{r}$. In all cases, we find that $\gamma_{\pi}=\frac{\alpha_{\pi}}{1-\rho_{r}}>1$,
which satisfies the Taylor principle. \foreignlanguage{english}{This
is in line with previous contributions on determinacy. Carlstrom et
al. (2006) show that $\gamma_{\pi}>1$ is a sufficient and necessary
condition for determinacy to hold in a two-sector model with both
perfect or no sectoral labor mobility, and both with symmetric price
stickiness and when one sector displays flexible prices. More generally,
if the two sectors display different (non zero) degrees of price stickiness,
determinacy depends also on restrictions about relative price stickiness
and preference parameters, hence $\gamma_{\pi}>1$ is only a sufficient
condition. They conclude that the restriction on the reaction parameter
to inflation holds regardless of whether the central bank targets
aggregate or sectoral inflation rates. While Carlstrom et al. (2006)
focus on a Taylor rule that responds only to inflation, Ascari and
Ropele (2009) build on Woodford (2003) and consider a Taylor rule
that responds to both inflation and the output gap. They first demonstrate
that under zero trend inflation, $\gamma_{\pi}>1$ ensures determinacy
regardless of the value of the reaction parameter to the output gap.
Moreover, they show that including interest rate inertia makes the
determinacy region larger. We likewise find determinacy in the analysis
of the fully-fledged model reported in Section }\ref{subsec:Optimal-monetary-policy}\foreignlanguage{english}{
and Appendix} \ref{subsec:Robustness-to-alternative}.

\section{The fully-fledged two-sectors model\label{sec:The-fully-fledged-two-sectors}}

The next step is to extend our analysis to a fully-fledged two-sectors
New-Keynesian model, featuring a rich set of real and nominal frictions
and structural shocks and, crucially, the three sources of heterogeneity
studied above. \foreignlanguage{english}{The aim of using this medium-scale
model is twofold. First, we want to verify that our main result, namely
the inverse relationship between labor mobility and the optimal weight
$\tau$, does not hinge on the simplicity of the model presented in
Section 2. Second, we want to provide a quantitative assessment of
the welfare loss caused by not accounting for labor mobility. To do
so it is necessary to add real and nominal frictions that help the
model fit the data well (see e.g. }Christiano et al., 2005 and Smets
and Wouters, 2007). Fitting the data is crucial also for obtaining
a plausible estimate of the degree of labor mobility, which turns
out to be close to the estimates of Horvath (2000) and Iacoviello
and Neri (2010).\footnote{In Cantelmo and Melina (2018) we show that, in an analogous model
(but without limited labor mobility) estimated using similar data
over the same sample, removing the various frictions dramatically
worsens the model's fit.}

As shown in seminal contributions by Fuhrer (2000), Christiano et
al. (2005) and Smets and Wouters (2007), habit formation in \foreignlanguage{english}{consumption
of nondurable goods allows the model to generate hump-shaped responses
of consumption, in line with empirical evidence. The importance of
accounting for investment adjustment costs is stressed by }Smets and
Wouters (2007)\foreignlanguage{english}{, who show that it is the
most relevant real friction of their model. Moreover, }Iacoviello
and Neri (2010)\foreignlanguage{english}{, to which our model is very
close, report that removing real and nominal frictions worsens the
ability of the model to match the standard deviations and cross-correlations
of model's variables with the data. In addition, they show that their
estimates of sectoral labor mobility is the most affected parameter.}

\selectlanguage{english}%
As far as nominal rigidities are concerned, both papers employing
one sector (e.g. Chris- tiano et al., 2005 and Smets and Wouters,
2007) and two-sector models (e.g. Iacoviello and Neri, 2010 and Cantelmo
and Melina, 2018) show their empirical relevance.

Moreover, the addition of the three forms of heterogeneity studied
in the stylized model of Section 2 (regarding the degree of nominal
rigidities, size and durability of the final goods produced) are relevant
for the design of optimal monetary policy in a two-sector economy.
\foreignlanguage{american}{Erceg and Levin (2006)} demonstrate how
different durability of the final goods produced makes the central
bank's trade-off between stabilizing inflation and output more severe
than in a model without durables. In a similar context, \foreignlanguage{american}{Petrella
et al. (2019)} show that it is optimal to attach less weight to inflation
in the durables sector. Finally, nominal rigidities and sectoral size
matter for the conduct of optimal monetary policy, as demonstrated
by Aoki (2001), Benigno (2004) and Bragoli et al. (2016), with the
general prescription that the sector with lower nominal rigidities
and/or smaller in size should receive less weight in the inflation
aggregator to target, but the weight does not necessarily coincide
with the sector's size. \foreignlanguage{american}{We next lay out
the extensions to the model described in Section 2 with durable goods.
Then we bring the model to the data and finally we analyze optimal
monetary policy.}
\selectlanguage{american}%

\subsection{Model extensions}

Households still aggregate nondurables and durables consumption according
to $X_{i,t}=C_{i,t}^{1-\alpha}D_{i,t}^{\alpha}$, but we allow for
external habit formation (with persistence, as in Fuhrer, 2000) in
the former and investment adjustment costs in the latter (as in Christiano
et al., 2005). In particular, we add the following equations:
\begin{align}
C_{i,t} & =Z_{i,t}-\zeta S_{t-1},\label{Habits}\\
S_{t} & =\rho_{c}S_{t-1}+(1-\rho_{c})Z_{t},\label{motion_c}
\end{align}
where $Z_{i,t}$ is the level of the household's nondurable consumption;
$S_{t}$, $\zeta\in\left(0,1\right)$ and $\rho_{c}\in\left(0,1\right)$
are the stock, the degree and the persistence of habit formation,
respectively, while $Z_{t}$ represents average consumption across
all households. Investment adjustment costs in durables imply that
the law of motion of durable goods \eqref{Durables LOM} now reads
as
\begin{equation}
D_{i,t+1}=(1-\delta)D_{i,t}+e_{t}^{I}I_{i,t}^{D}\left[1-S\left(\frac{I_{i,t}^{D}}{I_{i,t-1}^{D}}\right)\right],\label{Durables LOM-1}
\end{equation}
where $e_{t}^{I}$ represents an investment-specific shock. The adjustment
costs function $S\left(\cdot\right)$ satisfies $S\left(1\right)=S^{'}\left(1\right)=0$
and $S^{''}\left(1\right)>0$, which we assume to be quadratic: $S\left(\frac{I_{t}^{D}}{I_{t-1}^{D}}\right)=\frac{\phi}{2}\left(\frac{I_{t}^{D}}{I_{t-1}^{D}}-1\right)^{2}$
, $\phi>0$ (Christiano et al., 2005). We also introduce nominal wage
stickiness at the sectoral level in the form of quadratic adjustment
costs $\Phi_{t}^{j}=\frac{\vartheta_{j}^{w}}{2}\left(\frac{w_{i,t}^{j}}{w_{i,t-1}^{j}}\Pi_{t}^{C}-\Pi^{C}\right)^{2}w_{t}^{j}N_{t}^{j}$
as in Rotemberg (1982), where $w_{i,t}^{j}$ is the aggregate real
wage earned by the household in sector \textit{$j=C,D$}. Therefore
the left-hand side of the budget constraint \eqref{BC-1} features
the additional terms $\Phi_{t}^{j}$, while we add the following first-order
conditions with respect to investment in durables $I_{i,t}^{D}$ and
the real wages $w_{i,t}^{j}$:\\
\begin{align}
1 & =\psi_{t}e_{t}^{I}\left[1-S\left(\frac{I_{t}^{D}}{I_{t-1}^{D}}\right)-S^{'}\left(\frac{I_{t}^{D}}{I_{t-1}^{D}}\right)\frac{I_{t}^{D}}{I_{t-1}^{D}}\right]+\nonumber \\
 & +E_{t}\left\{ \Lambda_{t,t+1}\psi_{t+1}\frac{Q_{t+1}}{Q_{t}}e_{t+1}^{I}\left[S^{'}\left(\frac{I_{t+1}^{D}}{I_{t}^{D}}\right)\left(\frac{I_{t+1}^{D}}{I_{t}^{D}}\right)^{2}\right]\right\} ,\\
0 & =\left[1-e_{t}^{w,j}\eta\right]+\frac{e_{t}^{w,j}\eta}{\tilde{\mu_{t}}^{j}}-\vartheta_{j}^{w}\left(\Pi_{t}^{w,j}-\Pi^{C}\right)\Pi_{t}^{w,j}+\nonumber \\
 & +E_{t}\left[\Lambda_{t,t+1}\vartheta_{j}^{w}\left(\Pi_{t+1}^{w,j}-\Pi^{C}\right)\Pi_{t+1}^{w,j}\frac{w_{t+1}^{j}N_{t+1}^{j}}{w_{t}^{j}N_{t}^{j}}\right],\label{Wage}
\end{align}
where $\psi_{t}$ is the Lagrange multiplier attached to constraint
(\ref{Durables LOM-1}). Equation (\ref{Wage}) is the wage setting
equation in sector $j=C,D$, in which $\tilde{\mu_{t}}^{j}\equiv\frac{w_{t}^{j}}{MRS_{t}^{j}}$
is the sectoral wage markup, $MRS_{t}^{j}\equiv-\frac{U_{N,t}^{j}}{U_{C,t}}$
is the marginal rate of substitution between consumption and leisure
in sector $j$, $U_{N,t}^{j}$ is the marginal disutility of work
in sector $j$, $\Pi_{t}^{w,j}$ is the gross sectoral wage inflation
rate and $e_{t}^{w,j}$ is a sector-specific wage markup shock.

The supply-side of the economy is essentially unaltered, except for
the shocks. Indeed, we add sectoral price markup (or cost-push) shocks
$e_{t}^{j},j=C,D$, which are shocks to the sectoral intratemporal
elasticity of substitution across goods $\epsilon_{j}$. Moreover,
to be consistent with our observables, we remove the sectoral shocks
to labor productivity.\footnote{We use data on aggregate GDP to identify the aggregate shock to labor
productivity, while data on sectoral inflation rates allow us identify
the cost-push shocks and estimate the parameters of sectoral price
stickiness.} Therefore, the sectoral production functions \eqref{Prod func} are
replaced by
\begin{flalign}
Y_{\omega,t}^{j} & =e_{t}^{A}N_{\omega,t}^{j},
\end{flalign}
while in the sectoral price setting equations \eqref{pricesettC}
and \eqref{Price_settD-2} the parameters $\epsilon_{j}$ are multiplied
by the exogenous disturbances $e_{t}^{j}$. Moreover, following Erceg
and Levin (2006) we assume that the government purchases nondurable
goods. By allowing also for sectoral wage stickiness, the sectoral
market clearing conditions \eqref{MC1} and \eqref{eq:MCD2} now read
as
\begin{align}
Y_{t}^{C} & =C_{t}+e_{t}^{G}+\frac{\vartheta_{c}}{2}\left(\Pi_{t}^{C}-\Pi^{C}\right)^{2}Y_{t}^{C}+\Phi_{t}^{C},\\
Y_{t}^{D} & =I_{t}^{D}+\frac{\vartheta_{d}}{2}\left(\Pi_{t}^{D}-\Pi^{D}\right)^{2}Y_{t}^{D}+\Phi_{t}^{D}.
\end{align}
We still employ the monetary policy rule \eqref{SW Rule-1-1-2}, however
now $Y_{t}^{f}$ is the output that would prevail without nominal
rigidities and markup shocks. Finally, as in Smets and Wouters (2007),
the wage markup and the price markup shocks follow ARMA (1,1) processes,
while the remaining shocks move according to an AR (1) process.

\subsection{Bayesian estimation}

The model is estimated with Bayesian methods. The Kalman filter is
used to evaluate the likelihood function that, combined with the prior
distribution of the parameters, yields the posterior distribution.
Then, the Monte-Carlo-Markov-Chain Metropolis-Hastings (MCMC-MH) algorithm
with two parallel chains of 150,000 draws each is used to generate
a sample from the posterior distribution in order to perform inference.
We estimate the model over the sample 1969Q2-2007Q4, leaving aside
the Great Recession and the zero lower bound regime, by using US data
on: GDP, consumption of durable goods, consumption of nondurable goods,
sectoral real wages and hours worked, inflation in the nondurables
sector, inflation in the durables sector and the nominal interest
rate. Given the importance of the sectoral price stickiness parameters
in our analysis, we choose the same sample and observables (except
for sectoral wages) as in Cantelmo and Melina (2018), so that we can
verify that our results are in line with their evidence.

The following measurement equations link the data to the endogenous
variables of the model:
\begin{align}
\Delta Y_{t}^{o} & =\gamma+\hat{Y}_{t}-\hat{Y}_{t-1},\\
\Delta I_{D,t}^{o} & =\gamma+\hat{I}_{D,t}-\hat{I}_{D,t-1},\\
\Delta C_{t}^{o} & =\gamma+\hat{C}_{t}-\hat{C}_{t-1},\\
\Delta W_{t}^{C,o} & =\gamma+\hat{W}_{t}^{C}-\hat{W}_{t-1}^{C},\\
\Delta W_{t}^{D,o} & =\gamma+\hat{W}_{t}^{D}-\hat{W}_{t-1}^{D},
\end{align}
\begin{align}
N_{t}^{C,o} & =\hat{N}_{t}^{C},\label{labmeas-1}\\
N_{t}^{D,o} & =\hat{N}_{t}^{D},\label{labDmeas}\\
\Pi_{C,t}^{o} & =\bar{\pi}_{C}+\hat{\Pi}_{t}^{C},\\
\Pi_{D,t}^{o} & =\bar{\pi}_{D}+\hat{\Pi}_{t}^{D},\\
R_{t}^{o} & =\bar{r}+\hat{R}_{t},
\end{align}
where $\gamma$ is the common quarterly trend growth rate of GDP,
consumption of durables, consumption of nondurables and the real wage;
$\bar{\pi}_{C}$ and $\bar{\pi}_{D}$ are the average quarterly inflation
rates in nondurable and durable sectors respectively; $\bar{r}$ is
the average quarterly Federal funds rate. Hours worked are demeaned
so no constant is required in the corresponding measurement equations
(\ref{labmeas-1}) and \eqref{labDmeas}. Variables with a $\hat{}$
are in log-deviations from their own steady state.

\paragraph*{Calibration and priors.}

Table \ref{calibration} presents the structural parameters calibrated
at a quarterly frequency. The discount factor $\beta$ is equal to
the conventional value of $0.99$, implying an annual steady-state
gross interest rate of $4\%$. Following Monacelli (2009), we calibrate
the depreciation rate of durable goods $\delta$ at $0.010$ amounting
to an annual depreciation of $4$\%, and the durables share of total
expenditure $\alpha$ is set at $0.20$. The sectoral elasticities
of substitution across different varieties $\epsilon_{c}$ and $\epsilon_{d}$
equal $6$ in order to target a steady-state gross mark-up of $1.20$
in both sectors. We target a 5\% steady-state gross wage mark-up hence
we set the elasticity of substitution in the labor market $\eta$
equal to $21$ as in Zubairy (2014). The preference parameter $\nu$
is set to target steady-state total hours of work of $0.33$. The
government-output ratio $g_{y}$ is calibrated at $0.20$, in line
with the data.

Prior and posterior distributions of the parameters and the shocks
are reported in Table \ref{estimation_table-1}. We set the prior
mean of the inverse Frisch elasticity{\scriptsize{} }$\varphi$ to
0.5, broadly in line with Smets and Wouters (2007, SW henceforth)
who estimate a Frisch elasticity of 1.92. We also follow SW in setting
the prior means of the habit parameter, $\zeta$, to 0.7, the interest
rate smoothing parameter, $\rho_{r}$, to 0.80 and in assuming a stronger
response of the central bank to inflation than output. We set the
prior means of the constants in the measurement equations equal to
the average values in the dataset. In general, we use the Beta distribution
for all parameters bounded between 0 and 1. We use the Inverse Gamma
(IG) distribution for the standard deviation of the shocks for which
we set a loose prior with 2 degrees of freedom. We choose a Gamma
distribution for the Rotemberg parameters for both prices and wages,
given that these are non-negative. The price stickiness parameters
are assigned the same prior distribution corresponding to firms resetting
prices around 1.5 quarters on average in a Calvo world. Finally, we
follow Iacoviello and Neri (2010) who choose a Normal distribution
for the intra-temporal elasticity of substitution in labor supply
$\lambda$, with a prior mean of 1 which implies a limited degree
of labor mobility, and a standard deviation of 0.1.
\begin{table}[!t]
\centering%
\begin{tabular}{lll}
\hline 
Parameter &  & Value/target\tabularnewline
\hline 
Discount factor & $\beta$ & $0.99$\tabularnewline
Durables depreciation rate & $\delta$ & $0.010$\tabularnewline
\textcolor{black}{Durables share of total expenditure} & $\alpha$ & $0.20$\tabularnewline
Elasticity of substitution nondurable goods & $\epsilon_{c}$ & $6$\tabularnewline
Elasticity of substitution durable goods & $\epsilon_{d}$ & $6$\tabularnewline
Elasticity of substitution in labor & $\eta$ & $21$\tabularnewline
Preference parameter & $\nu$ & $\bar{N}=0.33$\tabularnewline
Government share of output & $g_{y}$ & $0.20$\tabularnewline
\hline 
\end{tabular}

\caption{Calibrated parameters}

\label{calibration}
\end{table}

\paragraph*{Estimation results.}

We report the posterior mean of the parameters together with the $90\%$
probability intervals in square brackets in Table \ref{estimation_table-1}.
In line with the literature, the labor mobility parameter $\lambda$
is estimated to be 1.2250 implying a non-negligible degree of friction
in the labor market. Indeed, Horvath (2000) estimates a regression
equation to find a value of 0.999 whereas Iacoviello and Neri (2010)
estimate values of 1.51 and 1.03 for savers and borrowers, respectively.\footnote{Iacoviello and Neri (2010) specify the CES aggregator such that the
labor mobility parameter is the inverse of $\lambda$. They find values
of 0.66 and 0.97 for savers and borrowers respectively hence the values
of 1/0.66=1.51 and 1/0.97=1.03 we reported to ease the comparison.} The estimated low sectoral labor mobility is also in line with the
microeconometric evidence reported by Jovanovic and Moffitt (1990)
and Lee and Wolpin (2006), who estimate a high cost of switching sectors
in the U.S. Moreover, in calibrated models, limited labor mobility
is typically set at a value of $\lambda=1$ (see Bouakez et al., 2009;
Petrella and Santoro, 2011; and Petrella et al., 2019; among others)
except Bouakez et al. (2011) who explore values between 0.5 and 1.5.
Our estimate is remarkably close to values estimated by Horvath (2000)
and Iacoviello and Neri (2010), and to those employed in calibrated
models.
\begin{table}[!tp]
\renewcommand{\arraystretch}{0.90} \centering%
\begin{tabular}{llllll}
\hline 
{\scriptsize{}Parameter} &  & \multicolumn{3}{c}{{\scriptsize{}Prior}} & \multicolumn{1}{c}{{\scriptsize{}Posterior Mean}}\tabularnewline
 &  & {\scriptsize{}Distrib.} & {\scriptsize{}Mean} & {\scriptsize{}Std/df} & \tabularnewline
\hline 
\emph{\scriptsize{}Structural} &  &  &  &  & \tabularnewline
{\scriptsize{}Labor mobility} & {\scriptsize{}$\lambda$} & {\scriptsize{}Normal} & {\scriptsize{}1.00} & {\scriptsize{}0.10} & {\scriptsize{}1.2250 {[}1.0966;1.3591{]}}\tabularnewline
{\scriptsize{}Inverse Frisch elasticity} & {\scriptsize{}$\varphi$} & {\scriptsize{}Normal} & {\scriptsize{}0.50} & {\scriptsize{}0.10} & {\scriptsize{}0.2320 {[}0.1077;0.3377{]}}\tabularnewline
{\scriptsize{}Habit in nondurables consumption} & {\scriptsize{}$\zeta$} & {\scriptsize{}Beta} & {\scriptsize{}0.70} & {\scriptsize{}0.10} & {\scriptsize{}0.6919 {[}0.6546;0.7317{]}}\tabularnewline
{\scriptsize{}Habit persist. nondurables consumption} & {\scriptsize{}$\rho_{c}$} & {\scriptsize{}Beta} & {\scriptsize{}0.70} & {\scriptsize{}0.10} & {\scriptsize{}0.4384 {[}0.3374;0.5399{]}}\tabularnewline
{\scriptsize{}Price stickiness nondurables} & {\scriptsize{}$\vartheta_{c}$} & {\scriptsize{}Gamma} & {\scriptsize{}15.0} & {\scriptsize{}5.00} & {\scriptsize{}20.424 {[}12.901;27.730{]}}\tabularnewline
{\scriptsize{}Price stickiness durables} & {\scriptsize{}$\vartheta_{d}$} & {\scriptsize{}Gamma} & {\scriptsize{}15.0} & {\scriptsize{}5.00} & {\scriptsize{}29.194 {[}19.865;38.531{]}}\tabularnewline
{\scriptsize{}Wage stickiness nondurables} & {\scriptsize{}$\vartheta_{C}^{w}$} & {\scriptsize{}Gamma} & {\scriptsize{}100.0} & {\scriptsize{}10.00} & {\scriptsize{}122.04 {[}105.11;139.16{]}}\tabularnewline
{\scriptsize{}Wage stickiness durables} & {\scriptsize{}$\vartheta_{D}^{w}$} & {\scriptsize{}Gamma} & {\scriptsize{}100.0} & {\scriptsize{}10.00} & {\scriptsize{}132.45 {[}119.05;149.17{]}}\tabularnewline
{\scriptsize{}Invest. adjust. costs durable goods} & {\scriptsize{}$\phi$} & {\scriptsize{}Normal} & {\scriptsize{}1.5} & {\scriptsize{}0.50} & {\scriptsize{}2.3028 {[}1.7563;2.8491{]}}\tabularnewline
{\scriptsize{}Share of durables inflation in aggregator} & {\scriptsize{}$\tau$} & {\scriptsize{}Beta} & {\scriptsize{}0.20} & {\scriptsize{}0.10} & {\scriptsize{}0.2264 {[}0.1400;0.3080{]}}\tabularnewline
{\scriptsize{}Inflation -Taylor rule} & {\scriptsize{}$\rho_{\pi}$} & {\scriptsize{}Normal} & {\scriptsize{}1.50} & {\scriptsize{}0.20} & {\scriptsize{}1.4761 {[}1.3061;1.6365{]}}\tabularnewline
{\scriptsize{}Output -Taylor rule} & {\scriptsize{}$\rho_{y}$} & {\scriptsize{}Gamma} & {\scriptsize{}0.10} & {\scriptsize{}0.05} & {\scriptsize{}0.0225 {[}0.0137;0.0309{]}}\tabularnewline
{\scriptsize{}Output growth -Taylor rule} & $\rho_{\Delta y}$ & {\scriptsize{}Gamma} & {\scriptsize{}0.10} & {\scriptsize{}0.05} & {\scriptsize{}0.3525 {[}0.1598;0.5392{]}}\tabularnewline
{\scriptsize{}Interest rate smoothing} & {\scriptsize{}$\rho_{r}$} & {\scriptsize{}Beta} & {\scriptsize{}0.80} & {\scriptsize{}0.10} & {\scriptsize{}0.6334 {[}0.5843;0.6854{]}}\tabularnewline
\hline 
\emph{\scriptsize{}Averages} &  &  &  &  & \tabularnewline
{\scriptsize{}Trend growth rate} & {\scriptsize{}$\gamma$} & {\scriptsize{}Normal} & {\scriptsize{}0.49} & {\scriptsize{}0.10} & {\scriptsize{}0.2120 {[}0.1854;0.2400{]}}\tabularnewline
{\scriptsize{}Inflation rate nondurables} & {\scriptsize{}$\bar{\pi}_{C}$} & {\scriptsize{}Gamma} & {\scriptsize{}1.05} & {\scriptsize{}0.10} & {\scriptsize{}1.0908 {[}1.0008;1.1768{]}}\tabularnewline
{\scriptsize{}Inflation rate durables} & {\scriptsize{}$\bar{\pi}_{D}$} & {\scriptsize{}Gamma} & {\scriptsize{}0.55} & {\scriptsize{}0.10} & {\scriptsize{}0.5327 {[}0.4414;0.6199{]}}\tabularnewline
{\scriptsize{}Interest rate} & {\scriptsize{}$\bar{r}$} & {\scriptsize{}Gamma} & {\scriptsize{}1.65} & {\scriptsize{}0.10} & {\scriptsize{}1.6241 {[}1.5096;1.7380{]}}\tabularnewline
\hline 
\emph{\scriptsize{}Exogenous processes} &  &  &  &  & \tabularnewline
{\scriptsize{}Technology} & {\scriptsize{}$\rho_{e^{A}}$} & {\scriptsize{}Beta} & {\scriptsize{}0.50} & {\scriptsize{}0.20} & {\scriptsize{}0.9713 {[}0.9584;0.9849{]}}\tabularnewline
 & {\scriptsize{}$\sigma_{e^{A}}$} & {\scriptsize{}IG} & {\scriptsize{}0.10} & {\scriptsize{}2.0} & {\scriptsize{}0.0047 {[}0.0040;0.0055{]}}\tabularnewline
{\scriptsize{}Monetary Policy} & {\scriptsize{}$\rho_{e^{R}}$} & {\scriptsize{}Beta} & {\scriptsize{}0.50} & {\scriptsize{}0.20} & {\scriptsize{}0.1273 {[}0.0447;0.2130{]}}\tabularnewline
 & {\scriptsize{}$\sigma_{e^{R}}$} & {\scriptsize{}IG} & {\scriptsize{}0.10} & {\scriptsize{}2.0} & {\scriptsize{}0.0031 {[}0.0027;0.0034{]}}\tabularnewline
{\scriptsize{}Investment Durables} & {\scriptsize{}$\rho_{e^{I}}$} & {\scriptsize{}Beta} & {\scriptsize{}0.50} & {\scriptsize{}0.20} & {\scriptsize{}0.2787 {[}0.1437;0.4046{]}}\tabularnewline
 & {\scriptsize{}$\sigma_{e^{I}}$} & {\scriptsize{}IG} & {\scriptsize{}0.10} & {\scriptsize{}2.0} & {\scriptsize{}0.0597 {[}0.0424;0.0770{]}}\tabularnewline
{\scriptsize{}Preference} & {\scriptsize{}$\rho_{e^{B}}$} & {\scriptsize{}Beta} & {\scriptsize{}0.50} & {\scriptsize{}0.20} & {\scriptsize{}0.7133 {[}0.6393;0.7965{]}}\tabularnewline
 & {\scriptsize{}$\sigma_{e^{B}}$} & {\scriptsize{}IG} & {\scriptsize{}0.10} & {\scriptsize{}2.0} & {\scriptsize{}0.0124 {[}0.0107;0.0141{]}}\tabularnewline
{\scriptsize{}Price mark-up nondurables} & {\scriptsize{}$\rho_{e^{C}}$} & {\scriptsize{}Beta} & {\scriptsize{}0.50} & {\scriptsize{}0.20} & {\scriptsize{}0.9859 {[}0.9762;0.9955{]}}\tabularnewline
 & {\scriptsize{}$\theta_{C}$} & {\scriptsize{}Beta} & {\scriptsize{}0.50} & {\scriptsize{}0.20} & {\scriptsize{}0.3046 {[}0.1367;0.4707{]}}\tabularnewline
 & {\scriptsize{}$\sigma_{e^{C}}$} & {\scriptsize{}IG} & {\scriptsize{}0.10} & {\scriptsize{}2.0} & {\scriptsize{}0.0141 {[}0.0103;0.0178{]}}\tabularnewline
{\scriptsize{}Price mark-up durables} & {\scriptsize{}$\rho_{e^{D}}$} & {\scriptsize{}Beta} & {\scriptsize{}0.50} & {\scriptsize{}0.20} & {\scriptsize{}0.9762 {[}0.9569;0.9955{]}}\tabularnewline
 & {\scriptsize{}$\theta_{D}$} & {\scriptsize{}Beta} & {\scriptsize{}0.50} & {\scriptsize{}0.20} & {\scriptsize{}0.1840 {[}0.0452;0.3094{]}}\tabularnewline
 & {\scriptsize{}$\sigma_{e^{D}}$} & {\scriptsize{}IG} & {\scriptsize{}0.10} & {\scriptsize{}2.0} & {\scriptsize{}0.0455 {[}0.0360;0.0551{]}}\tabularnewline
{\scriptsize{}Wage mark-up nondurables} & {\scriptsize{}$\rho_{e^{w,C}}$} & {\scriptsize{}Beta} & {\scriptsize{}0.50} & {\scriptsize{}0.20} & {\scriptsize{}0.9962 {[}0.9933;0.9992{]}}\tabularnewline
 & {\scriptsize{}$\theta_{w,C}$} & {\scriptsize{}Beta} & {\scriptsize{}0.50} & {\scriptsize{}0.20} & {\scriptsize{}0.2170 {[}0.0780;0.3539{]}}\tabularnewline
 & {\scriptsize{}$\sigma_{e^{w,C}}$} & {\scriptsize{}IG} & {\scriptsize{}0.10} & {\scriptsize{}2.0} & {\scriptsize{}0.0165 {[}0.0139;0.0190{]}}\tabularnewline
{\scriptsize{}Wage mark-up durables} & {\scriptsize{}$\rho_{e^{w,D}}$} & {\scriptsize{}Beta} & {\scriptsize{}0.50} & {\scriptsize{}0.20} & {\scriptsize{}0.9746 {[}0.9598;0.9902{]}}\tabularnewline
 & {\scriptsize{}$\theta_{w,D}$} & {\scriptsize{}Beta} & {\scriptsize{}0.50} & {\scriptsize{}0.20} & {\scriptsize{}0.1909 {[}0.0510;0.3180{]}}\tabularnewline
 & {\scriptsize{}$\sigma_{e^{w,D}}$} & {\scriptsize{}IG} & {\scriptsize{}0.10} & {\scriptsize{}2.0} & {\scriptsize{}0.0444 {[}0.0373;0.0512{]}}\tabularnewline
{\scriptsize{}Government spending} & {\scriptsize{}$\rho_{e^{G}}$} & {\scriptsize{}Beta} & {\scriptsize{}0.50} & {\scriptsize{}0.20} & {\scriptsize{}0.9201 {[}0.8751;0.9657{]}}\tabularnewline
 & {\scriptsize{}$\sigma_{e^{G}}$} & {\scriptsize{}IG} & {\scriptsize{}0.10} & {\scriptsize{}2.0} & {\scriptsize{}0.0347 {[}0.0314;0.0380{]}}\tabularnewline
{\scriptsize{}Log-marginal likelihood} &  &  &  &  & {\scriptsize{}-2349.865}\tabularnewline
\hline 
\end{tabular}

\caption{Prior and posterior distributions of estimated parameters (90\% confidence
bands in square brackets)}

\label{estimation_table-1}
\end{table}

Prices are estimated to be slightly stickier in the durables sector,
with no statistically significant difference between the two sectors,
as already implied by the macroeconometric estimates of Cantelmo and
Melina (2018). However, also in the microeconometric literature there
is no decisive evidence that prices of nondurable goods are much stickier
than those of many durables (see Bils and Klenow, 2004 and Nakamura
and Steinsson, 2008, among others). Wage stickiness is also not significantly
different across the two sectors, with wages in the durables sector
exhibiting a higher point estimate. Having said this, it is true that
prices of new houses are generally rather flexible, as usually assumed
in the literature (see Barsky et al., 2007; Iacoviello and Neri, 2010;
and our estimates in Cantelmo and Melina, 2018; among many others).
Therefore, given the sensitivity of the optimal monetary policy results
to the degree of price stickiness of durable goods, in the remainder
of the paper we use both the estimated value of durables price stickiness
and an alternative calibration, implying completely flexible durables
prices. Similarly, although wages in the durables sector are estimated
to be sticky, we also explore the counterfactual of flexible wages.

The remaining parameters are broadly in line with the literature and
suggest a relevance of the real frictions (IAC in durable goods and
habits in consumption of nondurables) and a stronger response of monetary
policy to inflation with respect to output, with a high degree of
policy inertia as, e.g., in the estimates of Smets and Wouters (2007)
and Smets and Villa (2016), which we follow in setting the monetary
policy rule, the latter covering a similar sample.

In sum, our estimation delivers results consistent with a wide range
of New-Keynesian models estimated with Bayesian methods and serves
as the starting point for our analysis of optimal monetary policy.
The estimated model exhibits well-behaved macroeconomic dynamics (see,
e.g., the bayesian impulse responses to a technology shock reported
in Figure \ref{BIRFs} in Appendix \ref{sec:Bayesian-impulse-responses}).
In the remainder of the paper, parameters are set according to the
calibration of Table \ref{calibration} and the posterior means reported
in Table \ref{estimation_table-1}, unless otherwise stated.

\subsection{Optimal monetary policy\label{subsec:Optimal-monetary-policy}}

\begin{table}[!t]
\setlength\tabcolsep{5 pt}
\renewcommand{\arraystretch}{1.5}

\centering%
\begin{tabular}{ccccccc}
\hline 
{\footnotesize{}$\lambda$} & {\footnotesize{}$\rho_{r}$} & {\footnotesize{}$\alpha_{\pi}$} & {\footnotesize{}$\alpha_{y}$} & {\footnotesize{}$\alpha_{\Delta y}$} & {\footnotesize{}$\tau$} & {\footnotesize{}$100\times\omega$}\tabularnewline
\hline 
\multicolumn{7}{c}{\textit{\footnotesize{}Sticky durables prices}}\tabularnewline
{\footnotesize{}$\infty$} & {\footnotesize{}0.0050} & {\footnotesize{}2.3150} & {\footnotesize{}0.0000} & {\footnotesize{}0.3388} & {\footnotesize{}0.0187} & {\footnotesize{}0.0888}\tabularnewline
{\footnotesize{}1.2250} & {\footnotesize{}0.4900} & {\footnotesize{}1.0615} & {\footnotesize{}0.0000} & {\footnotesize{}0.2553} & {\footnotesize{}0.1500} & {\footnotesize{}0.1364}\tabularnewline
{\footnotesize{}0.10} & {\footnotesize{}0.9174} & {\footnotesize{}0.8917} & {\footnotesize{}0.0014} & {\footnotesize{}0.0000} & {\footnotesize{}0.7724} & {\footnotesize{}0.2754}\tabularnewline
\multicolumn{7}{c}{\textit{\footnotesize{}Flexible durables prices}}\tabularnewline
{\footnotesize{}$\infty$} & {\footnotesize{}0.0136} & {\footnotesize{}2.3240} & {\footnotesize{}0.0000} & {\footnotesize{}0.0099} & {\footnotesize{}0.0000} & {\footnotesize{}0.0877}\tabularnewline
{\footnotesize{}1.2250} & {\footnotesize{}0.9954} & {\footnotesize{}0.0158} & {\footnotesize{}0.0000} & {\footnotesize{}0.0000} & {\footnotesize{}0.0000} & {\footnotesize{}0.1092}\tabularnewline
{\footnotesize{}0.10} & {\footnotesize{}0.9598} & {\footnotesize{}0.1752} & {\footnotesize{}0.0009} & {\footnotesize{}0.0000} & {\footnotesize{}0.5883} & {\footnotesize{}0.2255}\tabularnewline
\hline 
\end{tabular}

\caption{\foreignlanguage{english}{Optimized monetary policy rule: sticky vs flexible durables prices}}

\label{Optimal_1-2-2-1-1}
\end{table}

\begin{figure}[!t]
\centering\includegraphics[width=8cm,height=6cm]{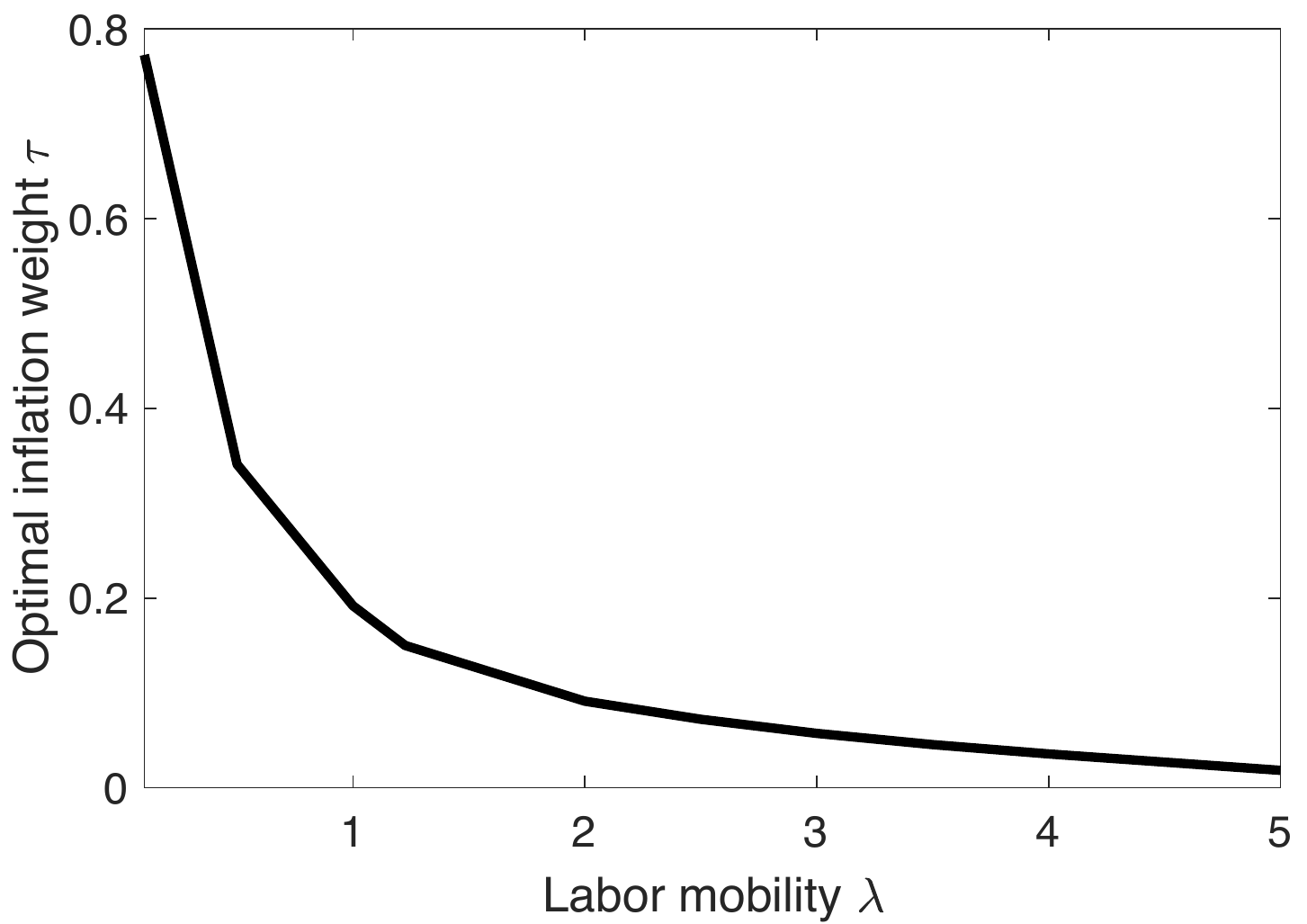}

\caption{\label{Figure_tau-lambda}Optimal inflation weight $\tau$ for different
degrees of labor mobility $\lambda$.}
\end{figure}

We now turn to the optimal monetary policy results in the fully-fledged
model (Table \ref{Optimal_1-2-2-1-1}). We first notice that regardless
of the degree of labor mobility, the central bank response to the
output gap is almost zero and that to output gap growth is usually
low, whereas a stronger reaction is devoted to inflation, a result
in line with the findings of Schmitt-Grohe and Uribe (2007) and Cantore
et al. (2019) in one-sector models. Crucially, the primary novel finding
obtained in the simple two-sectors model\textemdash namely the inverse
relationship that arises between the optimal weight placed on durables
inflation $\tau$ and sectoral labor mobility $\lambda$\textemdash carries
over to the fully-fledged model. The intuition developed in the smaller
two-sectors model still holds in this richer environment, which is
useful to provide quantitative insights. The top panel of Table \ref{Optimal_1-2-2-1-1}
shows that, in the range of $\lambda$ considered (including the estimated
labor mobility parameter, $\lambda=1.2250$), we highlight an inverse
relationship between sectoral labor mobility and the optimal weight
placed on durables inflation. As labor becomes less mobile (i.e. $\lambda$
decreases) the central bank finds it optimal to place more weight
on durables inflation (i.e. the optimal $\tau$ increases). Indeed,
when $\lambda$ drops from the estimated value of 1.2250 to 0.10,
$\tau$ increases from a value slightly below the sector's share (0.1500)
to a value well above it (0.7724). In contrast, when labor is perfectly
mobile ($\lambda\rightarrow\infty$), the weight on durables inflation
approaches zero. \foreignlanguage{english}{Figure \ref{Figure_tau-lambda}
plots this inverse relationship for a continuum of degrees of labor
mobility between 0 and 5, showing that the relationship is monotonically
negative.}

\begin{figure}[!t]
\centering\includegraphics[width=16cm,height=10cm]{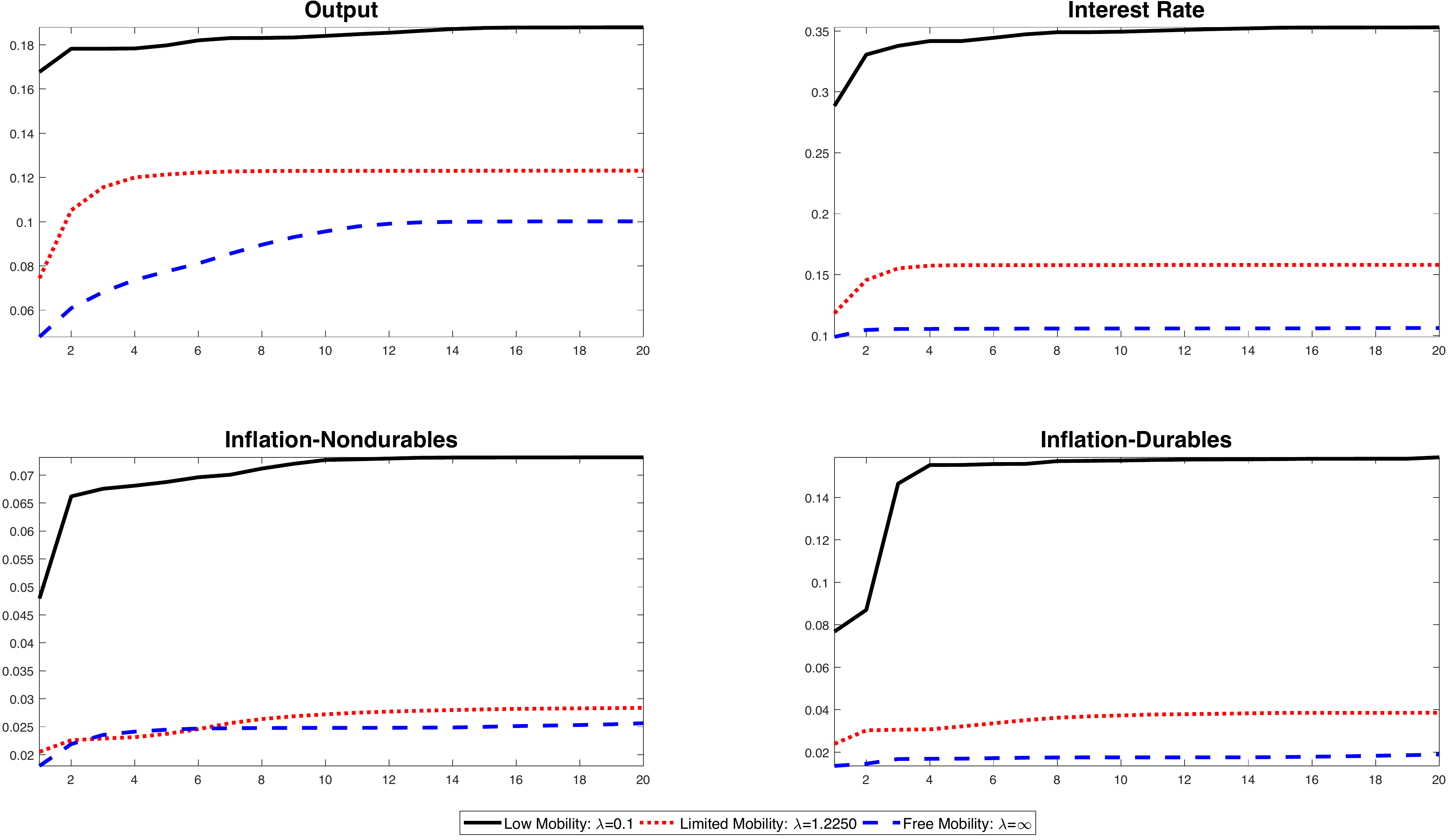}

\caption{\label{Opt_IRFS}\foreignlanguage{english}{Impulse responses to a
durables price markup shock in Ramsey policy and optimized rule.}}
\end{figure}

Welfare losses increase as labor becomes less mobile across sectors.
This is driven by the presence of the price markup shock in the durables
sector, which makes it more difficult for the central bank to replicate
the Ramsey policy when labor mobility decreases (see also Table \ref{Optimal_shocks}
in Section \ref{subsec:Robustness-to-alternative} of the Appendix).
While this shock is empirically important and has a material impact
on the magnitude of the welfare losses, the main result of the paper,
i.e. that the optimal weight on durables inflation increases as labor
mobility decreases, holds regardless of its presence\textsf{ }(on
this, see Appendix \ref{subsec:Robustness-to-alternative}). Figure
\ref{Opt_IRFS} reports\foreignlanguage{english}{, for the three degrees
of labor mobility under scrutiny,} the root cumulated squared difference
of the impulse responses of key variables to a durables price markup
shock under Ramsey and the optimized Taylor rule, i.e. \foreignlanguage{english}{$100\times\sqrt{\sum_{t=0}^{H}\left(x_{t}^{R}-x_{t}^{O}\right)^{2}}$,
where $H=1,2,...$ and $x_{t}^{R}$ and $x_{t}^{O}$ denote the impulse
response of variable $x$ under Ramsey and the optimized rule, respectively.}
In general, as labor mobility decreases, the difference between the
interest rate responses under Ramsey and the optimized rules widens,
causing a larger welfare loss. \foreignlanguage{english}{This is in
line with Petrella et al. (2019), who find that welfare losses are
larger for lower degrees of labor mobility, in a two-sector economy
subject only to sectoral technology shocks.}

When prices of durables are assumed to be flexible ($\vartheta^{d}=0$,
lower panel of Table \ref{Optimal_1-2-2-1-1}), as in the case of
new house prices (see Cantelmo and Melina, 2018, for a detailed discussion
on sectoral price stickiness), the optimal weight the central bank
attaches to durables inflation drops to a large extent. At the estimated
value of the degree of labor mobility and above, the optimal weight
is already zero. However, $\tau$ is still nonzero ($\tau=0.5883$)
for a sufficiently limited degree of labor mobility, this result being
mainly driven by nominal wage stickiness. In fact, wage stickiness
affects firms' marginal costs and their price setting behavior. The
pass-through of sticky wages to the durables sector's marginal cost
induces the central bank to place some weight on inflation in this
sector despite price flexibility. We isolate the contribution of wage
rigidity in Section \ref{subsec:Nominal-rigidities} of the Appendix.
Finally, comparing the welfare losses with respect to the Ramsey policy
(Table \ref{Optimal_1-2-2-1-1}), these are comparable to those calculated
by Cantore et al. (2019) in a one-sector model and Petrella et al.
(2019) in a two-sector model with limited labor mobility.

Our results survive a battery of robustness checks reported in Appendix
\ref{subsec:Robustness-to-alternative}. In particular, we show that
they are robust to: i) the elimination of sectoral shocks, one at
a time (\ref{subsec:Sectoral-shocks}); ii) various assumptions on
nominal rigidities (\ref{subsec:Nominal-rigidities}); iii) the elimination
of real frictions, one at a time (\ref{subsec:Real-frictions}); iv)
different depreciation rates of durable goods (\ref{subsec:Depreciation-rate-of});
and v) alternative interest rate rules (\ref{subsec:Alternative-interest-rate}),
including those that respond to wages.

\subsection{The importance of optimizing the weight on sectoral inflation}

Our results challenge standard practice used in central banks that
weight sectoral inflation rates only by the sectors\textquoteright{}
shares in the economy. In this section we ask: ``what are the welfare
implications of weighting or not weighting sectoral inflation optimally?''
The numerical procedure implemented to reach our results (both for
the simple and the fully-fledged models) by construction ensures that
the value taken by $\tau$, along with the other parameters of the
interest rate rule, is the one that maximizes social welfare (or equivalently,
minimizes the losses relative to the Ramsey policy). Although it is
obvious that deviating from the optimization of all parameters would
deliver higher welfare losses, it is interesting to quantify them.
We perform two exercises, both under the estimated degree of labor
mobility $(\lambda=1.2250)$. Table \ref{Optimal_notau} reports the
rule parameters, the welfare losses relative to the Ramsey policy
$\omega$, and the percent change in the welfare loss relative to
the benchmark case, i.e. $100\times\frac{\omega^{B}-\omega^{A}}{\omega^{A}}$.
In the first exercise, we keep all the parameters of the interest
rate rule \eqref{SW Rule-1-1-2} at their estimated values while optimizing
only the weight on durables inflation $\tau$, and compare the welfare
loss with that obtained under the estimated Taylor rule and the estimated
inflation weight. When all parameters of the interest rate rule are
constrained at the estimated values, the central bank would optimally
set $\tau=0.2988$. In this case, households would experience a welfare
gain of 0.7 percent. The second exercise follows the opposite logic:
we optimize the parameters of the interest rate rule \eqref{SW Rule-1-1-2}
while keeping the weight on durables inflation at the estimated value
of $\tau=0.2264$, and compare the welfare loss with that obtained
when all parameters are optimized . In other words, here we take the
viewpoint of a policymaker who is able to optimize the interest rate
rule but cannot review the inflation weights. In this case, households
would suffer a welfare loss of about 10 percent, relative to optimizing
all parameters. Both experiments show the importance of optimizing
the weight on durables inflation and that failing to do so brings
sizable welfare costs.

\begin{table}[!t]
\setlength\tabcolsep{5 pt}
\renewcommand{\arraystretch}{1.25}

\centering%
\begin{tabular}{cccccccc}
\hline 
{\footnotesize{}$\lambda$} & {\footnotesize{}$\rho_{r}$} & {\footnotesize{}$\alpha_{\pi}$} & {\footnotesize{}$\alpha_{y}$} & {\footnotesize{}$\alpha_{\Delta y}$} & {\footnotesize{}$\tau$} & {\footnotesize{}$100\times\omega$} & {\footnotesize{}$100\times\left(\frac{\omega^{B}-\omega^{A}}{\omega^{A}}\right)$}\tabularnewline
\hline 
\multicolumn{8}{c}{\textit{\emph{\footnotesize{}{[}A{]}}}\textit{\footnotesize{} Benchmark:
Estimated Taylor rule}}\tabularnewline
{\footnotesize{}1.2250} & {\footnotesize{}0.6334} & {\footnotesize{}0.5411} & {\footnotesize{}0.0082} & {\footnotesize{}0.1292} & {\footnotesize{}0.2264} & {\footnotesize{}0.6419} & {\footnotesize{}/}\tabularnewline
\multicolumn{8}{c}{\textit{\emph{\footnotesize{}{[}B{]}}}\textit{\footnotesize{} Optimizing
only $\tau$ in the estimated Taylor rule}}\tabularnewline
{\footnotesize{}1.2250} & {\footnotesize{}0.6334} & {\footnotesize{}0.5411} & {\footnotesize{}0.0082} & {\footnotesize{}0.1292} & {\footnotesize{}0.2988} & {\footnotesize{}0.6375} & {\footnotesize{}-0.70}\tabularnewline
\hline 
\multicolumn{8}{c}{\textit{\emph{\footnotesize{}{[}A{]}}}\textit{\footnotesize{} Benchmark:
Fully optimized Taylor rule}}\tabularnewline
{\footnotesize{}1.2250} & {\footnotesize{}0.4900} & {\footnotesize{}1.0615} & {\footnotesize{}0.0000} & {\footnotesize{}0.2553} & {\footnotesize{}0.1500} & {\footnotesize{}0.1364} & {\footnotesize{}/}\tabularnewline
\multicolumn{8}{c}{\textit{\emph{\footnotesize{}{[}B{]}}}\textit{\footnotesize{} Empirical
$\tau$ within the optimized Taylor rule}}\tabularnewline
{\footnotesize{}1.2250} & {\footnotesize{}0.7379} & {\footnotesize{}0.5061} & {\footnotesize{}0.0000} & {\footnotesize{}0.1039} & {\footnotesize{}0.2264} & {\footnotesize{}0.1502} & {\footnotesize{}10.1}\tabularnewline
\hline 
\end{tabular}

\caption{\foreignlanguage{english}{Optimized monetary policy rules: the importance of optimizing the
inflation weight}}

\label{Optimal_notau}
\end{table}

\section{Conclusions\label{sec:Conclusion}}

As the New-Keynesian literature on two-sector models has demonstrated,
setting the optimal weights on sectoral inflation rates is a crucial
task for a central bank to maximize social welfare. Importantly, these
weights generally differ from the sectoral shares in total consumption
expenditures. We analyze this issue from a perspective the literature
has so far overlooked, that is, the extent to which labor can move
across sectors.

We first look at a stylized two-sector model. Our main result is that
whenever the model allows for sectoral heterogeneity (namely in price
stickiness, sector size or goods' durability), we unveil an inverse
relationship between the degree of sectoral labor mobility and the
optimal weight on inflation in the sector that would otherwise deserve
less weight (that is, the sector with more flexible prices, or smaller
in size, or producing durable goods). We rationalize this result by
noticing that lower degrees of sectoral labor mobility are associated
with a more volatile relative price. In fact, with more limited labor
mobility, adjustments to asymmetric shocks do not easily occur through
quantities (via the reallocation of labor itself), but rather through
wages. We analytically show that the lower the degree of labor mobility,
the more the volatility in wage differentials translates into higher
relative price volatility. We also show that the effect of the degree
of labor mobility on the computation of optimal sectoral inflation
weights is magnified when one of the two sectors produces durable
goods. This finding can also be rationalized via simple analytics
showing that goods' durability enhances the effect that the degree
of sectoral labor mobility has on the relative price.

We then compute the welfare loss suffered by the economy because of
the adoption of suboptimal weights. To do this, we construct and estimate
a fully-fledged two-sector New-Keynesian model with durable and nondurable
goods, conventional real and nominal frictions and shocks, and imperfect
sectoral labor mobility. The Bayesian estimation confirms, \emph{inter
alia}, the evidence of a limited sectoral labor mobility. An inflation
weight set in line with either the posterior estimate or sectoral
expenditure shares (which mirrors the common practice of central banks)
imply a decrease in welfare up to 10 percent relative to the case
of an optimized weight. In line with the results obtained with the
stylized model, also in the fully-fledged model we detect an inverse
relationship between labor mobility and the weight optimally attached
to inflation in the durables sector, which is also smaller in size
and exhibits mildly more flexible prices relative to the nondurables
sector. These results survive a large array of robustness checks.

In sum, our findings echo previous contributions in the literature
that challenge standard practice of central banks, which weight sectoral
inflation merely based on sectoral economic size. From a welfare-maximizing
viewpoint the central bank should take a number of features into account.
Our contribution shows that, in a context of increased importance
of sectoral shocks, the extent to which labor can be reallocated across
sectors should be among central banks' decision factors.

\section*{References}

\begin{singlespace}
Andreasen, M. M., Fernandez-Villaverde, J., and Rubio-Ramirez, J.
F. (2018). The pruned state-space system for non-linear DSGE models:
theory and empirical applications. \textsl{The Review of Economic
Studies}, 85(1):1\textendash 49.\\
Aoki, K. (2001). Optimal monetary policy responses to relative-price
changes. \textit{Journal of Monetary Economics}, 48(1):55\textendash 80.
\\
Ascari, G. and Ropele, T. (2007). Optimal monetary policy under low
trend inflation. \textit{Journal of Monetary Economics}, 54(8):2568
\textendash{} 2583. \\
Ascari, G. and Ropele, T. (2009). Trend inflation, taylor principle,
and indeterminacy. \textit{Journal of Money, Credit and Banking},
41(8):1557\textendash 1584. \\
Ashournia, D. (2018). Labour market effects of international trade
when mobility is costly. \textit{The Economic Journal}, 128(616):3008\textendash 3038.\\
Barsky, R., Boehm, C. E., House, C. L., and Kimball, M. (2016). Monetary
policy and durable goods. Working Paper Series WP-2016-18, Federal
Reserve Bank of Chicago. Barsky, R. B., House, C. L., and Kimball,
M. S. (2007). Sticky-price models and durable goods. \textit{American
Economic Review}, 97(3):984\textendash 998. \\
Bauducco, S. and Caputo, R. (2020). Wicksellian rules and the taylor
principle: Some practical implications. \textit{The Scandinavian Journal
of Economics}, 122(1):340\textendash 368.\\
Benigno, P. (2004). Optimal monetary policy in a currency area. \textit{Journal
of International Economics}, 63(2):293\textendash 320. \\
Bernanke, B. S. and Gertler, M. (1995). Inside the black box: The
credit channel of monetary policy transmission. \textit{Journal of
Economic Perspectives}, 9(4):27\textendash 48. \\
Bils, M. and Klenow, P. J. (2004). Some evidence on the importance
of sticky prices. \textit{Journal of Political Economy}, 112(5):947\textendash 985.
\\
Botero, J. C., Djankov, S., Porta, R. L., de Silanes, F. L., and Shleifer,
A. (2004). The Regulation of labor. \textit{The Quarterly Journal
of Economics}, 119(4):1339\textendash 1382. \\
Bouakez, H., Cardia, E., and Ruge-Murcia, F. (2014). Sectoral price
rigidity and aggregate dynamics. \textit{European Economic Review},
65(C):1\textendash 22. \\
Bouakez, H., Cardia, E., and Ruge-Murcia, F. J. (2009). The transmission
of monetary policy in a multisector economy. \textit{International
Economic Review}, 50(4):1243\textendash 1266.\\
Bouakez, H., Cardia, E., and Ruge-Murcia, F. J. (2011). Durable goods,
inter-sectoral linkages and monetary policy. \textit{Journal of Economic
Dynamics and Control}, 35(5):730\textendash{} 745. \\
Bragoli, D., Rigon, M., and Zanetti, F. (2016). Optimal inflation
weights in the Euro Area. \textit{International Journal of Central
Banking}, 12(2):357\textendash 383. \\
Caliendo, L., Dvorkin, M., and Parro, F. (2019). Trade and labor market
dynamics: General equilibrium analysis of the china trade shock. \textit{Econometrica},
87(3):741\textendash 835. \\
Cantelmo, A. and Melina, G. (2018). Monetary policy and the relative
price of durable goods. \textit{Journal of Economic Dynamics and Control},
86(C):1\textendash 48. \\
Cantore, C., Ferroni, F., and Leon-Ledesma, M. A. (2017a). The dynamics
of hours worked and technology. \textit{Journal of Economic Dynamics
and Control}, 82(C):67\textendash 82. \\
Cantore, C., Leon-Ledesma, M., McAdam, P., and Willman, A. (2014).
Shocking stuff: technology, hours, and factor substitution. \textit{Journal
of the European Economic Association}, 12(1):108\textendash 128. \\
Cantore, C. and Levine, P. (2012). Getting normalization right: dealing
with dimensional constants in macroeconomics. \textit{Journal of Economic
Dynamics and Control}, 36(12):1931\textendash{} 1949. \\
Cantore, C., Levine, P., Melina, G., and Pearlman, J. (2019). Optimal
fiscal and monetary policy, debt crisis, and management. \textit{Macroeconomic
Dynamics}, 23(3):1166\textendash 1204.\\
Cantore, C., Levine, P., Melina, G., and Yang, B. (2012). A fiscal
stimulus with deep habits and optimal monetary policy. \textit{Economics
Letters}, 117(1):348\textendash 353. \\
Cantore, C., Levine, P., Pearlman, J., and Yang, B. (2015). CES technology
and business cycle fluctuations. \textit{Journal of Economic Dynamics
and Control}, 61(C):133\textendash 151. \\
Cardi, O. and Restout, R. (2015). Imperfect mobility of labor across
sectors: a reappraisal of the Balassa-Samuelson effect. \textit{Journal
of International Economics}, 97(2):249\textendash 265. \\
Carlstrom, C. T., Fuerst, T. S., and Ghironi, F. (2006). Does it matter
(for equilibrium determinacy) what price index the central bank targets?
\textit{Journal of Economic Theory}, 128(1):214\textendash 231. \\
Christiano, L. J., Eichenbaum, M., and Evans, C. L. (2005). Nominal
rigidities and the dynamic effects of a shock to monetary policy.
\textit{Journal of Political Economy}, 113(1):1\textendash 45. \\
Davis, S. J. and Haltiwanger, J. (2001). Sectoral job creation and
destruction responses to oil price changes. \textit{Journal of Monetary
Economics}, 48(3):465\textendash 512. \\
Di Pace, F. and Villa, S. (2016). Factor complementarity and labour
market dynamics. \textit{European Economic Review}, 82(C):70\textendash 112.
\\
Dix-Carneiro, R. (2014). Trade liberalization and labor market dynamics.
\textit{Econometrica}, 82(3):825\textendash 885. \\
Erceg, C. and Levin, A. (2006). Optimal monetary policy with durable
consumption goods. \textit{Journal of Monetary Economics}, 53(7):1341\textendash 1359.
\\
Faia, E. (2008). Optimal monetary policy rules with labor market frictions.
\textit{Journal of Economic Dynamics and Control}, 32(5):1600\textendash 1621.
\\
Foerster, A. T., Sarte, P.-D. G., and Watson, M. W. (2011). Sectoral
versus Aggregate Shocks: A Structural Factor Analysis of Industrial
Production. \textit{Journal of Political Econ- omy}, 119(1):1\textendash 38.
\\
Fuhrer, J. C. (2000). Habit formation in consumption and its implications
for monetary policy models. \textit{American Economic Review}, 90(3):367\textendash 390.
\\
Gallipoli, G. and Pelloni, G. (2013). Macroeconomic Effects of Job
Reallocations: A Survey. \textit{Review of Economic Analysis}, 5(2):127\textendash 176.
\\
Gerberding, C., Gerke, R., and Hammermann, F. (2012). Price-level
targeting when there is price-level drift. \textit{Journal of Macroeconomics},
34(3):757\textendash 768. \\
Giannoni, M. P. (2014). Optimal interest-rate rules and inflation
stabilization versus price-level stabilization. \textit{Journal of
Economic Dynamics and Control}, 41(C):110\textendash 129. \\
Horvath, M. (2000). Sectoral shocks and aggregate fluctuations. \textit{Journal
of Monetary Economics}, 45(1):69\textendash 106. \\
Huang, K. X. and Liu, Z. (2005). Inflation targeting: What inflation
rate to target? \textit{Journal of Monetary Economics}, 52(8):1435\textendash 1462.
\\
Iacoviello, M. and Neri, S. (2010). Housing market spillovers: Evidence
from an estimated DSGE model. \textit{American Economic Journal: Macroeconomics},
2(2):125\textendash 64. \\
Jeske, K. and Liu, Z. (2013). Should the central bank be concerned
about housing prices? \textit{Macroeconomic Dynamics}, 17(01):29\textendash 53.
\\
Jovanovic, B. and Moffitt, R. (1990). An estimate of a sectoral model
of labor mobility. \textit{Journal of Political Economy}, 98(4):827\textendash 852.
\\
Kara, E. (2010). Optimal monetary policy in the generalized Taylor
economy. \textit{Journal of Economic Dynamics and Control}, 34(10):2023\textendash 2037.
\\
Katayama, M. and Kim, K. H. (2018). Intersectoral labor immobility,
sectoral comovement, and news shocks. \textit{Journal of Money, Credit
and Banking}, 50(1):77\textendash 114. \\
Kim, K. H. and Katayama, M. (2013). Non-separability and sectoral
comovement in a sticky price model. \textit{Journal of Economic Dynamics
and Control}, 37(9):1715\textendash 1735. \\
La\textquoteright O, J. and Tahbaz-Salehi, A. (2020). Optimal Monetary
Policy in Production Networks. NBER Working Papers 27464, National
Bureau of Economic Research, Inc. \\
Lee, D. and Wolpin, K. I. (2006). Intersectoral Labor Mobility and
the Growth of the Service Sector. \textit{Econometrica}, 74(1):1\textendash 46.
\\
Levin, A. T., Onatski, A., Williams, J., and Williams, N. M. (2006).
Monetary Policy Under Uncertainty in Micro-Founded Macroeconometric
Models. In \textit{NBER Macroeconomics Annual 2005, Volume 20}, NBER
Chapters, pages 229\textendash 312. National Bureau of Economic Research,
Inc. \\
Levine, P., McAdam, P., and Pearlman, J. (2008). Quantifying and sustaining
welfare gains from monetary commitment. \textit{Journal of Monetary
Economics}, 55(7):1253\textendash 1276. \\
Lopez-Salido, D. and Levin, A. T. (2004). Optimal Monetary Policy
with Endogenous Capital Accumulation. Technical report. \\
Mankiw, N. G. and Reis, R. (2003). What measure of inflation should
a central bank target? \textit{Journal of the European Economic Association},
1(5):1058\textendash 1086. \\
McCully, C. P., Moyer, B. C., and Stewart, K. J. (2007). A Reconciliation
between the Consumer Price Index and the Personal Consumption Expenditures
Price Index. BEA Papers 0079, Bureau of Economic Analysis. \\
McKnight, S. (2018). Investment And Forward-Looking Monetary Policy:
A Wicksellian Solution To The Problem Of Indeterminacy. \textit{Macroeconomic
Dynamics}, 22(05):1345\textendash 1369. \\
Melina, G. and Villa, S. (2018). Leaning against windy bank lending.
\textit{Economic Inquiry}, 56(1):460\textendash 482. \\
Monacelli, T. (2008). Optimal monetary policy with collateralized
household debt and bor- rowing constraints. In \textit{Asset Prices
and Monetary Policy}, NBER Chapters, pages 103\textendash 146. National
Bureau of Economic Research, Inc. \\
Monacelli, T. (2009). New Keynesian models, durable goods, and collateral
constraints. \textit{Journal of Monetary Economics}, 56(2):242\textendash 254.
\\
Nakamura, E. and Steinsson, J. (2008). Five facts about prices: A
reevaluation of menu cost models. \textit{The Quarterly Journal of
Economics}, 123(4):1415\textendash 1464. \\
Nisticò, S. (2007). The welfare loss from unstable inflation. \textit{Economics
Letters}, 96(1):51\textendash 57. \\
Pasten, E., Schoenle, R., and Weber, M. (2020). The propagation of
monetary policy shocks in a heterogeneous production economy. \textit{Journal
of Monetary Economics}, Forthcoming. \\
Petrella, I., Rossi, R., and Santoro, E. (2019). Monetary policy with
sectoral trade-offs. \textit{The Scandinavian Journal of Economics},
121(1):55\textendash 88.\\
Petrella, I. and Santoro, E. (2011). Input\textendash output interactions
and optimal monetary policy. \textit{Journal of Economic Dynamics
and Control}, Elsevier, 35(11):1817\textendash 1830. \\
Rotemberg, J. J. (1982). Monopolistic price adjustment and aggregate
output. \textit{Review of Economic Studies}, 49(4):517\textendash 31.
\\
Rubbo, E. (2020). Networks, Phillips Curves, and Monetary Policy.
Mimeo.\\
Schmitt-Grohe, S. and Uribe, M. (2007). Optimal simple and implementable
monetary and fiscal rules. \textit{Journal of Monetary Economics},
54(6):1702\textendash 1725. \\
Smets, F. and Villa, S. (2016). Slow recoveries: Any role for corporate
leverage? \textit{Journal of Economic Dynamics and Control}, 70(C):54\textendash 85.
\\
Smets, F. and Wouters, R. (2007). Shocks and frictions in US business
cycles: A Bayesian DSGE approach. \textit{American Economic Review},
97(3):586\textendash 606. \\
Sterk, V. and Tenreyro, S. (2018). The transmission of monetary policy
through redistributions and durable purchases. \textit{Journal of
Monetary Economics}, 99:124 \textendash{} 137. \\
Strum, B. E. (2009). Monetary policy in a forward-looking input-output
economy. \textit{Journal of Money, Credit and Banking}, 41(4):619\textendash 650.
\\
Woodford, M. (2003). \textit{Interest and Prices. Foundations of a
Theory of Monetary Policy}. Princeton University Press, Princeton,
NJ. \\
Zubairy, S. (2014). On fiscal multipliers: Estimates from a medium
scale DSGE model. \textit{International Economic Review}, 55:169\textendash 195.
\end{singlespace}

\newpage{}

\appendix
%dummy comment inserted by tex2lyx to ensure that this paragraph is not empty
%dummy comment inserted by tex2lyx to ensure that this paragraph is not empty
\pagenumbering{Roman}
\numberwithin{table}{section}
\numberwithin{figure}{section}
\numberwithin{equation}{section}

\section*{Appendix}

\section{Symmetric equilibrium\label{sec:Symmetric-equilibrium}}

\subsection{Baseline model\label{subsec:Baseline-model}}

\begin{eqnarray}
X_{t} & = & C_{t}^{1-\alpha}D_{t}^{\alpha}\label{Aggregator-1}\\
U\left(X_{t},N_{t}\right) & = & \text{log}\left(X_{t}\right)-\nu\frac{N_{t}^{1+\varphi}}{1+\varphi}\label{Utility-1}\\
U_{C,t} & = & \frac{\left(1-\alpha\right)}{C_{t}}\label{UZ}\\
U_{D,t} & = & \frac{\alpha}{D_{t}}\label{UD}\\
\frac{C_{t}}{D_{t}} & = & \frac{1-\alpha}{\alpha}Q_{t}\label{reldemands}\\
w_{t}^{C} & = & -\dfrac{-\nu\left(\chi^{C}\right)^{-\dfrac{1}{\lambda}}\left(N_{t}^{C}\right)^{\dfrac{1}{\lambda}}N_{t}^{\varphi-\frac{1}{\lambda}}}{U_{C,t}},\label{UN}\\
w_{t}^{D} & = & -\dfrac{-\nu\left(1-\chi^{C}\right)^{-\dfrac{1}{\lambda}}\left(N_{t}^{D}\right)^{\dfrac{1}{\lambda}}N_{t}^{\varphi-\frac{1}{\lambda}}}{U_{C,t}}.\\
N_{t} & = & \left[\left(\chi^{C}\right)^{-\frac{1}{\lambda}}\left(N_{t}^{C}\right)^{\frac{1+\lambda}{\lambda}}+\left(1-\chi^{C}\right)^{-\frac{1}{\lambda}}\left(N_{t}^{D}\right)^{\frac{1+\lambda}{\lambda}}\right]^{\frac{\lambda}{1+\lambda}}\\
\Lambda_{t,t+1} & \equiv & \beta\frac{U_{C,t+1}}{U_{C,t}}\frac{e_{t+1}^{B}}{e_{t}^{B}}\label{Disc}\\
1 & = & E_{t}\left[\Lambda_{t,t+1}\frac{R_{t}}{\Pi_{t+1}^{C}}\right]\\
\Pi_{t}^{D} & = & \Pi_{t}^{C}\frac{Q_{t}}{Q_{t-1}}\\
Y_{t}^{C} & = & e_{t}^{A}e_{t}^{A,C}N_{t}^{C}\label{Wage markup}\\
Y_{t}^{D} & = & e_{t}^{A}e_{t}^{A,D}N_{t}^{D}\\
\epsilon_{c}MC_{t}^{C} & = & \left(\epsilon_{c}-1\right)+\vartheta_{c}\left(\Pi_{t}^{C}-\Pi^{C}\right)\Pi_{t}^{C}-\nonumber \\
 & - & \vartheta_{c}E_{t}\left[\Lambda_{t,t+1}\frac{Y_{t+1}^{C}}{Y_{t}^{C}}\left(\Pi_{t+1}^{C}-\Pi^{C}\right)\Pi_{t+1}^{C}\right]\\
MC_{t}^{C} & = & \frac{w_{t}^{C}}{e_{t}^{A}e_{t}^{A,C}}\label{Inv}
\end{eqnarray}
\begin{eqnarray}
\epsilon_{d}MC_{t}^{D} & = & \left(\epsilon_{d}-1\right)+\vartheta_{d}\left(\Pi_{t}^{D}-\Pi^{D}\right)\Pi_{t}^{D}-\nonumber \\
 & - & \vartheta_{d}E_{t}\left[\Lambda_{t,t+1}\frac{Q_{t+1}}{Q_{t}}\frac{Y_{t+1}^{D}}{Y_{t}^{D}}\left(\Pi_{t+1}^{D}-\Pi^{D}\right)\Pi_{t+1}^{D}\right]\\
MC_{t}^{D} & = & \frac{w_{t}^{D}}{e_{t}^{A}e_{t}^{A,D}Q_{t}}\\
\widetilde{\Pi}_{t} & = & \left(\Pi_{t}^{C}\right)^{1-\tau}\left(\Pi_{t}^{D}\right)^{\tau}\\
\log\left(\frac{R_{t}}{\bar{R}}\right) & = & \rho_{r}\log\left(\frac{R_{t-1}}{\bar{R}}\right)+\alpha_{\pi}\log\left(\frac{\tilde{\Pi}_{t}}{\tilde{\Pi}}\right)+\alpha_{y}\log\left(\frac{Y_{t}}{Y_{t}^{f}}\right)+\nonumber \\
 & + & \alpha_{\Delta y}\left[\log\left(\frac{Y_{t}}{Y_{t}^{f}}\right)-\log\left(\frac{Y_{t-1}}{Y_{t-1}^{f}}\right)\right],\\
Y_{t}^{C} & = & C_{t}+\frac{\vartheta_{c}}{2}\left(\Pi_{t}^{C}-\Pi^{C}\right)^{2}Y_{t}^{C}\\
Y_{t}^{D} & = & D_{t}+\frac{\vartheta_{d}}{2}\left(\Pi_{t}^{D}-\Pi^{D}\right)^{2}Y_{t}^{D}\label{YD MC}\\
Y_{t} & = & Y_{t}^{C}+Q_{t}Y_{t}^{D}\label{Ybaseline}
\end{eqnarray}

\subsection{Two-sector model with durable goods}

The symmetric equilibrium changes as follows. Durable goods follow
the low of motion
\begin{equation}
D_{t+1}=(1-\delta)D_{t}+I_{t}^{D}.\label{Durables LOM-2}
\end{equation}
Equation \eqref{reldemands} now reads as
\begin{equation}
Q_{t}=\frac{U_{D,t}}{U_{C,t}}+\left(1-\delta\right)E_{t}\left[\Lambda_{t,t+1}Q_{t+1}\right].\label{eq:RelPriceD-1}
\end{equation}
Finally, the market clearing condition \eqref{YD MC} in sector $D$
becomes
\begin{equation}
Y_{t}^{D}=I_{t}^{D}+\frac{\vartheta_{d}}{2}\left(\Pi_{t}^{D}-\Pi^{D}\right)^{2}Y_{t}^{D}.
\end{equation}

\subsection{Fully-fledged two-sector model}

\begin{eqnarray}
X_{t} & = & C_{t}^{1-\alpha}D_{t}^{\alpha}\label{Aggregator-1-1}\\
C_{t} & = & Z_{t}-\zeta S_{t-1}\label{Habits-1-1}\\
S_{t} & = & \rho_{c}S_{t-1}+(1-\rho_{c})Z_{t}\label{motion_c-1-1}\\
U\left(X_{t},N_{t}\right) & = & \text{log}\left(X_{t}\right)-\nu\frac{N_{t}^{1+\varphi}}{1+\varphi}\label{Utility-1-1}\\
U_{C,t} & = & \frac{\left(1-\alpha\right)}{C_{t}}\label{UZ-1}\\
U_{D,t} & = & \frac{\alpha}{D_{t}}\label{UD-1}\\
U_{N^{C},it} & = & -\nu\left(\chi^{C}\right)^{-\frac{1}{\lambda}}\left(N_{i,t}^{C}\right)^{\frac{1}{\lambda}}\left[\left(\chi^{C}\right)^{-\frac{1}{\lambda}}\left(N_{i,t}^{C}\right)^{\frac{1+\lambda}{\lambda}}+\left(1-\chi^{C}\right)^{-\frac{1}{\lambda}}\left(N_{i,t}^{D}\right)^{\frac{1+\lambda}{\lambda}}\right]^{\frac{\lambda\varphi-1}{1+\lambda}}\label{UN-1}\\
U_{N^{D},it} & = & -\nu\left(1-\chi^{C}\right)^{-\frac{1}{\lambda}}\left(N_{i,t}^{D}\right)^{\frac{1}{\lambda}}\left[\left(\chi^{C}\right)^{-\frac{1}{\lambda}}\left(N_{i,t}^{C}\right)^{\frac{1+\lambda}{\lambda}}+\left(1-\chi^{C}\right)^{-\frac{1}{\lambda}}\left(N_{i,t}^{D}\right)^{\frac{1+\lambda}{\lambda}}\right]^{\frac{\lambda\varphi-1}{1+\lambda}}\\
N_{t} & = & \left[\left(\chi^{C}\right)^{-\frac{1}{\lambda}}\left(N_{t}^{C}\right)^{\frac{1+\lambda}{\lambda}}+\left(1-\chi^{C}\right)^{-\frac{1}{\lambda}}\left(N_{t}^{D}\right)^{\frac{1+\lambda}{\lambda}}\right]^{\frac{\lambda}{1+\lambda}}\\
w_{t}^{C} & = & -\dfrac{-\nu\left(\chi^{C}\right)^{-\dfrac{1}{\lambda}}\left(N_{t}^{C}\right)^{\dfrac{1}{\lambda}}N_{t}^{\varphi-\frac{1}{\lambda}}}{U_{C,t}},\\
w_{t}^{D} & = & -\dfrac{-\nu\left(1-\chi^{C}\right)^{-\dfrac{1}{\lambda}}\left(N_{t}^{D}\right)^{\dfrac{1}{\lambda}}N_{t}^{\varphi-\frac{1}{\lambda}}}{U_{C,t}}.\\
\Lambda_{t,t+1} & \equiv & \beta\frac{U_{C,t+1}}{U_{C,t}}\frac{e_{t+1}^{B}}{e_{t}^{B}}\label{Disc-1}\\
0 & = & \left[1-e_{t}^{w,C}\eta\right]+\frac{e_{t}^{w,C}\eta}{\tilde{\mu_{t}}^{C}}-\vartheta_{C}^{w}\left(\Pi_{t}^{w,C}-\Pi^{C}\right)\Pi_{t}^{w,C}+\nonumber \\
 & + & E_{t}\left[\Lambda_{t,t+1}\vartheta_{C}^{w}\left(\Pi_{t+1}^{w,C}-\Pi^{C}\right)\Pi_{t+1}^{w,C}\frac{w_{t+1}^{C}N_{t+1}^{C}}{w_{t}^{C}N_{t}^{C}}\right]\\
\tilde{\mu_{t}}^{C} & = & -\frac{U_{C,t}}{U_{N,t}^{C}}w_{t}^{C}\label{Wage markup-1}\\
0 & = & \left[1-e_{t}^{w,D}\eta\right]+\frac{e_{t}^{w,D}\eta}{\tilde{\mu_{t}}^{D}}-\vartheta_{D}^{w}\left(\Pi_{t}^{w,D}-\Pi^{C}\right)\Pi_{t}^{w,D}+\nonumber \\
 & + & E_{t}\left[\Lambda_{t,t+1}\vartheta_{D}^{w}\left(\Pi_{t+1}^{w,D}-\Pi^{C}\right)\Pi_{t+1}^{w,D}\frac{w_{t+1}^{D}N_{t+1}^{D}}{w_{t}^{D}N_{t}^{D}}\right]\\
\tilde{\mu_{t}}^{D} & = & -\frac{U_{C,t}}{U_{N,t}^{D}}w_{t}^{D}\\
Q_{t}\psi_{t} & = & \frac{U_{D,t}}{U_{Z,t}}+\left(1-\delta\right)E_{t}\left[\Lambda_{t,t+1}Q_{t+1}\psi_{t+1}\right]\label{Q}
\end{eqnarray}

\begin{singlespace}
\begin{eqnarray}
1 & = & E_{t}\left\{ \Lambda_{t,t+1}\psi_{t+1}\frac{Q_{t+1}}{Q_{t}}e_{t+1}^{I}\left[S^{'}\left(\frac{I_{t+1}^{D}}{I_{t}^{D}}\right)\left(\frac{I_{t+1}^{D}}{I_{t}^{D}}\right)^{2}\right]\right\} +\nonumber \\
 & + & \psi_{t}e_{t}^{I}\left[1-S\left(\frac{I_{t}^{D}}{I_{t-1}^{D}}\right)-S^{'}\left(\frac{I_{t}^{D}}{I_{t-1}^{D}}\right)\frac{I_{t}^{D}}{I_{t-1}^{D}}\right]\\
S\left(\frac{I_{t}^{D}}{I_{t-1}^{D}}\right) & = & \frac{\phi}{2}\left(\frac{I_{t}^{D}}{I_{t-1}^{D}}-1\right)^{2}\\
S^{'}\left(\frac{I_{t}^{D}}{I_{t-1}^{D}}\right) & = & \phi\left(\frac{I_{t}^{D}}{I_{t-1}^{D}}-1\right)\\
1 & = & E_{t}\left[\Lambda_{t,t+1}\frac{R_{t}}{\Pi_{t+1}^{C}}\right]\\
\Pi_{t}^{D} & = & \Pi_{t}^{C}\frac{Q_{t}}{Q_{t-1}}\\
Y_{t}^{C} & = & e_{t}^{A}N_{t}^{C}\\
Y_{t}^{D} & = & e_{t}^{A}N_{t}^{D}\\
e_{t}^{C}\epsilon_{c}MC_{t}^{C} & = & \left(e_{t}^{C}\epsilon_{c}-1\right)+\vartheta_{c}\left(\Pi_{t}^{C}-\Pi^{C}\right)\Pi_{t}^{C}-\nonumber \\
 & - & \vartheta_{c}E_{t}\left[\Lambda_{t,t+1}\frac{Y_{t+1}^{C}}{Y_{t}^{C}}\left(\Pi_{t+1}^{C}-\Pi^{C}\right)\Pi_{t+1}^{C}\right]\label{Price_settC-1-1}\\
MC_{t}^{C} & = & \frac{w_{t}^{C}}{e_{t}^{A}}\label{MCC-1-1}\\
e_{t}^{D}\epsilon_{d}MC_{t}^{D} & = & \left(e_{t}^{D}\epsilon_{d}-1\right)+\vartheta_{d}\left(\Pi_{t}^{D}-\Pi^{D}\right)\Pi_{t}^{D}-\nonumber \\
 & - & \vartheta_{d}E_{t}\left[\Lambda_{t,t+1}\frac{Q_{t+1}}{Q_{t}}\frac{Y_{t+1}^{D}}{Y_{t}^{D}}\left(\Pi_{t+1}^{D}-\Pi^{D}\right)\Pi_{t+1}^{D}\right]\label{Price_settD-1-1}\\
MC_{t}^{D} & = & \frac{w_{t}^{D}}{e_{t}^{A}Q_{t}}\label{MCD-1-1}\\
\widetilde{\Pi}_{t} & = & \left(\Pi_{t}^{C}\right)^{1-\tau}\left(\Pi_{t}^{D}\right)^{\tau}\label{InflationAgg-1-1}\\
\log\left(\frac{R_{t}}{\bar{R}}\right) & = & \rho_{r}\log\left(\frac{R_{t-1}}{\bar{R}}\right)+\alpha_{\pi}\log\left(\frac{\tilde{\Pi}_{t}}{\tilde{\Pi}}\right)+\alpha_{y}\log\left(\frac{Y_{t}}{Y_{t}^{f}}\right)+\nonumber \\
 & + & \alpha_{\Delta y}\left[\log\left(\frac{Y_{t}}{Y_{t}^{f}}\right)-\log\left(\frac{Y_{t-1}}{Y_{t-1}^{f}}\right)\right],\\
Y_{t}^{C} & = & C_{t}+e_{t}^{G}+\frac{\vartheta_{c}}{2}\left(\Pi_{t}^{C}-\Pi^{C}\right)^{2}Y_{t}^{C}+\frac{\vartheta_{c}^{w}}{2}\left(\frac{w_{i,t}^{C}}{w_{i,t-1}^{C}}\Pi_{t}^{C}-\Pi^{C}\right)^{2}w_{t}^{C}N_{t}^{C}\label{eq:YC-1}\\
Y_{t}^{D} & = & I_{t}^{D}+\frac{\vartheta_{d}}{2}\left(\Pi_{t}^{D}-\Pi^{D}\right)^{2}Y_{t}^{D}+\frac{\vartheta_{d}^{w}}{2}\left(\frac{w_{i,t}^{D}}{w_{i,t-1}^{D}}\Pi_{t}^{C}-\Pi^{C}\right)^{2}w_{t}^{D}N_{t}^{D}\\
Y_{t} & = & Y_{t}^{C}+Q_{t}Y_{t}^{D}\label{AggY-1-1}
\end{eqnarray}

\end{singlespace}

\pagebreak{}

\section{Ramsey problem}

In this section we outline the Ramsey problem in the stylized model
of Section 2. For illustrative purposes, we focus on the simpler version
of the model, e.g. without durable goods. The symmetric equilibrium
of the model is reported in Appendix \ref{subsec:Baseline-model}.
The social planner maximizes the present value of households' utility
subject to the equilibrium conditions of the model, but does not have
to follow an interest rate rule. We now report the Lagrangian function
of the optimization problem (\ref{subsec:Lagrangian-function}), the
first-order conditions (\ref{subsec:Ramsey-planner=002019s-first-order})
and the steady state procedure (\ref{subsec: RamseySteady-state}).\footnote{To derive the Ramsey first-order conditions we use the toolbox provided
by Lopez-Salido and Levin (2004) and Levin et al. (2006).}

\subsection{Lagrangian function\label{subsec:Lagrangian-function}}

{\tiny{}
\[
\mathcal{L}_{t}=E_{t}\sum_{t=0}^{\infty}\beta^{t}\begin{Bmatrix} & e_{t}^{B}\text{log}\left(X_{t}\right)-\nu\frac{N_{t}^{1+\varphi}}{1+\varphi}-w_{r}\left(R_{t}-R\right)^{2}\\
+\lambda_{1,t} & \left[X_{t}-C_{t}^{1-\alpha}D_{t}^{\alpha}\right]\\
+\lambda_{2,t} & \left[U_{C,t}-\frac{\left(1-\alpha\right)}{C_{t}}\right]\\
+\lambda_{3,t} & \left[U_{D,t}-\frac{\alpha}{D_{t}}\right]\\
+\lambda_{4,t} & \left[\frac{C_{t}}{D_{t}}-\frac{1-\alpha}{\alpha}Q_{t}\right]\\
+\lambda_{5,t} & \left[w_{t}^{C}+\frac{U_{N,t}}{U_{C,t}}\right]\\
+\lambda_{6,t} & \left[U_{N,t}+\nu\left(\chi^{C}\right)^{-\frac{1}{\lambda}}\left(N_{t}^{C}\right)^{\frac{1}{\lambda}}N_{t}^{\varphi-\frac{1}{\lambda}}\right]\\
+\lambda_{7,t} & \left[\frac{w_{t}^{C}}{w_{t}^{D}}-\left(\frac{\chi^{C}}{1-\chi^{C}}\right)^{-\frac{1}{\lambda}}\left(\frac{N_{t}^{C}}{N_{t}^{D}}\right)^{\frac{1}{\lambda}}\right]\\
+\lambda_{8,t} & \left[\Lambda_{t,t+1}-\beta\frac{U_{C,t+1}}{U_{C,t}}\frac{e_{t+1}^{B}}{e_{t}^{B}}\right]\\
+\lambda_{9,t} & \left[1-\Lambda_{t,t+1}\frac{R_{t}}{\Pi_{t+1}^{C}}\right]\\
+\lambda_{10,t} & \left[\Pi_{t}^{D}-\Pi_{t}^{C}\frac{Q_{t}}{Q_{t-1}}\right]\\
+\lambda_{11,t} & \left[R_{t}^{\text{real}}-\frac{R_{t}}{\Pi_{t+1}^{C}}\right]\\
+\lambda_{12,t} & \left[Y_{t}^{C}-e_{t}^{A}e_{t}^{A,C}N_{t}^{C}\right]\\
+\lambda_{13,t} & \left[Y_{t}^{D}-e_{t}^{A}e_{t}^{A,D}N_{t}^{D}\right]\\
+\lambda_{14,t} & \Biggl[\epsilon_{c}MC_{t}^{C}-\left(\epsilon_{c}-1\right)-\vartheta_{c}\left(\Pi_{t}^{C}-\Pi^{C}\right)\Pi_{t}^{C}+\vartheta_{c}E_{t}\left[\Lambda_{t,t+1}\frac{Y_{t+1}^{C}}{Y_{t}^{C}}\left(\Pi_{t+1}^{C}-\Pi^{C}\right)\Pi_{t+1}^{C}\right]\\
+\lambda_{15,t} & \Biggl[\epsilon_{d}MC_{t}^{D}-\left(\epsilon_{d}-1\right)-\vartheta_{d}\left(\Pi_{t}^{D}-\Pi^{D}\right)\Pi_{t}^{D}+\vartheta_{d}E_{t}\left[\Lambda_{t,t+1}\frac{Q_{t+1}}{Q_{t}}\frac{Y_{t+1}^{D}}{Y_{t}^{D}}\left(\Pi_{t+1}^{D}-\Pi^{D}\right)\Pi_{t+1}^{D}\right]\Biggr]\\
+\lambda_{16,t} & \left[MC_{t}^{C}-\frac{w_{t}^{C}}{e_{t}^{A}e_{t}^{A,C}}\right]\\
+\lambda_{17,t} & \left[MC_{t}^{D}-\frac{w_{t}^{D}}{e_{t}^{A}Q_{t}}\right]\\
+\lambda_{18,t} & \left[Y_{t}^{C}-C_{t}-\frac{\vartheta_{c}}{2}\left(\Pi_{t}^{C}-\Pi^{C}\right)^{2}Y_{t}^{C}\right]\\
+\lambda_{19,t} & \left[Y_{t}^{D}-D_{t}-\frac{\vartheta_{d}}{2}\left(\Pi_{t}^{D}-\Pi^{D}\right)^{2}Y_{t}^{D}\right]\\
+\lambda_{20,t} & \left[Y_{t}-Y_{t}^{C}-Q_{t}Y_{t}^{D}\right]\\
+\lambda_{21,t} & \left[N_{t}-\left[\left(\chi^{C}\right)^{-\frac{1}{\lambda}}\left(N_{t}^{C}\right)^{\frac{1+\lambda}{\lambda}}+\left(1-\chi^{C}\right)^{-\frac{1}{\lambda}}\left(N_{t}^{D}\right)^{\frac{1+\lambda}{\lambda}}\right]^{\frac{\lambda}{1+\lambda}}\right]
\end{Bmatrix}
\]
}{\tiny\par}

\subsection{Ramsey planner\textquoteright s first-order conditions\label{subsec:Ramsey-planner=002019s-first-order}}

Differentiating the Lagrangian function reported in Section \ref{subsec:Lagrangian-function}
with respect to all the endogenous variables $X_{t},C_{t},D_{t},U_{C,t},U_{D,t},Q_{t},U_{N,t},W_{t}^{C},W_{t}^{D},N_{t},\Lambda_{t,t+1},$$R_{t},R_{t}^{\text{real}},\Pi_{t}^{C},\Pi_{t}^{D},Y_{t}^{C},$$N_{t}^{C},$$Y_{t}^{D},$$N_{t}^{D},$
$MC_{t}^{C},$$MC_{t}^{D}$$,Y_{t}$ and setting the first derivatives
equal to zero yields the following first-order conditions:

{\footnotesize{}
\begin{align}
0= & \lambda_{3,t}\frac{\alpha}{D_{t}^{2}}-\lambda_{4,t}\frac{C_{t}}{D_{t}^{2}}-\lambda_{19,t}-\lambda_{1,t}\alpha D_{t}^{\alpha-1}C_{t}^{1-\alpha}\\
0= & \lambda_{8,t}-\lambda_{9,t}R_{t}^{\text{real}}+\frac{\lambda_{14,t}\vartheta_{c}\left(\Pi_{t+1}^{C}-\Pi^{C}\right)\Pi_{t+1}^{C}Y_{t}^{C}}{Y_{t}^{C}}+\frac{\lambda_{15,t}\vartheta_{d}\left(\Pi_{t+1}^{D}-\Pi^{C}\right)\Pi_{t+1}^{D}Q_{t+1}Y_{t}^{D}}{Q_{t}Y_{t}^{C}}\\
0= & \lambda_{16,t}+\lambda_{14,t}\epsilon_{c}\\
0= & \lambda_{17,t}+\lambda_{15,t}\epsilon_{d}\\
0= & \lambda_{21,t}-\nu N_{t}^{\varphi}+\frac{\lambda_{6,t}\nu\left(\varphi-\frac{1}{\lambda}\right)\left(N_{t}^{C}\right)^{\frac{1}{\lambda}}N_{t}^{\varphi-\frac{1}{\lambda}-1}}{\left(\chi^{C}\right)^{\frac{1}{\lambda}}}\\
0= & \frac{\lambda_{6,t}\nu\left(N_{t}^{C}\right)^{\frac{1}{\lambda}-1}N_{t}^{\varphi-\frac{1}{\lambda}}}{\lambda\left(\chi^{C}\right)^{\frac{1}{\lambda}}}-\frac{\lambda_{21,t}\left(N_{t}^{C}\right)^{\frac{1}{\lambda}}\left(\frac{\left(N_{t}^{C}\right)^{\frac{1+\lambda}{\lambda}}}{\left(\chi^{C}\right)^{\frac{1}{\lambda}}}+\frac{\left(N_{t}^{D}\right)^{\frac{1+\lambda}{\lambda}}}{\left(1-\chi^{C}\right)^{\frac{1}{\lambda}}}\right)^{-\frac{1}{1+\lambda}}}{\left(\chi^{C}\right)^{\frac{1}{\lambda}}}-\nonumber \\
 & -\lambda_{12,t}e_{t}^{A}e_{t}^{A,C}+\frac{\lambda_{7,t}N_{t}^{D}\left(\frac{N_{t}^{D}}{N_{t}^{C}}\right)^{\frac{1}{\lambda}-1}}{\lambda\left(N_{t}^{C}\right)^{2}\left(-\frac{\chi^{C}-1}{\chi^{C}}\right)^{\frac{1}{\lambda}}}\\
0= & -\lambda_{13,t}e_{t}^{A}e_{t}^{A,D}-\frac{\lambda_{21,t}\left(N_{t}^{D}\right)^{\frac{1}{\lambda}}\left(\frac{\left(N_{t}^{C}\right)^{\frac{1+\lambda}{\lambda}}}{\left(\chi^{C}\right)^{\frac{1}{\lambda}}}+\frac{\left(N_{t}^{D}\right)^{\frac{1+\lambda}{\lambda}}}{\left(1-\chi^{C}\right)^{\frac{1}{\lambda}}}\right)^{-\frac{1}{1+\lambda}}}{\left(1-\chi^{C}\right)^{\frac{1}{\lambda}}}-\lambda_{7,t}\frac{\lambda_{7,t}\left(\frac{N_{t}^{D}}{N_{t}^{C}}\right)^{\frac{1}{\lambda}-1}}{\lambda N_{t}^{C}\left(-\frac{\chi^{C}-1}{\chi^{C}}\right)^{\frac{1}{\lambda}}}\\
0= & \frac{\lambda_{14,t-1}\left[\frac{\vartheta_{c}\Lambda_{t-1,t}\Pi_{t}^{C}Y_{t}^{C}}{Y_{t-1}^{C}}+\frac{\vartheta_{c}\Lambda_{t-1,t}\left(\Pi_{t}^{C}-\Pi^{C}\right)Y_{t}^{C}}{Y_{t-1}^{C}}\right]}{\beta}-\lambda_{14,t}\left(\vartheta_{c}\Pi_{t}^{C}+\vartheta_{c}\left(\Pi_{t}^{C}-\Pi^{C}\right)\right)-\nonumber \\
 & -\lambda_{10,t}\frac{Q_{t}}{Q_{t-1}}-\lambda_{18,t}\vartheta_{c}\left(\Pi_{t}^{C}-\Pi^{C}\right)Y_{t}^{C}+\frac{\lambda_{11,t-1}R_{t-1}}{\beta\left(\Pi_{t}^{C}\right)^{2}}\\
0= & \lambda_{10,t}-\lambda_{15,t}\left(\vartheta_{d}\Pi_{t}^{D}+\vartheta_{d}\left(\Pi_{t}^{D}-\Pi^{C}\right)\right)-\frac{\lambda_{15,t-1}\left[\frac{\vartheta_{d}\Lambda_{t-1,t}\Pi_{t}^{D}Q_{t}Y_{t}^{D}}{Q_{t-1}Y_{t-1}^{D}}+\frac{\vartheta_{d}\Lambda_{t-1,t}\left(\Pi_{t}^{D}-\Pi^{C}\right)Y_{t}^{D}}{Q_{t-1}Y_{t-1}^{D}}\right]}{\beta}-\nonumber \\
 & -\lambda_{19,t}\left(\vartheta_{d}Y_{t}^{D}\left(\Pi_{t}^{D}-\Pi^{C}\right)\right)\\
0= & \lambda_{4,t}\frac{\alpha-1}{\alpha}-\lambda_{10,t}\frac{\Pi_{t}^{C}}{Q_{t-1}}+\lambda_{17,t}\frac{w_{t}^{D}}{e_{t}^{A}e_{t}^{A,D}Q_{t}^{2}}+\lambda_{10,t}\frac{\beta\Pi_{t+1}^{C}Q_{t+1}}{Q_{t}^{2}}
\end{align}
\begin{align}
0= & -\frac{\lambda_{11,t}}{\Pi_{t+1}^{C}}-2w_{r}\left(R_{t}-R\right)\label{eq:wr}\\
0= & \lambda_{11,t}-\lambda_{9,t}\Lambda_{t,t+1}\\
0= & \lambda_{3,t}\\
0= & \lambda_{6,t}+\frac{\lambda_{5,t}}{U_{C,t}}\\
0= & \lambda_{2,t}-\lambda_{5,t}\frac{U_{N,t}}{U_{C,t}^{2}}+\lambda_{8,t}\beta\frac{e_{t+1}^{B}}{e_{t}^{B}}\frac{U_{C,t+1}}{U_{C,t}^{2}}-\lambda_{8,t-1}\frac{e_{t}^{B}}{e_{t-1}^{B}U_{C,t-1}}\\
0= & \lambda_{5,t}-\frac{\lambda_{16,t}}{e_{t}^{A}e_{t}^{A,C}}-\lambda_{7,t}\frac{w_{t}^{D}}{\left(w_{t}^{C}\right)^{2}}\\
0= & \frac{\lambda_{7,t}}{w_{t}^{C}}-\frac{\lambda_{17,t}}{e_{t}^{A}e_{t}^{A,D}Q_{t}}\\
0= & \lambda_{1,t}+\frac{1}{X_{t}}\\
0= & \lambda_{20,t}\\
0= & \lambda_{12,t}-\lambda_{20,t}-\lambda_{18,t}\left[\frac{\vartheta_{c}\left(\Pi_{t}^{C}-\Pi^{C}\right)^{2}}{2}-1\right]-\lambda_{14,t-1}\left[\frac{\vartheta_{c}\Lambda_{t,t+1}\left(\Pi_{t+1}^{C}-\Pi^{C}\right)\Pi_{t+1}^{C}Y_{t+1}^{C}}{\left(Y_{t}^{C}\right)^{2}}\right]+\nonumber \\
 & +\frac{\lambda_{14,t-1}\left[\vartheta_{c}\Lambda_{t-1,t}\left(\Pi_{t}^{C}-\Pi^{C}\right)\Pi_{t}^{C}\right]}{\beta Y_{t-1}^{C}}\\
0= & \lambda_{13,t}-\lambda_{20,t}Q-\lambda_{19,t}\left[\frac{\vartheta_{d}\left(\Pi_{t}^{D}-\Pi^{C}\right)^{2}}{2}-1\right]-\lambda_{15,t-1}\left[\frac{\vartheta_{d}\Lambda_{t,t+1}\left(\Pi_{t+1}^{D}-\Pi^{C}\right)\Pi_{t+1}^{D}Q_{t+1}Y_{t+1}^{D}}{Q_{t}\left(Y_{t}^{D}\right)^{2}}\right]+\nonumber \\
 & +\frac{\lambda_{15,t-1}\left[\vartheta_{d}\Lambda_{t-1,t}\left(\Pi_{t}^{D}-\Pi^{C}\right)\Pi_{t}^{D}Q_{t}\right]}{\beta Q_{t-1}Y_{t-1}^{D}}\\
0= & \frac{\lambda_{4,t}}{D_{t}}-\lambda_{18,t}-\lambda_{2,t}\frac{\alpha-1}{C_{t}^{2}}+\lambda_{1,t}\frac{\left(\alpha-1\right)D_{t}^{\alpha}}{C_{t}^{\alpha}}
\end{align}
}The Ramsey\textquoteright s first-order conditions together with
the 21 equations characterizing the symmetric equilibrium reported
in Section \ref{subsec:Baseline-model} (excluding the Taylor rule,
the inflation aggregator and the processes for the exogenous shocks)
make a system of 43 dynamic equations in 43 unknowns (22 endogenous
variables and 21 Lagrange multipliers). We approximate the solution
to this system by using the Dynare solver that takes a second-order
Taylor expansion around the Ramsey-optimal steady state, which we
compute numerically as described in Section \ref{subsec: RamseySteady-state}.

\subsection{Steady-state\label{subsec: RamseySteady-state}}

The steady-state values of all endogenous variables and Lagrange multipliers
in the Ramsey equilibrium are found simultaneously using a numerical
procedure. In particular, the procedure is designed to choose the
values of C and $\Pi^{C}$ that simultaneously solve equations \eqref{Ybaseline}
and \eqref{eq:wr} evaluated at the steady state. The value of the
remaining endogenous variables is found recursively by evaluating
equations in Section \ref{subsec:Baseline-model} at the steady state,
while the steady state values of the 21 Lagrange multipliers of the
Ramsey problem are found by solving the system of 21 equations (linear
in the Lagrange multipliers) in 21 unknowns, obtained by evaluating
equations in Section \eqref{subsec:Ramsey-planner=002019s-first-order}
at the steady state. Note that value of $\Pi^{C}$ defines the optimal
steady state inflation under the Ramsey policy.

\section{Robustness analysis in the stylized model\label{sec:Robustness-stylized}}

In Table \ref{Table:robust_Stylized} we provide an analysis of robustness
to different assumptions on the level and sectoral distribution of
price stickiness, and to different degrees of labor mobility in the
stylized model (building on cases \textit{(ii)} and \textit{(iii)}
reported in Table 1). The inverse relationship between the optimal
$\tau$ and $\lambda$ continues to hold.

\begin{table}[H]
\setlength\tabcolsep{6 pt}
\renewcommand{\arraystretch}{1.35}

\centering{\small{}}%
\begin{tabular}{ccccccc}
\hline 
{\small{}$\lambda$} & {\small{}$\rho_{r}$} & {\small{}$\alpha_{\pi}$} & {\small{}$\alpha_{y}$} & {\small{}$\alpha_{\Delta y}$} & {\small{}$\tau$} & {\small{}$100\times\omega$}\tabularnewline
\hline 
\multicolumn{7}{c}{\selectlanguage{english}%
\textit{\small{}(i) Heterogeneous price stickiness $\vartheta_{c}=90,\vartheta_{d}=30$}\selectlanguage{american}%
}\tabularnewline
\selectlanguage{english}%
{\small{}$\infty$}\selectlanguage{american}%
 & \selectlanguage{english}%
{\small{}1.0000}\selectlanguage{american}%
 & \selectlanguage{english}%
{\small{}0.0081}\selectlanguage{american}%
 & \selectlanguage{english}%
{\small{}0.0216}\selectlanguage{american}%
 & \selectlanguage{english}%
{\small{}0.0000}\selectlanguage{american}%
 & \selectlanguage{english}%
{\small{}0.3143}\selectlanguage{american}%
 & \selectlanguage{english}%
{\small{}0.0003}\selectlanguage{american}%
\tabularnewline
\selectlanguage{english}%
{\small{}3}\selectlanguage{american}%
 & \selectlanguage{english}%
{\small{}1.0000}\selectlanguage{american}%
 & \selectlanguage{english}%
{\small{}0.0080}\selectlanguage{american}%
 & \selectlanguage{english}%
{\small{}0.0227}\selectlanguage{american}%
 & \selectlanguage{english}%
{\small{}0.0000}\selectlanguage{american}%
 & \selectlanguage{english}%
{\small{}0.3295}\selectlanguage{american}%
 & \selectlanguage{english}%
{\small{}0.0003}\selectlanguage{american}%
\tabularnewline
\selectlanguage{english}%
{\small{}1}\selectlanguage{american}%
 & \selectlanguage{english}%
{\small{}1.0000}\selectlanguage{american}%
 & \selectlanguage{english}%
{\small{}0.0084}\selectlanguage{american}%
 & \selectlanguage{english}%
{\small{}0.0210}\selectlanguage{american}%
 & \selectlanguage{english}%
{\small{}0.0000}\selectlanguage{american}%
 & \selectlanguage{english}%
{\small{}0.3847}\selectlanguage{american}%
 & \selectlanguage{english}%
{\small{}0.0005}\selectlanguage{american}%
\tabularnewline
\selectlanguage{english}%
{\small{}0.5}\selectlanguage{american}%
 & \selectlanguage{english}%
{\small{}1.0000}\selectlanguage{american}%
 & \selectlanguage{english}%
{\small{}0.0090}\selectlanguage{american}%
 & \selectlanguage{english}%
{\small{}0.0211}\selectlanguage{american}%
 & \selectlanguage{english}%
{\small{}0.0000}\selectlanguage{american}%
 & \selectlanguage{english}%
{\small{}0.4402}\selectlanguage{american}%
 & \selectlanguage{english}%
{\small{}0.0005}\selectlanguage{american}%
\tabularnewline
\selectlanguage{english}%
{\small{}0.10}\selectlanguage{american}%
 & \selectlanguage{english}%
{\small{}1.0000}\selectlanguage{american}%
 & \selectlanguage{english}%
{\small{}0.0102}\selectlanguage{american}%
 & \selectlanguage{english}%
{\small{}0.0202}\selectlanguage{american}%
 & \selectlanguage{english}%
{\small{}0.0000}\selectlanguage{american}%
 & \selectlanguage{english}%
{\small{}0.5779}\selectlanguage{american}%
 & \selectlanguage{english}%
{\small{}0.0009}\selectlanguage{american}%
\tabularnewline
\multicolumn{7}{c}{\textit{\small{}(ii) Heterogeneous price stickiness }\foreignlanguage{english}{\textit{\small{}$\vartheta_{c}=60,\vartheta_{d}=0$}}}\tabularnewline
{\small{}$\infty$} & {\small{}1.0000} & {\small{}0.0040} & {\small{}0.0215} & {\small{}0.0000} & {\small{}0.0000} & {\small{}0.0002}\tabularnewline
\selectlanguage{english}%
3\selectlanguage{american}%
 & {\small{}1.0000} & {\small{}0.0041} & {\small{}0.0217} & {\small{}0.0000} & {\small{}0.0225} & {\small{}0.0002}\tabularnewline
{\small{}1} & {\small{}1.0000} & {\small{}0.0042} & {\small{}0.0221} & {\small{}0.0000} & {\small{}0.0373} & {\small{}0.0002}\tabularnewline
\selectlanguage{english}%
0.5\selectlanguage{american}%
 & {\small{}1.0000} & {\small{}0.0043} & {\small{}0.0225} & {\small{}0.0000} & {\small{}0.0514} & {\small{}0.0003}\tabularnewline
{\small{}0.10} & {\small{}1.0000} & {\small{}0.0044} & {\small{}0.0231} & {\small{}0.0000} & {\small{}0.0709} & {\small{}0.0003}\tabularnewline
\multicolumn{7}{c}{\selectlanguage{english}%
\textit{\small{}(iii) Heterogeneous price stickiness $\vartheta_{c}=120,\vartheta_{d}=0$}\selectlanguage{american}%
}\tabularnewline
\selectlanguage{english}%
{\small{}$\infty$}\selectlanguage{american}%
 & \selectlanguage{english}%
{\small{}1.0000}\selectlanguage{american}%
 & \selectlanguage{english}%
{\small{}0.0076}\selectlanguage{american}%
 & \selectlanguage{english}%
{\small{}0.0197}\selectlanguage{american}%
 & \selectlanguage{english}%
{\small{}0.0000}\selectlanguage{american}%
 & \selectlanguage{english}%
{\small{}0.0000}\selectlanguage{american}%
 & \selectlanguage{english}%
{\small{}0.0002}\selectlanguage{american}%
\tabularnewline
\selectlanguage{english}%
3\selectlanguage{american}%
 & \selectlanguage{english}%
{\small{}1.0000}\selectlanguage{american}%
 & \selectlanguage{english}%
{\small{}0.0077}\selectlanguage{american}%
 & \selectlanguage{english}%
{\small{}0.0198}\selectlanguage{american}%
 & \selectlanguage{english}%
{\small{}0.0000}\selectlanguage{american}%
 & \selectlanguage{english}%
{\small{}0.0000}\selectlanguage{american}%
 & \selectlanguage{english}%
{\small{}0.0003}\selectlanguage{american}%
\tabularnewline
\selectlanguage{english}%
{\small{}1}\selectlanguage{american}%
 & \selectlanguage{english}%
{\small{}1.0000}\selectlanguage{american}%
 & \selectlanguage{english}%
{\small{}0.0079}\selectlanguage{american}%
 & \selectlanguage{english}%
{\small{}0.0202}\selectlanguage{american}%
 & \selectlanguage{english}%
{\small{}0.0000}\selectlanguage{american}%
 & \selectlanguage{english}%
{\small{}0.0184}\selectlanguage{american}%
 & \selectlanguage{english}%
{\small{}0.0003}\selectlanguage{american}%
\tabularnewline
\selectlanguage{english}%
0.5\selectlanguage{american}%
 & \selectlanguage{english}%
{\small{}1.0000}\selectlanguage{american}%
 & \selectlanguage{english}%
{\small{}0.0082}\selectlanguage{american}%
 & \selectlanguage{english}%
{\small{}0.0206}\selectlanguage{american}%
 & \selectlanguage{english}%
{\small{}0.0000}\selectlanguage{american}%
 & \selectlanguage{english}%
{\small{}0.0375}\selectlanguage{american}%
 & \selectlanguage{english}%
{\small{}0.0003}\selectlanguage{american}%
\tabularnewline
\selectlanguage{english}%
{\small{}0.10}\selectlanguage{american}%
 & \selectlanguage{english}%
{\small{}1.0000}\selectlanguage{american}%
 & \selectlanguage{english}%
{\small{}0.0085}\selectlanguage{american}%
 & \selectlanguage{english}%
{\small{}0.0213}\selectlanguage{american}%
 & \selectlanguage{english}%
{\small{}0.0000}\selectlanguage{american}%
 & \selectlanguage{english}%
{\small{}0.0710}\selectlanguage{american}%
 & \selectlanguage{english}%
{\small{}0.0003}\selectlanguage{american}%
\tabularnewline
\hline 
\end{tabular}{\small\par}

\caption{Optimized monetary policy rules: robustness in the stylized model}

\label{Table:robust_Stylized}
\end{table}

\pagebreak{}

\section{Data \label{sec:Data}}

We define the durables sector as the a composite of durable goods
and residential investments whereas the nondurables sector comprises
nondurables goods and services.

\begin{table}[H]
\setlength\tabcolsep{2 pt}
\renewcommand{\arraystretch}{1}\centering%
\begin{tabular}{clcl}
\hline 
Series & Definition & Source & Mnemonic\tabularnewline
\hline 
{\small{}$DUR^{N}$} & {\small{}Nominal Durable Goods} & {\small{}BEA} & {\small{}Table 2.3.5 Line 3}\tabularnewline
{\small{}$RI^{N}$} & {\small{}Nominal Residential Investment} & {\small{}BEA} & {\small{}Table 1.1.5 Line 13}\tabularnewline
{\small{}$ND^{N}$} & {\small{}Nominal Nondurable Goods} & {\small{}BEA} & {\small{}Table 2.3.5 Line 8}\tabularnewline
{\small{}$S^{N}$} & {\small{}Nominal Services} & {\small{}BEA} & {\small{}Table 2.3.5 Line 13}\tabularnewline
{\small{}$P_{DUR}$} & {\small{}Price Deflator, Durable Goods} & {\small{}BEA} & {\small{}Table 1.1.9 Line 4}\tabularnewline
{\small{}$P_{RI}$} & {\small{}Price Deflator, Residential Investment} & {\small{}BEA} & {\small{}Table 1.1.9 Line 13}\tabularnewline
{\small{}$P_{ND}$} & {\small{}Price Deflator, Nondurable Goods} & {\small{}BEA} & {\small{}Table 1.1.9 Line 5}\tabularnewline
{\small{}$P_{S}$} & {\small{}Price Deflator, Services} & {\small{}BEA} & {\small{}Table 1.1.9 Line 6}\tabularnewline
{\small{}$Y^{N}$} & {\small{}Nominal GDP} & {\small{}BEA} & {\small{}Table 1.1.5 Line 1}\tabularnewline
{\small{}$P_{Y}$} & {\small{}Price Deflator, GDP} & {\small{}BEA} & {\small{}Table 1.1.9 Line 1}\tabularnewline
{\small{}$FFR$} & {\small{}Effective Federal Funds Rate} & {\small{}FRED} & {\small{}FEDFUNDS}\tabularnewline
{\small{}$N^{C}$} & {\small{}Average Weekly Hours: Nondurable Goods and Services} & {\small{}FRED} & {\small{}CES3200000007-CES0800000007}\tabularnewline
{\small{}$N^{D}$} & {\small{}Average Weekly Hours: Durable Goods and Construction} & {\small{}FRED} & {\small{}CES3100000007-CES2000000007}\tabularnewline
{\small{}$W^{C}$} & {\small{}Average Hourly Earnings: Nondurable Goods and Services} & {\small{}FRED} & {\small{}CES3200000008-CES0800000008}\tabularnewline
{\small{}$W^{D}$} & {\small{}Average Hourly Earnings: Durable Goods and Construction} & {\small{}FRED} & {\small{}CES3100000008-CES2000000008}\tabularnewline
{\small{}$POP$} & {\small{}Civilian Non-institutional Population, over 16} & {\small{}FRED} & {\small{}CNP16OV}\tabularnewline
{\small{}$CE$} & {\small{}Civilian Employment, 16 over} & {\small{}FRED} & {\small{}CE16OV}\tabularnewline
\hline 
\end{tabular}

\caption{Data Sources}
\end{table}

\subsection{Durables and Residential Investments}
\begin{enumerate}
\item Sum nominal series: $DUR^{N}+RI^{N}=DR^{N}$
\item Calculate sectoral weights of deflators: $\omega^{D}=\frac{DUR^{N}}{DR^{N}};\;\omega^{RI}=\frac{RI^{N}}{DR^{N}}$
\item Calculate Deflator: $P_{D}=\omega^{D}P_{DUR}+\omega^{RI}P_{RI}$
\item Calculate Real Durable Consumption: $D=\frac{DUR^{N}+RI^{N}}{P_{D}}$
\end{enumerate}

\subsection{Nondurables and Services}
\begin{enumerate}
\begin{singlespace}
\item Sum nominal series: $ND^{N}+S^{N}=NS^{N}$
\item Calculate sectoral weights of deflators: $\omega^{ND}=\frac{ND^{N}}{NS^{N}};\;\omega^{S}=\frac{S^{N}}{NS^{N}}$
\item Calculate Deflator: $P_{C}=\omega^{ND}P_{ND}+\omega^{S}P_{S}$
\item Calculate Real Nondurable Consumption: $C=\frac{ND^{N}+S^{N}}{P_{C}}$
\end{singlespace}
\end{enumerate}

\subsection{Data transformation for Bayesian estimation}

\begin{table}[H]
\setlength\tabcolsep{2 pt}
\renewcommand{\arraystretch}{1.5}\centering%
\begin{tabular}{clc}
\hline 
Variable & Description & Construction\tabularnewline
\hline 
$POP_{index}$ & Population index & {\tiny{}$\frac{POP}{POP_{2009:1}}$}\tabularnewline
$CE_{index}$ & Employment index & {\tiny{}$\frac{CE}{CE_{2009:1}}$}\tabularnewline
$Y^{o}$ & Real per capita GDP & {\tiny{}$\ln\left(\frac{\frac{Y^{N}}{P_{Y}}}{POP_{index}}\right)100$}\tabularnewline
$I_{D}^{o}$ & Real per capita consumption: durables & {\tiny{}$\ln\left(\frac{D}{POP_{index}}\right)100$}\tabularnewline
$C^{o}$ & Real per capita consumption: nondurables & {\tiny{}$\ln\left(\frac{C}{POP_{index}}\right)100$}\tabularnewline
$W^{o,j}$ & Real wage sector $j=C,D$ & {\tiny{}$\ln\left(\frac{W^{j}}{P_{Y}}\right)100$}\tabularnewline
$N^{o,j}$ & Hours worked per capita sector $j=C,D$ & {\tiny{}$\ln\left(\frac{H^{j}\times CE_{index}}{POP_{index}}\right)100$}\tabularnewline
$\Pi_{C}^{o}$ & Inflation: nondurables sector & {\tiny{}$\Delta\left(\ln P_{C}\right)100$}\tabularnewline
$\Pi_{D}^{o}$ & Inflation: durables sector & {\tiny{}$\Delta\left(\ln P_{D}\right)100$}\tabularnewline
$R^{o}$ & Quarterly Federal Funds Rate & {\tiny{}$\frac{FFR}{4}$}\tabularnewline
\hline 
\end{tabular}

\caption{Data transformation - Observables}
\end{table}

\section{Bayesian impulse responses in the estimated model\label{sec:Bayesian-impulse-responses}}

\begin{figure}[!t]
\centering\includegraphics[width=14cm,height=8cm]{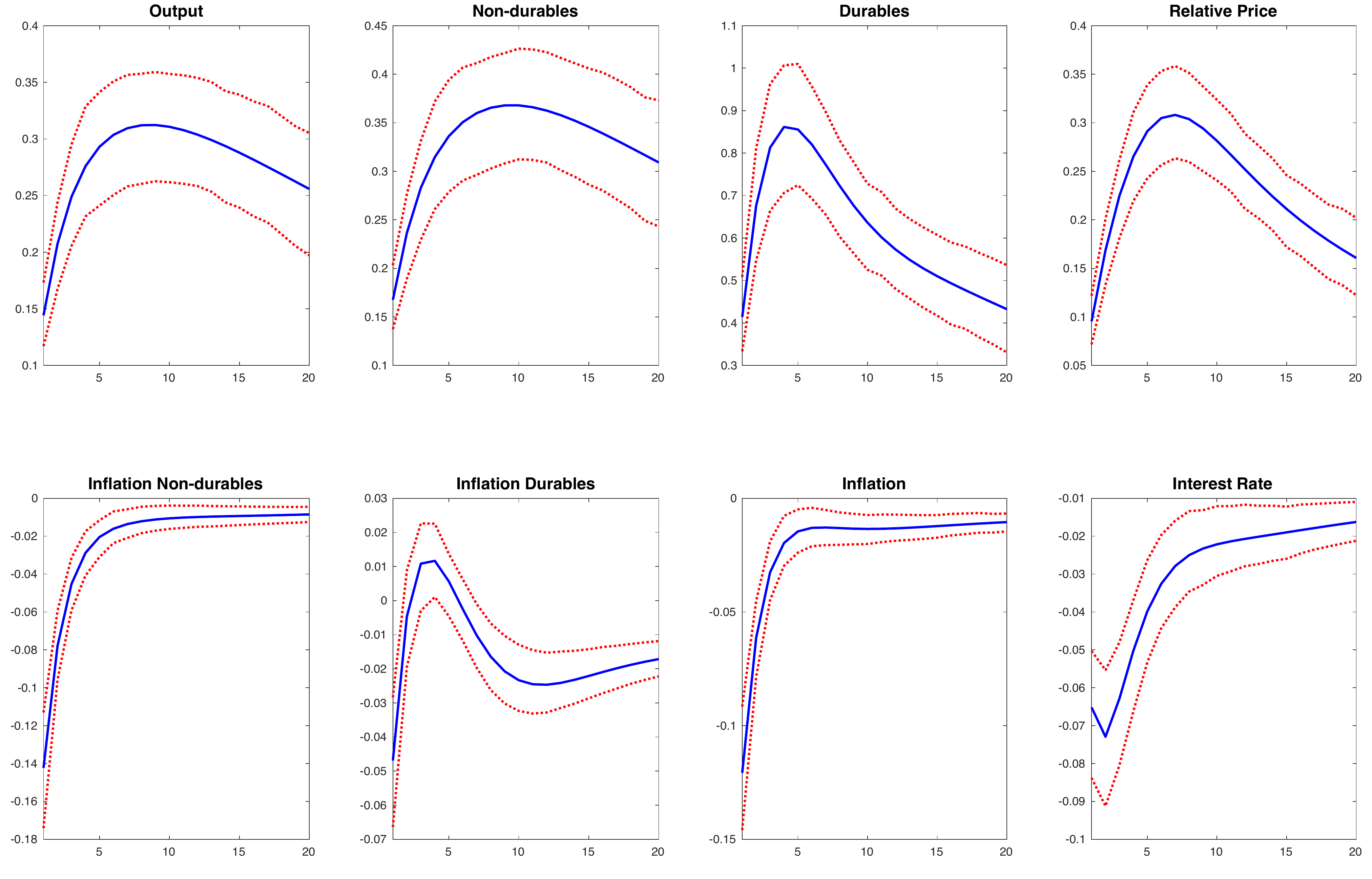}

\caption{\foreignlanguage{english}{\label{BIRFs}Bayesian impulse responses to aggregate technology shock.
Blue solid lines represent mean responses. Red dotted lines represent
90\% confidence bands.}}
\end{figure}

The estimated model exhibits well-behaved macroeconomic dynamics.
For instance, Figure \ref{BIRFs} shows that Bayesian impulse responses
of selected macroeconomic variables to an aggregate positive technology
shock are in line with the dynamics of standard models (see e.g. Kim
and Katayama, 2013, for an example of a two-sector model). Labor productivity
increases in both sectors, thus implying an expansion of sectoral
production and aggregate output, which leads to a decline in sectoral
and aggregate inflation to which the central bank responds by cutting
the interest rate. Responses to the other shocks are likewise standard
and are available upon request.

\section{The role of durability\label{sec:The-role-of-durables}}

Erceg and Levin (2006) show why durable goods are particularly important
for optimal monetary policy. In general, in a two sector model, following
a sector-specific shock, demand in the two sectors moves in opposite
direction. The central bank should therefore increase the interest
rate to stabilize the output gap in one sector while decreasing it
to stabilize the output gap in the other sector. This trade-off is
particularly severe when one sector produces durable goods for two
reasons. First, the demand for durables is for a stock, so also small
changes in the demand for the stock generate large changes in the
flow of newly produced durables. Then, the presence of sectoral price
stickiness prevents prices from adjusting and insulate the durables
sector from the shocks. Together, these two intrinsic features imply
that durables are much more sensitive to the interest rate than nondurables.
Therefore the same magnitude of the interest rate change generates
a larger response of output in the durables sector, hence the more
severe trade-off. To see this, we follow the reasoning made by Erceg
and Levin (2006). The asset price equation of durables (21) requires
that the marginal rate of substitution between durables and nondurables
$\frac{U_{D,t}}{U_{C,t}}=\frac{\alpha}{1-\alpha}\frac{C_{t}}{D_{t}}$
equals the \textit{user cost} of durable goods $\Theta_{t}$:
\begin{equation}
\frac{U_{D,t}}{U_{C,t}}=\Theta_{t}\equiv Q_{t}-\beta\left(1-\delta\right)E_{t}\left[\frac{U_{C,t+1}}{U_{C,t}}Q_{t+1}\right],\label{assetprice eq}
\end{equation}
which implies that 
\begin{equation}
D_{t}=\frac{\alpha}{1-\alpha}\frac{C_{t}}{\Theta_{t}},
\end{equation}
or in log-linear form:
\begin{equation}
\hat{D}_{t}=\hat{C}_{t}-\hat{\Theta}_{t}.\label{MRSloglin}
\end{equation}
Log-linearizing also the user cost of durables \eqref{assetprice eq}
and the Euler equation (4) around the steady state yields, respectively
\begin{align}
\hat{\Theta}_{t} & =\frac{\hat{Q}_{t}-\left(1-\delta\right)\beta E_{t}\left[\hat{U}_{C,t}-\hat{U}_{C,t+1}-\hat{Q}_{t+1}\right]}{1-\left(1-\delta\right)\beta},\label{User cost-loglin}\\
\hat{U}_{C,t}-\hat{U}_{C,t+1} & =\hat{R}_{t}-E_{t}\hat{\Pi}_{t+1}^{C},
\end{align}
combining which yields
\begin{equation}
\hat{\Theta}_{t}=\frac{\hat{Q}_{t}-\left(1-\delta\right)\beta E_{t}\left[\hat{R}_{r,t}-\hat{Q}_{t+1}\right]}{1-\left(1-\delta\right)\beta},\label{User cost-loglin-1}
\end{equation}
where $\hat{R}_{r,t}=\hat{R}_{t}-E_{t}\hat{\Pi}_{t+1}^{C}$ is the
real interest rate. Equation \eqref{User cost-loglin-1} shows that
the user cost of durables depends on the relative price and the real
interest rate. When prices are sticky, the relative price will adjust
slowly to shocks so that the user cost and hence, for a sufficiently
low depreciation rate $\delta$, the demand of durables is very sensitive
to the real interest rate. Note also that when there is no durability
($\delta=1$), the output gap in the two sectors is entirely determined
by the relative price. Finally, rearranging \eqref{User cost-loglin-1}
yields equation (28).

\section{Robustness analysis in the fully-fledged model\label{subsec:Robustness-to-alternative}}

In this section we perform several robustness checks. We first look
at the role of sectoral shocks, nominal and real frictions, and the
depreciation rate of durables. Then, we replace the monetary policy
rule (13) with alternative rules and compare the results with the
baseline model (top panel of Table 4). Our main findings continue
to hold under all the robustness checks.
\begin{table}[!t]
\setlength\tabcolsep{6 pt}
\renewcommand{\arraystretch}{1.25}

\centering%
\begin{tabular}{ccccccc}
\hline 
{\footnotesize{}$\lambda$} & {\footnotesize{}$\rho_{r}$} & {\footnotesize{}$\alpha_{\pi}$} & {\footnotesize{}$\alpha_{y}$} & {\footnotesize{}$\alpha_{\Delta y}$} & {\footnotesize{}$\tau$} & {\footnotesize{}$100\times\omega$}\tabularnewline
\hline 
\multicolumn{7}{c}{\textit{\footnotesize{}Excluding price markup shocks in nondurables}}\tabularnewline
{\footnotesize{}$\infty$} & {\footnotesize{}0.7586} & {\footnotesize{}0.7578} & {\footnotesize{}0.0000} & {\footnotesize{}0.0000} & {\footnotesize{}0.0000} & {\footnotesize{}0.0091}\tabularnewline
{\footnotesize{}1.2250} & {\footnotesize{}0.2087} & {\footnotesize{}2.2056} & {\footnotesize{}0.0000} & {\footnotesize{}0.0000} & {\footnotesize{}0.0771} & {\footnotesize{}0.0633}\tabularnewline
{\footnotesize{}0.1} & {\footnotesize{}0.8920} & {\footnotesize{}1.2426} & {\footnotesize{}0.0011} & {\footnotesize{}0.0000} & {\footnotesize{}0.6090} & {\footnotesize{}0.2345}\tabularnewline
\multicolumn{7}{c}{\textit{\footnotesize{}Excluding price markup shocks in durables}}\tabularnewline
{\footnotesize{}$\infty$} & {\footnotesize{}0.0237} & {\footnotesize{}2.3965} & {\footnotesize{}0.0000} & {\footnotesize{}0.0000} & {\footnotesize{}0.0935} & {\footnotesize{}0.0744}\tabularnewline
{\footnotesize{}1.2250} & {\footnotesize{}0.4514} & {\footnotesize{}1.2501} & {\footnotesize{}0.0000} & {\footnotesize{}0.1539} & {\footnotesize{}0.2387} & {\footnotesize{}0.1165}\tabularnewline
{\footnotesize{}0.1} & {\footnotesize{}1.0000} & {\footnotesize{}1.2106} & {\footnotesize{}0.0026} & {\footnotesize{}0.0000} & {\footnotesize{}0.7451} & {\footnotesize{}0.0551}\tabularnewline
\multicolumn{7}{c}{\textit{\footnotesize{}Excluding wage markup shocks in nondurables}}\tabularnewline
{\footnotesize{}$\infty$} & {\footnotesize{}1.0000} & {\footnotesize{}0.0672} & {\footnotesize{}0.0003} & {\footnotesize{}0.0702} & {\footnotesize{}0.2000} & {\footnotesize{}0.0656}\tabularnewline
{\footnotesize{}1.2250} & {\footnotesize{}1.0000} & {\footnotesize{}0.2849} & {\footnotesize{}0.0000} & {\footnotesize{}0.3744} & {\footnotesize{}0.2212} & {\footnotesize{}0.0506}\tabularnewline
{\footnotesize{}0.1} & {\footnotesize{}0.9666} & {\footnotesize{}0.7299} & {\footnotesize{}0.0013} & {\footnotesize{}0.3498} & {\footnotesize{}0.9904} & {\footnotesize{}0.1454}\tabularnewline
\multicolumn{7}{c}{\textit{\footnotesize{}Excluding wage markup shocks in durables}}\tabularnewline
{\footnotesize{}$\infty$} & {\footnotesize{}0.0409} & {\footnotesize{}2.4523} & {\footnotesize{}0.0000} & {\footnotesize{}0.5041} & {\footnotesize{}0.0021} & {\footnotesize{}0.1044}\tabularnewline
{\footnotesize{}1.2250} & {\footnotesize{}0.7124} & {\footnotesize{}0.5995} & {\footnotesize{}0.0000} & {\footnotesize{}0.2049} & {\footnotesize{}0.1193} & {\footnotesize{}0.1454}\tabularnewline
{\footnotesize{}0.1} & {\footnotesize{}0.9141} & {\footnotesize{}0.8887} & {\footnotesize{}0.0014} & {\footnotesize{}0.0000} & {\footnotesize{}0.7726} & {\footnotesize{}0.2755}\tabularnewline
\multicolumn{7}{c}{\textit{\footnotesize{}Excluding government spending shocks}}\tabularnewline
{\footnotesize{}$\infty$} & {\footnotesize{}0.0004} & {\footnotesize{}2.3028} & {\footnotesize{}0.0000} & {\footnotesize{}0.0352} & {\footnotesize{}0.0189} & {\footnotesize{}0.0886}\tabularnewline
{\footnotesize{}1.2250} & {\footnotesize{}0.4759} & {\footnotesize{}1.0841} & {\footnotesize{}0.0000} & {\footnotesize{}0.2577} & {\footnotesize{}0.1481} & {\footnotesize{}0.1353}\tabularnewline
{\footnotesize{}0.1} & {\footnotesize{}0.9164} & {\footnotesize{}0.8832} & {\footnotesize{}0.0014} & {\footnotesize{}0.0000} & {\footnotesize{}0.7725} & {\footnotesize{}0.2761}\tabularnewline
\multicolumn{7}{c}{\textit{\footnotesize{}Excluding durables investment specific shocks}}\tabularnewline
{\footnotesize{}$\infty$} & {\footnotesize{}0.0005} & {\footnotesize{}2.3139} & {\footnotesize{}0.0000} & {\footnotesize{}0.0411} & {\footnotesize{}0.0179} & {\footnotesize{}0.0882}\tabularnewline
{\footnotesize{}1.2250} & {\footnotesize{}0.3614} & {\footnotesize{}1.2647} & {\footnotesize{}0.0000} & {\footnotesize{}0.2239} & {\footnotesize{}0.1652} & {\footnotesize{}0.1264}\tabularnewline
{\footnotesize{}0.1} & {\footnotesize{}0.3086} & {\footnotesize{}0.9622} & {\footnotesize{}0.0000} & {\footnotesize{}0.0000} & {\footnotesize{}0.6721} & {\footnotesize{}0.1477}\tabularnewline
\hline 
\end{tabular}

\caption{Optimized monetary policy rules: robustness to the absence of sectoral
shocks}

\label{Optimal_shocks}
\end{table}

\subsection{Sectoral shocks\label{subsec:Sectoral-shocks}}

Our model includes both aggregate (or symmetric) and sector-specific
shocks. In particular, technology and preference shocks fall in the
former category, while durables investment, nondurables and durables
price markup and wage markup and government spending shocks fall in
the latter category. In multi-sector models, aggregate shocks typically
generate a comovement across sector thus inducing little labor reallocation.\footnote{The sectoral comovement in response to aggregate shocks is evident
in our model from Figure \ref{BIRFs} (Appendix \ref{sec:Bayesian-impulse-responses}),
where we consider an economy-wide technology shock and, more generally,
in the vast literature on the sectoral responses to a monetary policy
innovation (see Cantelmo and Melina, 2018, for a detailed review).} Conversely, sectoral disturbances have the potential to generate
larger labor reallocation since demand or supply in different sectors
move in opposite direction. It is therefore natural to assess whether
the inverse relationship between the optimal weight on durables inflation
and labor mobility is driven by any specific sectoral disturbance.
We thus eliminate each sectoral shock one at a time and verify that
our results still hold. Table \ref{Optimal_shocks} shows that our
findings do not hinge on a specific sectoral disturbance. In each
case, the weight placed on durables inflation is inversely related
to the degree of labor mobility, while welfare losses are comparable
to the baseline results. As already noted in Section 4.3, the price
markup shock in the durables sector matters only for the magnitude
of the welfare loss, but not for the inverse relationship between
labor mobility and the optimal durables inflation weight.

\subsection{Nominal rigidities\label{subsec:Nominal-rigidities}}

\begin{table}[t]
\setlength\tabcolsep{5 pt}
\renewcommand{\arraystretch}{1.25}

\centering%
\begin{tabular}{ccccccc}
\hline 
{\footnotesize{}$\lambda$} & {\footnotesize{}$\rho_{r}$} & {\footnotesize{}$\alpha_{\pi}$} & {\footnotesize{}$\alpha_{y}$} & {\footnotesize{}$\alpha_{\Delta y}$} & {\footnotesize{}$\tau$} & {\footnotesize{}$100\times\omega$}\tabularnewline
\hline 
\multicolumn{7}{c}{\textit{\footnotesize{}Flexible wages in durables sector}}\tabularnewline
{\footnotesize{}$\infty$} & {\footnotesize{}0.0254} & {\footnotesize{}0.3397} & {\footnotesize{}0.0000} & {\footnotesize{}0.0000} & {\footnotesize{}0.0000} & {\footnotesize{}0.0722}\tabularnewline
{\footnotesize{}1.2250} & {\footnotesize{}0.2229} & {\footnotesize{}1.6563} & {\footnotesize{}0.0000} & {\footnotesize{}0.2437} & {\footnotesize{}0.0617} & {\footnotesize{}0.1151}\tabularnewline
{\footnotesize{}0.1} & {\footnotesize{}0.9832} & {\footnotesize{}0.1289} & {\footnotesize{}0.0001} & {\footnotesize{}0.0000} & {\footnotesize{}0.5143} & {\footnotesize{}0.1220}\tabularnewline
\multicolumn{7}{c}{\textit{\footnotesize{}Flexible prices and wages in durables sector}}\tabularnewline
{\footnotesize{}$\infty$} & {\footnotesize{}0.0370} & {\footnotesize{}2.3748} & {\footnotesize{}0.0000} & {\footnotesize{}0.0000} & {\footnotesize{}0.0000} & {\footnotesize{}0.0742}\tabularnewline
{\footnotesize{}1.2250} & {\footnotesize{}0.2204} & {\footnotesize{}1.6011} & {\footnotesize{}0.0000} & {\footnotesize{}0.1471} & {\footnotesize{}0.0000} & {\footnotesize{}0.1150}\tabularnewline
{\footnotesize{}0.1} & {\footnotesize{}0.7326} & {\footnotesize{}0.5162} & {\footnotesize{}0.0000} & {\footnotesize{}0.1901} & {\footnotesize{}0.0589} & {\footnotesize{}0.1491}\tabularnewline
\hline 
\end{tabular}

\caption{\foreignlanguage{english}{Optimized monetary policy rules: robustness to nominal rigidities}}

\label{Optimal_1-2-3-2}
\end{table}

We next verify whether our results still hold in counterfactual economies
without nominal rigidities in prices and nominal wages in the durables
sector, although the estimation suggests that both are substantially
sticky. The top panel of Table \ref{Optimal_1-2-3-2} shows the case
of flexible wages ($\vartheta_{d}^{w}=0$) whereas in the lower panel
both durables prices and wages are flexible ($\vartheta_{d}=\vartheta_{d}^{w}=0$).
Relative to the baseline model, at the estimated limited degree of
labor mobility, the optimal weight on durables inflation drops as
wages become flexible in the durables sector ($\tau$ falls from 0.15
to 0.0617) and it becomes zero as both nominal frictions are removed.
Nevertheless, a sufficiently low degree of labor mobility (e.g. $\lambda=0.10$)
still entails a positive weight on durables inflation both with flexible
wages and sticky prices ($\tau=0.5143$) and with both flexible wages
and prices ($\tau=0.0589)$, meaning that imperfect sectoral labor
mobility creates scope for a positive weight on durables inflation
even if nominal rigidities are absent in that sector. Overall, the
main conclusions drawn in the previous section are carried over with
these two counterfactual economies: i) $\tau$ and $\lambda$ are
negatively related, hence a higher weight is assigned to durables
inflation as labor becomes less mobile across sectors; ii) the interaction
between labor mobility and wage stickiness is key in that sticky wages
and limited labor mobility entails a higher weight on durables inflation,
but flexible wages alone do not necessarily imply a zero weight on
durables inflation if labor is sufficiently non-mobile.

\subsection{Real frictions\label{subsec:Real-frictions}}

The model employed in this paper features two sources of real frictions,
important to bring it to the data. In particular, households display
habit formation in consumption of nondurable goods, while changing
investment plans in durables goods entails a quadratic cost. In this
section we verify that the inverse relationship between $\lambda$
and $\tau$ continues to hold in restricted models in which we remove
one real friction at a time. Table \ref{Optimal_real frictions} demonstrates
that all the results are robust both to a calibration of the model
which excludes habits in nondurables consumption $(\zeta=\rho_{c}=0)$,
and to a model without investment adjustment costs in durables $\left(\phi=0\right)$,
cases in which the inverse relationship between the optimal weight
on durables and sectoral labor mobility still exists.

\begin{table}[!t]
\setlength\tabcolsep{5 pt}
\renewcommand{\arraystretch}{1.5}

\centering%
\begin{tabular}{ccccccc}
\hline 
{\footnotesize{}$\lambda$} & {\footnotesize{}$\rho_{r}$} & {\footnotesize{}$\alpha_{\pi}$} & {\footnotesize{}$\alpha_{y}$} & {\footnotesize{}$\alpha_{\Delta y}$} & {\footnotesize{}$\tau$} & {\footnotesize{}$100\times\omega$}\tabularnewline
\hline 
\multicolumn{7}{c}{\textit{\footnotesize{}Excluding habit formation in nondurables consumption}}\tabularnewline
{\footnotesize{}$\infty$} & {\footnotesize{}0.1588} & {\footnotesize{}2.5457} & {\footnotesize{}0.0000} & {\footnotesize{}0.0000} & {\footnotesize{}0.0000} & {\footnotesize{}0.0165}\tabularnewline
{\footnotesize{}1.2250} & {\footnotesize{}0.1988} & {\footnotesize{}1.6282} & {\footnotesize{}0.0000} & {\footnotesize{}0.0000} & {\footnotesize{}0.1636} & {\footnotesize{}0.0668}\tabularnewline
{\footnotesize{}0.10} & {\footnotesize{}0.8821} & {\footnotesize{}1.8159} & {\footnotesize{}0.0014} & {\footnotesize{}0.9923} & {\footnotesize{}0.4430} & {\footnotesize{}0.1257}\tabularnewline
\multicolumn{7}{c}{\textit{\footnotesize{}Excluding investment adjustment costs in durables}}\tabularnewline
{\footnotesize{}$\infty$} & {\footnotesize{}1.0000} & {\footnotesize{}2.0050} & {\footnotesize{}0.0033} & {\footnotesize{}1.2715} & {\footnotesize{}0.1859} & {\footnotesize{}0.1775}\tabularnewline
{\footnotesize{}1.2250} & {\footnotesize{}0.8790} & {\footnotesize{}5.0000} & {\footnotesize{}0.0069} & {\footnotesize{}2.9205} & {\footnotesize{}0.3918} & {\footnotesize{}0.2463}\tabularnewline
{\footnotesize{}0.10} & {\footnotesize{}0.5083} & {\footnotesize{}5.0000} & {\footnotesize{}0.0004} & {\footnotesize{}0.0000} & {\footnotesize{}1.0000} & {\footnotesize{}0.8665}\tabularnewline
\hline 
\end{tabular}

\caption{Optimized monetary policy rule: robustness to the absence of real
friction\foreignlanguage{english}{s}}

\label{Optimal_real frictions}
\end{table}

\subsection{Depreciation rate of durable goods\label{subsec:Depreciation-rate-of}}

Our baseline calibration, inspired by previous studies, assumes a
1\% quarterly depreciation rate of durable goods. Barsky et al. (2016)
study optimal monetary policy in a two-sector economy with durable
goods, with price stickiness as the only source of nominal rigidity,
no real frictions and a smaller set of shocks, and show how the optimal
weight on durables inflation is affected by the depreciation rate
of durables. We therefore check the robustness of our findings to
alternative rates of depreciation of durable goods. Table \ref{Optimal_delta}
reports the optimized parameters and welfare losses under higher (quarterly)
depreciation rates than that assumed in the baseline calibration.
We find that for higher depreciation rates, and even if durables would
fully depreciate each quarter $(\delta\rightarrow1)$, the inverse
relationship between the optimal weight on durables inflation and
sectoral labor mobility survives.

\begin{table}[!t]
\setlength\tabcolsep{5 pt}
\renewcommand{\arraystretch}{1.5}

\centering%
\begin{tabular}{ccccccc}
\hline 
{\footnotesize{}$\lambda$} & {\footnotesize{}$\rho_{r}$} & {\footnotesize{}$\alpha_{\pi}$} & {\footnotesize{}$\alpha_{y}$} & {\footnotesize{}$\alpha_{\Delta y}$} & {\footnotesize{}$\tau$} & {\footnotesize{}$100\times\omega$}\tabularnewline
\hline 
\multicolumn{7}{c}{\textit{\footnotesize{}$\delta=0.025$}}\tabularnewline
{\footnotesize{}$\infty$} & {\footnotesize{}0.7098} & {\footnotesize{}0.6090} & {\footnotesize{}0.0000} & {\footnotesize{}0.1303} & {\footnotesize{}0.0125} & {\footnotesize{}0.1908}\tabularnewline
{\footnotesize{}1.2250} & {\footnotesize{}0.7084} & {\footnotesize{}0.5799} & {\footnotesize{}0.0000} & {\footnotesize{}0.1977} & {\footnotesize{}0.1358} & {\footnotesize{}0.1627}\tabularnewline
{\footnotesize{}0.1} & {\footnotesize{}0.9070} & {\footnotesize{}1.0141} & {\footnotesize{}0.0016} & {\footnotesize{}0.1066} & {\footnotesize{}0.7638} & {\footnotesize{}0.4252}\tabularnewline
\multicolumn{7}{c}{\textit{\footnotesize{}$\delta=0.10$}}\tabularnewline
{\footnotesize{}$\infty$} & {\footnotesize{}0.9962} & {\footnotesize{}0.0133} & {\footnotesize{}0.0000} & {\footnotesize{}0.0000} & {\footnotesize{}0.0137} & {\footnotesize{}0.0994}\tabularnewline
{\footnotesize{}1.2250} & {\footnotesize{}0.7538} & {\footnotesize{}0.5580} & {\footnotesize{}0.0000} & {\footnotesize{}0.2736} & {\footnotesize{}0.0877} & {\footnotesize{}0.1701}\tabularnewline
{\footnotesize{}0.1} & {\footnotesize{}0.7558} & {\footnotesize{}2.1838} & {\footnotesize{}0.0016} & {\footnotesize{}0.9159} & {\footnotesize{}0.5698} & {\footnotesize{}0.4924}\tabularnewline
\multicolumn{7}{c}{\textit{\footnotesize{}$\delta\rightarrow1$}}\tabularnewline
{\footnotesize{}$\infty$} & {\footnotesize{}0.7425} & {\footnotesize{}0.5619} & {\footnotesize{}0.0000} & {\footnotesize{}0.2548} & {\footnotesize{}0.0289} & {\footnotesize{}0.1770}\tabularnewline
{\footnotesize{}1.2250} & {\footnotesize{}0.7469} & {\footnotesize{}0.5516} & {\footnotesize{}0.0000} & {\footnotesize{}0.2665} & {\footnotesize{}0.0523} & {\footnotesize{}0.1765}\tabularnewline
{\footnotesize{}0.1} & {\footnotesize{}0.7045} & {\footnotesize{}0.5617} & {\footnotesize{}0.0000} & {\footnotesize{}0.2765} & {\footnotesize{}0.3143} & {\footnotesize{}0.3170}\tabularnewline
\hline 
\end{tabular}

\caption{\foreignlanguage{english}{Optimized monetary policy rule: robustness to alternative depreciation
rates of durables}}

\label{Optimal_delta}
\end{table}

\subsection{Alternative interest rate rules\label{subsec:Alternative-interest-rate}}

\begin{table}[!t]
\setlength\tabcolsep{5 pt}
\renewcommand{\arraystretch}{1.5}

\centering%
\begin{tabular}{cccccccc}
\hline 
{\footnotesize{}$\lambda$} & {\footnotesize{}$\rho_{r}$} & {\footnotesize{}$\alpha_{\pi}$} & {\footnotesize{}$\alpha_{y}$} & {\footnotesize{}$\alpha_{\Delta y}$} & {\footnotesize{}$\alpha_{w}$} & {\footnotesize{}$\tau$} & {\footnotesize{}$100\times\omega$}\tabularnewline
\hline 
\multicolumn{8}{c}{\textit{\footnotesize{}Implementable rule}}\tabularnewline
{\footnotesize{}$\infty$} & {\footnotesize{}0.0004} & {\footnotesize{}2.3170} & {\footnotesize{}0.0000} & {\footnotesize{}/} & {\footnotesize{}/} & {\footnotesize{}0.0177} & {\footnotesize{}0.0868}\tabularnewline
{\footnotesize{}1.2250} & {\footnotesize{}0.2358} & {\footnotesize{}1.5019} & {\footnotesize{}0.0000} & {\footnotesize{}/} & {\footnotesize{}/} & {\footnotesize{}0.1579} & {\footnotesize{}0.1264}\tabularnewline
{\footnotesize{}0.1} & {\footnotesize{}0.9152} & {\footnotesize{}0.8904} & {\footnotesize{}0.0016} & {\footnotesize{}/} & {\footnotesize{}/} & {\footnotesize{}0.7729} & {\footnotesize{}0.2753}\tabularnewline
\multicolumn{8}{c}{\textit{\footnotesize{}Wage inflation}}\tabularnewline
{\footnotesize{}$\infty$} & {\footnotesize{}0.0667} & {\footnotesize{}2.0806} & {\footnotesize{}0.0000} & {\footnotesize{}0.0000} & {\footnotesize{}0.3821} & {\footnotesize{}0.0000} & {\footnotesize{}0.0893}\tabularnewline
{\footnotesize{}1.2250} & {\footnotesize{}0.8426} & {\footnotesize{}0.7527} & {\footnotesize{}0.0002} & {\footnotesize{}0.0740} & {\footnotesize{}0.4096} & {\footnotesize{}0.1407} & {\footnotesize{}0.1121}\tabularnewline
{\footnotesize{}0.1} & {\footnotesize{}1.0000} & {\footnotesize{}0.6682} & {\footnotesize{}0.0015} & {\footnotesize{}0.0000} & {\footnotesize{}0.4038} & {\footnotesize{}0.6441} & {\footnotesize{}0.0183}\tabularnewline
\multicolumn{8}{c}{\textit{\footnotesize{}Real wage growth}}\tabularnewline
{\footnotesize{}$\infty$} & {\footnotesize{}0.5931} & {\footnotesize{}0.9068} & {\footnotesize{}0.0000} & {\footnotesize{}0.0000} & {\footnotesize{}0.2061} & {\footnotesize{}0.0000} & {\footnotesize{}0.1401}\tabularnewline
{\footnotesize{}1.2250} & {\footnotesize{}1.0000} & {\footnotesize{}2.9147} & {\footnotesize{}0.0020} & {\footnotesize{}0.9829} & {\footnotesize{}0.8763} & {\footnotesize{}0.0559} & {\footnotesize{}0.1071}\tabularnewline
{\footnotesize{}0.1} & {\footnotesize{}1.0000} & {\footnotesize{}1.0715} & {\footnotesize{}0.0015} & {\footnotesize{}0.0000} & {\footnotesize{}0.4037} & {\footnotesize{}0.4017} & {\footnotesize{}0.0183}\tabularnewline
\multicolumn{8}{c}{\textit{\footnotesize{}Real sectoral wage growth differential}}\tabularnewline
{\footnotesize{}$\infty$} & {\footnotesize{}/} & {\footnotesize{}/} & {\footnotesize{}/} & {\footnotesize{}/} & {\footnotesize{}/} & {\footnotesize{}/} & {\footnotesize{}/}\tabularnewline
{\footnotesize{}1.2250} & {\footnotesize{}0.6612} & {\footnotesize{}0.8592} & {\footnotesize{}0.0000} & {\footnotesize{}0.3209} & {\footnotesize{}0.0231} & {\footnotesize{}0.1243} & {\footnotesize{}0.1434}\tabularnewline
{\footnotesize{}0.1} & {\footnotesize{}1.0000} & {\footnotesize{}0.7169} & {\footnotesize{}0.0016} & {\footnotesize{}0.0000} & {\footnotesize{}0.1095} & {\footnotesize{}0.6200} & {\footnotesize{}0.1167}\tabularnewline
\hline 
\end{tabular}

\caption{\foreignlanguage{english}{Robustness to alternative optimized monetary policy rule}}

\label{Optimal_1-2-3-1-1}
\end{table}

\paragraph*{Implementable rules.}

We replace rule (13) with an interest rate rule that responds only
to deviations of inflation and output from their respective steady
states. Following Schmitt- Grohe and Uribe (2007) this type of interest
rate rule is typically labeled as \textit{implementable rule} and
reads as follows:

\begin{equation}
\log\left(\frac{R_{t}}{\bar{R}}\right)=\rho_{r}\log\left(\frac{R_{t-1}}{\bar{R}}\right)+\alpha_{\pi}\log\left(\frac{\tilde{\Pi}_{t}}{\tilde{\Pi}}\right)+\alpha_{y}\log\left(\frac{Y_{t}}{\bar{Y}}\right).
\end{equation}
The top panel of Table \ref{Optimal_1-2-3-1-1} demonstrates that
despite these modifications, the inverse relationship between labor
mobility and the optimal weight on durables inflation still hold true.
In addition, the implied welfare losses are similar to the baseline
model.

\paragraph*{Responding to wages.}

Erceg and Levin (2006) find that rules targeting the output gap or
a weighted average of price and wage inflation represent good approximations
of the optimal rule. We therefore check whether the inverse relationship
between labor mobility and the optimal weight on durables inflation
continues to hold under interest rate rules that respond to wages.

We start by closely following Erceg and Levin (2006) by adding a term
to the interest rate rule \eqref{SW Rule-1-1-2-1} that responds to
nominal wage inflation and optimize $\alpha_{w}\in\left[0,5\right]$
along with the other policy parameters and the weight on durables
inflation.\footnote{Following Iacoviello and Neri (2010), we define an aggregate wage
index $W_{t}=\left(\left(W_{t}^{C}\right)^{\frac{1+\lambda}{\lambda}}+\left(W_{t}^{D}\right)^{\frac{1+\lambda}{\lambda}}\right)^{\frac{\lambda}{1+\lambda}}$
and wage inflation as $\Pi_{t}^{w}=\frac{W_{t}}{W_{t-1}}\Pi_{t}^{C}$.} In accordance with the findings in Erceg and Levin (2006), the second
panel of Table \ref{Optimal_1-2-3-1-1} shows that responding to wage
inflation is welfare enhancing.
\begin{eqnarray}
\log\left(\frac{R_{t}}{\bar{R}}\right) & = & \rho_{r}\log\left(\frac{R_{t-1}}{\bar{R}}\right)+\alpha_{\pi}\log\left(\frac{\tilde{\Pi}_{t}}{\tilde{\Pi}}\right)+\alpha_{w}\log\left(\frac{\Pi_{t}^{w}}{\Pi^{w}}\right)+\nonumber \\
 & + & \alpha_{y}\log\left(\frac{Y_{t}}{Y_{t}^{f}}\right)+\alpha_{\Delta y}\left[\log\left(\frac{Y_{t}}{Y_{t}^{f}}\right)-\log\left(\frac{Y_{t-1}}{Y_{t-1}^{f}}\right)\right].\label{SW Rule-1-1-2-1}
\end{eqnarray}

We then take a step further, following Faia (2008), by assessing whether
responding to real, rather than nominal, wage growth is welfare enhancing
in our model. Specifically we add a term to the interest rate rule
(13) that responds to real wage growth:
\begin{eqnarray}
\log\left(\frac{R_{t}}{\bar{R}}\right) & = & \rho_{r}\log\left(\frac{R_{t-1}}{\bar{R}}\right)+\alpha_{\pi}\log\left(\frac{\tilde{\Pi}_{t}}{\tilde{\Pi}}\right)+\alpha_{w}\log\left(\frac{w_{t}}{w_{t-1}}\right)\nonumber \\
 & + & \alpha_{y}\log\left(\frac{Y_{t}}{Y_{t}^{f}}\right)+\alpha_{\Delta y}\left[\log\left(\frac{Y_{t}}{Y_{t}^{f}}\right)-\log\left(\frac{Y_{t-1}}{Y_{t-1}^{f}}\right)\right].\label{SW Rule-1-1}
\end{eqnarray}
 The third panel of Table \ref{Optimal_1-2-3-1-1} shows that responding
to real wage growth slightly improves welfare relative to responding
to wage inflation for limited degrees of labor mobility.

Finally, the last check we perform is optimizing a monetary rule that
embeds a response to the change in the relative wage across sectors,
so that equation \eqref{SW Rule-1-1} becomes:
\begin{eqnarray}
\log\left(\frac{R_{t}}{\bar{R}}\right) & = & \rho_{r}\log\left(\frac{R_{t-1}}{\bar{R}}\right)+\alpha_{\pi}\log\left(\frac{\tilde{\Pi}_{t}}{\tilde{\Pi}}\right)+\alpha_{w}\left[\log\left(\frac{w_{t}^{d}}{w_{t}^{c}}\right)-\log\left(\frac{w_{t-1}^{d}}{w_{t-1}^{c}}\right)\right]\nonumber \\
 & + & \alpha_{y}\log\left(\frac{Y_{t}}{Y_{t}^{f}}\right)+\alpha_{\Delta y}\left[\log\left(\frac{Y_{t}}{Y_{t}^{f}}\right)-\log\left(\frac{Y_{t-1}}{Y_{t-1}^{f}}\right)\right].\label{SW Rule-1-1-1}
\end{eqnarray}
We only consider cases of limited labor mobility as, with perfect
labor mobility, wages in the two sectors are always the same by construction
and the interest rate rule \eqref{SW Rule-1-1-1} collapses to the
rule (13) studied in the main analysis. It turns out that it is optimal
for the central bank to respond to some extent to the change in the
wage differential.

Crucially, the main result on the negative relationship between sectoral
labor mobility and the optimal weight on durables inflation survives
in all cases considered.
\end{document}